\documentclass[fleqn,usenatbib]{mnras_draft}

\usepackage{newtxtext,newtxmath}

\usepackage[T1]{fontenc}
\usepackage{ae,aecompl}

\usepackage{graphicx}	
\usepackage{amsmath}	
\usepackage{amssymb}	
\usepackage[dvipsnames]{xcolor} 
\usepackage{ulem}

\newcommand{\p}{\partial}
\newcommand{\ccred}{black}
\newcommand{\ccorange}{black}
\usepackage{bm}

\makeatletter
\def\Left#1#2\Right{\begingroup%
  \def\ts@r{\nulldelimiterspace=0pt \mathsurround=0pt}%
  \let\@hat=#1%
  \def\sht@im{#2}%
  \def\@t{{\mathchoice{\def\@fen{\displaystyle}\k@fel}%
      {\def\@fen{\textstyle}\k@fel}%
      {\def\@fen{\scriptstyle}\k@fel}%
      {\def\@fen{\scriptscriptstyle}\k@fel}}}%
  \def\g@rin{\ts@r\left\@hat\vphantom{\sht@im}\right.}%
  \def\k@fel{\setbox0=\hbox{$\@fen\g@rin$}\hbox{%
      $\@fen \kern.3875\wd0 \copy0 \kern-.3875\wd0%
      \llap{\copy0}\kern.3875\wd0$}}%
  \def\pt@h{\mathopen\@t}\pt@h\sht@im%
  \Right}%
\def\Right#1{\let\@hat=#1%
  \def\st@m{\mathclose\@t}%
  \st@m\endgroup}
\makeatother

\title[Gravito-turbulence in protoplanetary discs]{Gravito-turbulence in irradiated protoplanetary discs}

\author[S. Hirose and J. Shi]{
Shigenobu Hirose,$^{1}$\thanks{E-mail: hirose.shigenobu@gmail.com (SH)}
and Ji-Ming Shi$^{2}$
\\
$^{1}$Department of Mathematical Science and Advanced Technology, JAMSTEC, Yokohama 236-0001, Japan\\
$^{2}$Department of Astrophysical Sciences, Princeton University, 4 Ivy Ln, Princeton, NJ 08544\\
}

\date{Accepted 2017 March 29. Received 2017 March 29 ; in original form 2016 December 22}
\pubyear{2016}

\begin{document}
\label{firstpage}
\pagerange{\pageref{firstpage}--\pageref{lastpage}}
\maketitle

\begin{abstract}
Using radiation hydrodynamics simulations in a local stratified shearing box {\color{\ccorange} with realistic equations of state and opacities}, we explored the outcome of self-gravity at $50$ AU
in a protoplanetary disc irradiated by the central star. 
We found that gravito-turbulence
is sustained for a finite range of the surface density,
{\color{\ccorange}from $\sim 80$ to $\sim$ 250 gcm$^{-2}$. The disk is laminar below the range while fragments above it. }
In the range of gravito-turbulence, the Toomre parameter decreases monotonically from $\sim 1$ to $\sim 0.7$ as the surface density increases
while an effective cooling time is almost constant at $\sim {\color{\ccorange}4}$ in terms of the inverse of the orbital frequency.
The turbulent motions are supersonic at all heights, which dissipates through both shock waves and compressional heating.
The compressional motions, occurring near the midplane, create upward flows,
which not only contribute to supporting the disc but also to transporting the dissipated energy to the disc surfaces.
The irradiation does not affect much the gravito-turbulence near the midplane unless the grazing angle is larger than 0.32.
We also show that a simple cooling function with a constant cooling time does not approximate the realistic cooling.
\end{abstract}

\begin{keywords}
protoplanetary discs --- gravitation --- hydrodynamics --- radiative transfer --- instabilities --- turbulence
\end{keywords}



\section{Introduction}

Angular momentum transport in protoplanetary discs controls their time evolution and thus strongly affects the planet formation process within them. In accretion discs where gas and magnetic field are well coupled, magnetic turbulence driven by the magneto-rotational instability (MRI) transports angular momentum quite efficiently \citep{Balbus:1991fi}. However, that would not be the case in protoplanetary discs, where temperatures are too low for thermal ionization to operate and thus MRI is generally suppressed by non-ideal magnetohydrodynamic effects \citep[e.g.][]{2014prpl.conf..411T}.

{\color{black}In some cold and massive} protoplanetary discs, angular momentum can be transported by shear stresses associated with the gravitational instability (GI) {\color{\ccred}\citep[see][for a recent comprehensive review]{Kratter:2016dwa}}.
A natural consequence of the long-range nature of gravity is formation of spiral arms as a result of GI, which globally transport angular momentum. 
On the other hand, \citet{Gammie:2001hv} showed another nonlinear outcome of GI, called gravito-turbulence,
in which angular momentum transport can be described locally as in the $\alpha$ disc model \citep{Shakura73}.
He used razor-thin, local shearing box simulations with a cooling function that has a constant cooling time $t_\text{cool}$ {\color{\ccorange}in terms of $\Omega^{-1}$, which is called the $\beta$ cooling prescription}.
He showed that fragmentation occurs when the cooling is rather fast as
\begin{align}
\beta \equiv t_\text{cool}\Omega < 3,\label{eq:gammie}
\end{align}
while quasi-steady gravito-turbulence of $Q \sim 1$ is sustained otherwise.
Here, $Q$, defined as
\begin{align}
Q \equiv \frac{c_\text{s}\Omega}{\pi G\Sigma},\label{eq:toomre}
\end{align}
is called Toomre parameter,
and $Q < 1$ is the condition for the linear axisymmetric GI in the Keplerian disc, 
where $c_\text{s}$, $\Omega$, and $\Sigma$ are, respectively, the sound velocity, the orbital frequency, and the surface density at the radius considered \citep{Toomre:1964fe}.
{\color{\ccorange}At this gravito-turbulent phase where dissipation of the turbulence balances with the cooling, the nominal $\alpha$ parameter and $\beta$ are simply related through $\alpha = (9/4)\gamma(\gamma-1)\beta $ \citep{Gammie:2001hv}.}

{\color{\ccred}

Since then, the fragmentation criteria have been extensively studied, especially in protoplanetary discs, by many authors adopting various types of numerical methods and cooling prescriptions in both local and global simulations \citep[e.g.][]{Johnson:2003kl,Stamatellos:2009kx,Cossins:2010gh,Baehr:2015bl,2016MNRAS.460.2223R}, mostly motivated by the idea of forming gas giants via GI \citep{1997Sci...276.1836B,1998ApJ...503..923B,2007prpl.conf..607D,Zhu:2012bm}.

However, the exact value of the critical $\beta$ for fragmentation is still an open question. The non-convergence of the fragmentation criterion may arise from numerical artifacts \citep{Meru:2010dn,Lodato:2011ea,Meru:2012hw}, inherent stochasticity of fragmentation \citep[][]{Paardekooper:2012dta,Hopkins:2013jt}, the dimension (i.e. 2D vs. 3D) \citep{Young:2015jq}, or the fact that there is no physical temperature floor in the $\beta$ cooling prescription \citep{Lin:2016cx}. The irradiation can be a main heating source in cool protoplanetary discs subject to GI, and thus may affect the fragmentation criterion \citep{Rice:2011bj}.
It has also been suggested that a maximum $\alpha \sim 0.1$ might be a more general criterion than the critical $\beta$ criterion \citep{Rice:2005fl}.  Alternatively, \citet{Takahashi:2016bx} found that fragmentation boundary determined by the Toomre parameter using global simulations with realistic thermodynamics.

}
Here, we present local 3D simulations of irradiated, self-gravitating discs with realistic opacities and self-consistent equation of states (EOSs). {\color{\ccorange}This is an extended work from \citet{Shi:2014gz}, who firstly performed 3D local shearing box simulations, using the $\beta$ cooling as well as simple optically-thin cooling prescriptions.}
We used the flux-limited diffusion approximation (FLD) in the transfer of the disc's own radiation field while we used a ray-tracing method to calculate the irradiation heating by the central star. The aim of this paper is to explore the physics involved in the gravito-turbulence and the condition for it to be sustained {\color{\ccorange}for a fixed radius}. We also clarify the effect of the irradiation as well as how different the realistic radiative transfer is than the simple cooling function.

This paper is organized as follows. After we describe our numerical methods in section \ref{sec:methods}, we present our numerical results in section \ref{sec:results}. In Section 4, we {\color{\ccorange}mainly discuss validity of our simulations}, and we conclude in section \ref{sec:conclusion}.

\section{Methods}\label{sec:methods}

\subsection{Basic equations and numerical schemes}

The basic equations solved in our simulations are hydrodynamics equations with the Poisson equation for self-gravity and the frequency-integrated angular-moment equations of radiative transfer:
\begin{align}
  \frac{\partial\rho}{\partial t} + \nabla\cdot(\rho\bm{v}) = 0, \\
  \frac{\partial(\rho\bm{v})}{\partial t} + \nabla\cdot(\rho\bm{v}\bm{v}) =
  -\nabla p -\rho\nabla\Phi + \frac{{\kappa}_\text{R}\rho}{c}\bm{F}, \label{eq:motion}\\
  \frac{\partial e}{\partial t} + \nabla\cdot(e\bm{v}) =
  -(\nabla\cdot\bm{v})p - \left(4\pi B(T) - cE\right){\kappa}_\text{P}\rho, \label{eq:energy_gas}\\
  \frac{\partial E}{\partial t} + \nabla\cdot(E\bm{v}) =
  -\nabla\bm{v}:\mathsf{P} + \left(4\pi B(T) - cE\right){\kappa}_\text{P}\rho - \nabla\cdot\bm{F}, \label{eq:energy_rad}\\
  \nabla^2\Phi = 4\pi G\rho, \label{eq:poisson}
\end{align}
where $\rho$ is the gas density, $e$ the gas internal energy, $p$ the gas pressure, $T$ the gas temperature (assumed to be the same as the dust temperature), $E$ the radiation
energy density, $\mathsf{P}$ the radiation pressure tensor, $\bm{F}$ the
radiation energy flux, $\bm{v}$ the velocity field vector, $B(T) = \sigma_\text{B}T^4/\pi$ the Planck
function ($\sigma_\text{B}$, the Stefan-Boltzmann constant), and $c$ the speed
of light.
Under the FLD approximation, $\bm{F}$ and $\mathsf{P}$ are
related to $E$ as $\bm{F} =
-(c\lambda(R)/\kappa_\text{R}\rho)\nabla E$ and $\mathsf{P} =
\mathsf{f}(R)E$. Here $\lambda(R) \equiv (2+R)/(6+3R+R^2)$ is a flux limiter
with $R \equiv |\nabla E|/(\kappa_\text{R}\rho E)$, and $\mathsf{f}(R) \equiv
(1/2)(1-f(R))\mathsf{I} + (1/2)(3-f(R))\bm{n}\bm{n}$ is the Eddington tensor
with $f(R) \equiv \lambda(R) + \lambda(R)^2R^2$ and $\bm{n}\equiv\nabla E/|\nabla E|$ \citep{Turner:2001gx}.

The EOSs ($p = p(e,\rho)$ and $T = T(e,\rho)$) and the Rosseland-mean and the Planck-mean opacities ($\kappa_\text{R}(\rho,T)$ and $\kappa_\text{P}(\rho,T)$) were tabulated beforehand. The EOS tables are updated versions of those used in \citet{Tomida:2012fb}. The opacity tables are the same as those used in \citet{Hirose:2015cv}, where dust opacities are taken from \citet{Semenov:2003hk} while low-temperatures gas opacities are taken from \citet{Ferguson:2005gn}. The EOS and opacity tables are plotted in figure \ref{fig:opacity_table} in Appendix \ref{sec:clump}.

We used the shearing box approximation to model a local patch of an accretion disc 
as a co-rotating Cartesian frame $(x,y,z)$ with a linearized Keplerian shear flow
$\bm{v}_\text{K} \equiv -(3/2)\Omega x\hat{\bm{y}}$, where the $x$, $y$, and $z$ directions correspond to the
radial, azimuthal, and vertical directions, respectively, and $\hat{\bm{y}}$ is the unit vector in the $y$ direction \citep{Hawley:1995gd}.
The inertial forces in the co-rotating frame and the vertical component of the central star's gravity,
$-2\Omega\hat{\bm{z}}\times\bm{v} +3\Omega^2x\hat{\bm{x}} - \Omega^2z\hat{\bm{z}}$, 
are added in the right hand side (RHS) of the equation of motion (\ref{eq:motion}), 
where $\hat{\bm{x}}$ and $\hat{\bm{z}}$ are the unit vectors in the $x$ and $z$ direction, respectively.
Shearing-periodic, periodic, and outflow boundary conditions are applied to the boundaries in the $x$, $y$, and $z$ direction, respectively. The outflow boundary condition is described in \citep{Hirose:2006gp}.

To solve the above equations, we employed ZEUS \citep{Stone92a}, where we also implemented an orbital advection algorithm \citep{Stone:2010bh} for accurate calculations in a wide shearing box.
The Poisson equation with the vacuum boundary condition in the $z$ direction was solved by Fast Fourier Transforms \citep{Koyama:2009fw}. To test these methods, we solved one of the disc models (the one with the constant cooling time of $10 \Omega^{-1}$) in \citet{Shi:2014gz} and got a quantitatively consistent result.

The radiative transfer part of the basic equations is extracted as
\begin{align}
  \frac{\partial e}{\partial t} =
  - {\kappa}_\text{P}\rho\left(4\pi B(T) - cE\right) + q_\text{irr}, \label{eq:gas_energy}\\
  \frac{\partial E}{\partial t} =
  {\kappa}_\text{P}\rho\left(4\pi B(T) - cE\right) + \nabla\cdot\left(\frac{c\lambda(R)}{\kappa_\text{R}\rho}\nabla E\right),
\end{align}
which was solved time-implicitly using a multi-grid algorithm with the Gauss-Seidel method as a smoother. During iteration in each time step, the irradiation heating rate $q_\text{irr}$ as well as the coefficients $\kappa_\text{P}\rho$, $\kappa_\text{R}\rho$ and $\lambda(R)$ are fixed. 

The irradiation heating rate $q_\text{irr}$ is evaluated by solving the time-independent radiative transfer equation that ignores scattering,
\begin{align}
0 = -\kappa_\text{P$_*$}\rho I - \frac{dI}{ds}, \label{eq:RT}
\end{align}
using a ray tracing method, with the coefficient $\kappa_\text{P$_*$}\rho$ being fixed. Here, $I$ and $s$ have the usual meanings while $\kappa_\text{P$_*$}=\kappa_\text{P$_*$}(\rho,T)$ is a mean opacity averaged over the Planck function of the stellar effective temperature $T_*$ \citep{Hirose:2011gi}. See Appendix \ref{sec:irradiation} for descritization of the equation (\ref{eq:RT}) to compute $q_\text{irr}$.

In our simulations, the kinetic energy may dissipates either numerically in the grid-scale or physically in shock waves. In any case, the dissipated energy is captured in the form of gas internal energy, effectively
resulting in additional source terms, $q_\text{num}$ and $q_\text{shock}$ respectively, in the gas energy equation (\ref{eq:energy_gas}). The shock heating rate $q_\text{shock}$ is computed as $-q \nabla\cdot\bm{v}$, where $q$ determines the strength of the ZEUS's shock capturing viscosity \citep{Stone92a}. The method to evaluate the grid-scale dissipation rate $q_\text{num}$ is described in detail in Appendix in \citet{Hirose:2006gp}. Thus the sum of the kinetic and internal energies is conserved in the simulation box.

\subsection{Parameters and the initial conditions}\label{sec:parameters}
Parameters in a stratified shearing box are the orbital frequency $\Omega$ [s$^{-1}$], which appears in the inertial force terms and the shearing periodic boundary condition, and the surface density $\Sigma$ [g cm$^{-2}$], which is the amount of gas in the box. On the other hand, parameters for the irradiation are the energy flux $F_\text{irr}$ [erg cm$^{-2}$ s$^{-1}$] and the grazing angle $\theta$ at the surfaces of the box. {\color{black}The surface density $\Sigma$ may vary during a simulation
due to outflows through the top and bottom boundaries or the density floor described below, but the variation was typically less than $1$ \% per one hundred orbits unless noted.}

In this paper, a shearing box was placed at $a = 50$ AU away from the central star that has the effective temperature of $T_* = 4000$ K, the mass of $M_* = 1M_\odot$, and the radius of $R_* = 1R_\odot$. Then, the orbital frequency of the shearing box is determined as $\Omega = \sqrt{GM_*/a^3} = 5.63\times10^{-10}$ s$^{-1}$ while the irradiation energy flux is determined as $F_\text{irr} = (R_*/a)^2\sigma_\text{B}T_*^4 = 1.26\times 10^2$ erg cm$^{-2}$ s$^{-1}$. The rest two parameters, the surface density $\Sigma$ and the grazing angle $\theta$, are free physical parameters in this study. 

The initial disc was assumed to be isothermal and hydrostatic ignoring self-gravity. The isothermal temperature $T_0$ was evaluated using \citet{Chiang97}'s radiative equilibrium disc model (their equation 12a), $T_0 = \left({\theta}/{4}\right)^\frac14\left({R_*}/{a}\right)^\frac12T_*$. We note that this temperature $T_0$ was used only in constructing the initial disc, in which the mean molecular weight $\mu = 2.38$ and the adiabatic exponent $\gamma = 5/3$ were also temporarily used. The initial radiation field $E_0$ was assumed to be in thermal equilibrium with gas, $E_0 = aT_0^4$. The initial velocity field was the linearized Keplerian shear flow $\bm{v}_0 = \bm{v}_\text{K}$, whose $x$ and $z$ components were perturbed randomly up to 0.5\% of the local sound speed.

From numerical reasons, we had to introduce floors for density, internal energy, and temperature. The density floor was set to $10^{-6}$ of the initial midplane density.
The internal energy floor was set basically to avoid negative values \citep[see Appendix A3 in][for details]{Hirose:2006gp}. The temperature floor was set to $5$ K, which is the lower limit of the temperature range in the EOS and opacity tables. The total artificial energy injection rate due to these floors was typically less than 1\% of the physical heating rate when volume-averaged (see the middle panel in figure \ref{fig:time_evolution}).

\section{Results}\label{sec:results}

\subsection{Diagnostics}
We present results of our simulations using diagnostics based on horizontally-averaged vertical profiles, which were recorded every $0.01$ orbits.\footnote{\color{\ccorange}The data for figure \ref{fig:kappa_vs_tmp} was recorded every single orbit.} The horizontally-averaged vertical profile of quantity $f$, for example, was computed as
\begin{align}
\left<f\right>(z,t) \equiv \dfrac{\int\!\!\int f(x,y,z,t) dxdy}{\int\!\!\int dxdy},\label{eq:ave}
\end{align}
where the integrations were done over the full extent of the box in the $x$ and $y$ directions.
In the diagnostics, it is time-averaged or vertically-integrated as
\begin{align}
&\overline{\left<f\right>} \equiv \frac{\int\left<f\right>dt}{\int dt},\\
&\left<\!\left<f\right>\!\right> \equiv \int\left<f\right>dz.
\end{align}
The vertical integration is done over the full extent of the box height, and the time averaging is done between $t = 20$ and $t = 80$ orbits unless noted.
In some diagnostics, we also use density-weighted, vertical averaging as
\begin{align}
&\left<\!\left<f\right>\!\right>_\rho \equiv \frac{\int\left<f\right>\left<\rho\right>dz}{\int\left<\rho\right>dz}.
\end{align}

\subsection{Fiducial run}\label{sec:fiducial}
The two physical parameters, the surface density $\Sigma$ and the grazing angle $\theta$, depend on the global modeling of protoplanetary discs. Therefore, in our local shearing box simulations, they may be chosen somewhat arbitrarily. We chose $\Sigma = 100$ g cm$^{-2}$
and $\theta = 0.02$ for the fiducial run, where the gravito-turbulence was sustained.

The box size and the number of cells in the fiducial run were $(L_x,L_y,L_z) = (24H,24H,12H)$ and $(N_x,N_y,N_z)=(128,128,64)$, respectively. Here and hereafter, the scale height of the initial isothermal disc $H \equiv \sqrt{2RT_0/(\mu\Omega^2)}$ is used as the unit length, where $R$ is the gas constant. It reads $H=4.75\times10^{13}$ cm = $3.18$ AU from the physical parameters chosen in the second paragraph in section \ref{sec:parameters}. We keep H as our unit length in all runs in this paper.

\begin{figure}
\includegraphics[width=7cm]{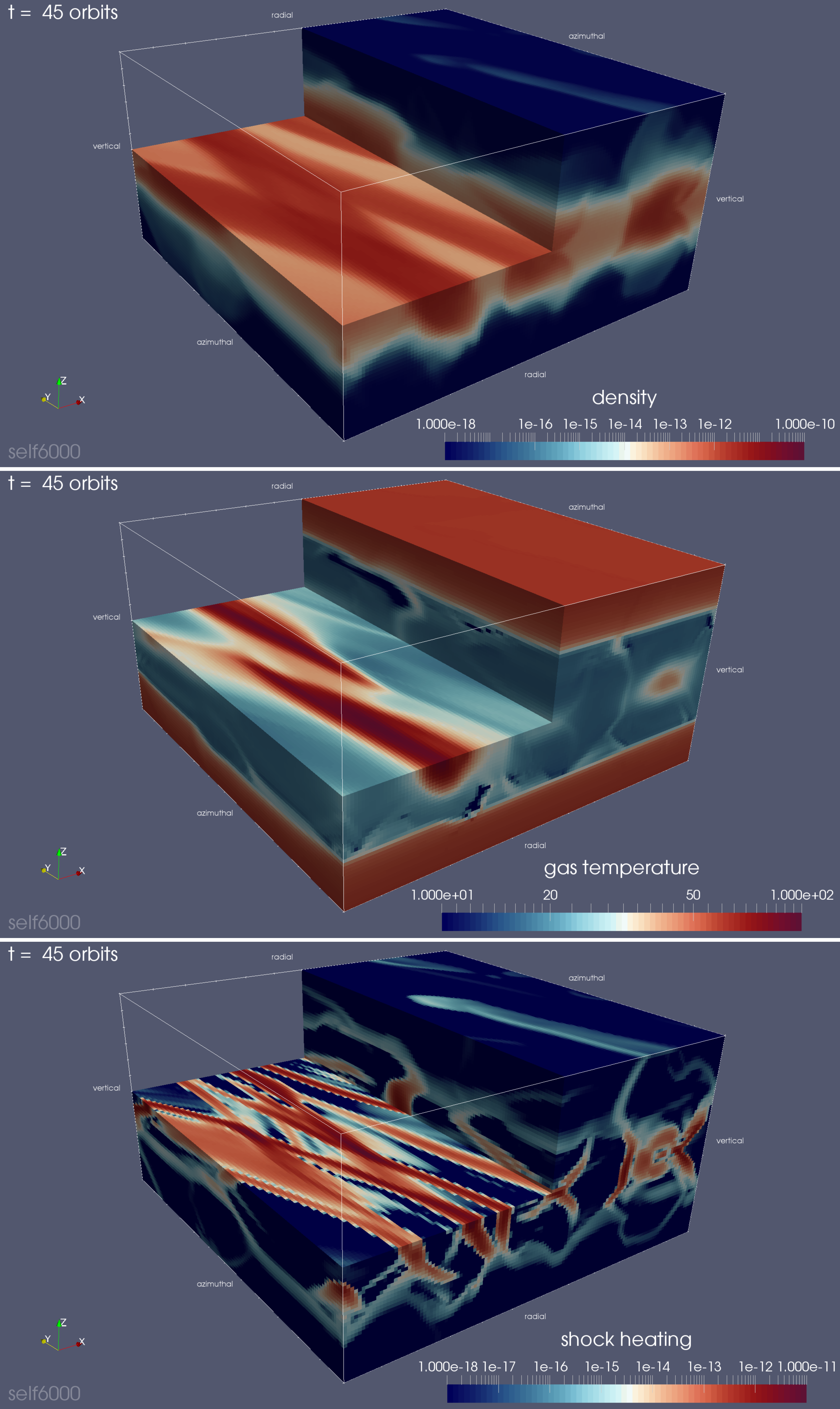}
\caption{Snapshots of density (top), gas temperature (middle), and shock heating rate (bottom) in the fiducial run. The colour scale ranges logarithmically from $10^{-18}$ to $10^{-10}$ [g cm$^{-3}$] in the top panel, from $10$ to $100$ [K] in the middle panel, and from $10^{-18}$ to $10^{-11}$ [erg cm$^{-3}$ s$^{-1}$] in the lower panel.}
\label{fig:snapshots}
\end{figure}

First of all, for readers to grasp what is happening in the fiducial run, we show in figure \ref{fig:snapshots} typical snapshots of density, gas temperature, and shock heating rate in the quasi-steady state. It is seen that many shock waves are excited by collisions of non-axisymmetric density waves driven by GI, raising gas temperatures near the midplane.

\subsubsection{Time evolution and cooling time}\label{sec:evolution}
\begin{figure}
\includegraphics[width=7cm]{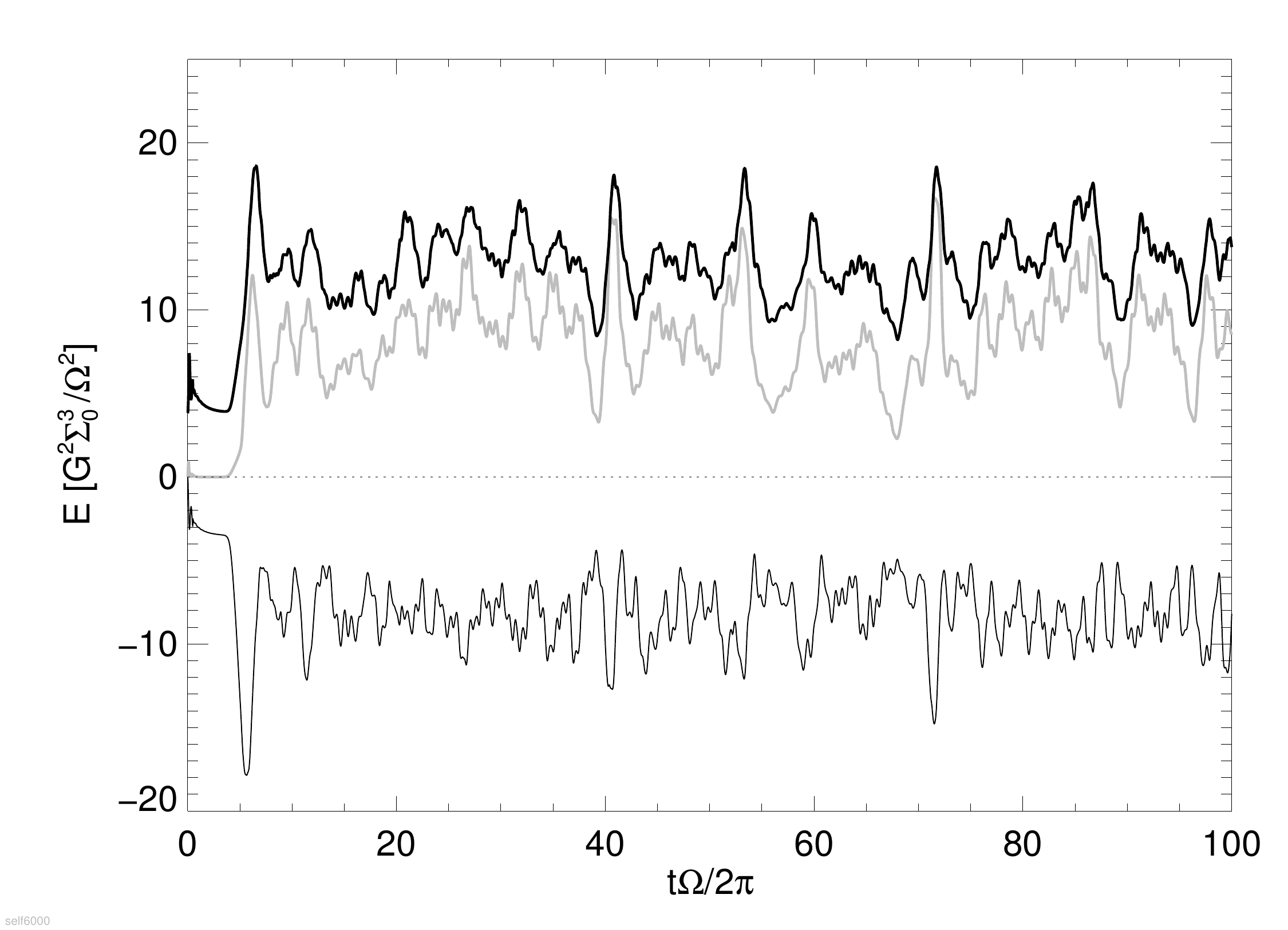}
\includegraphics[width=7cm]{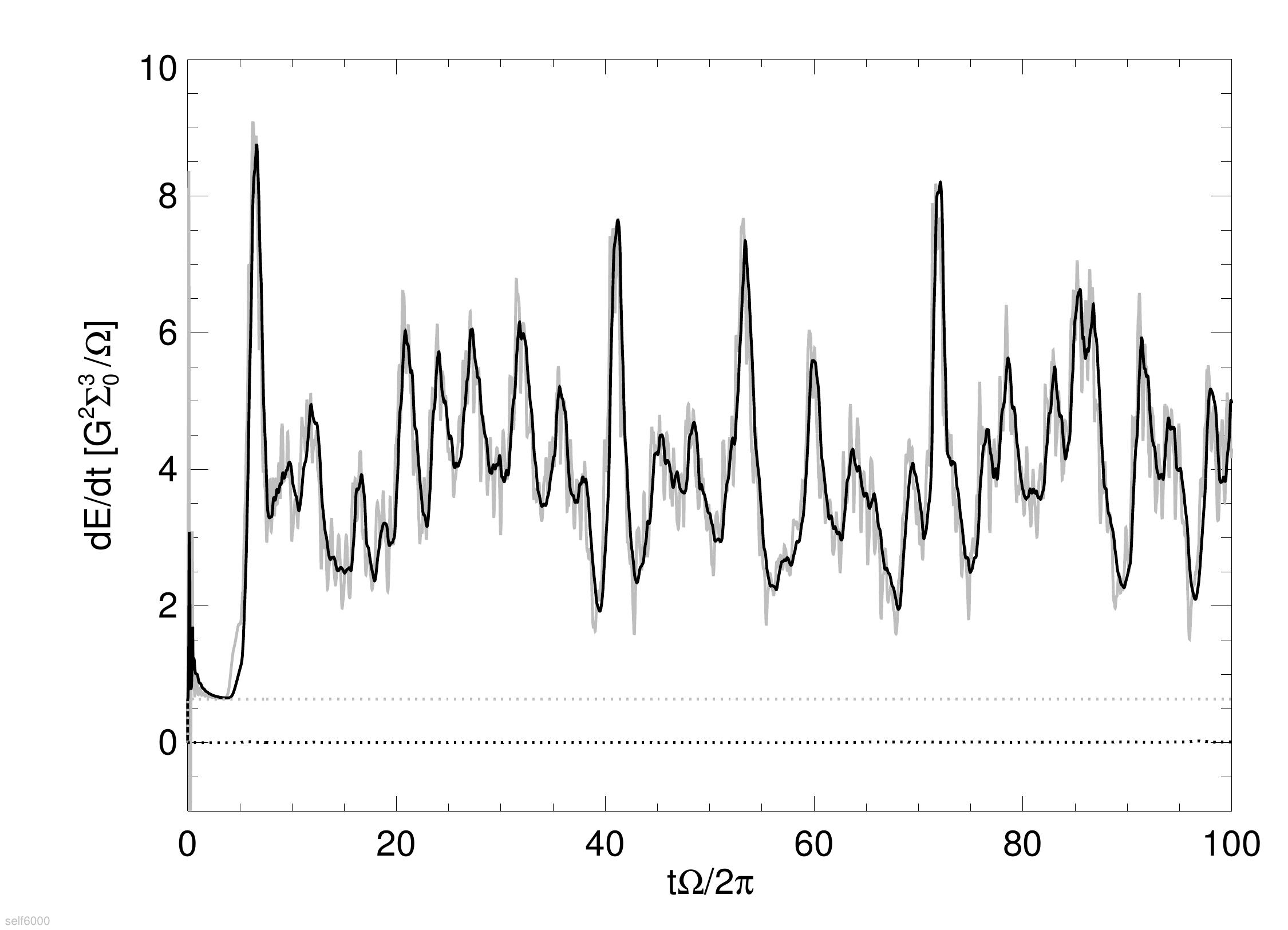}
\includegraphics[width=7cm]{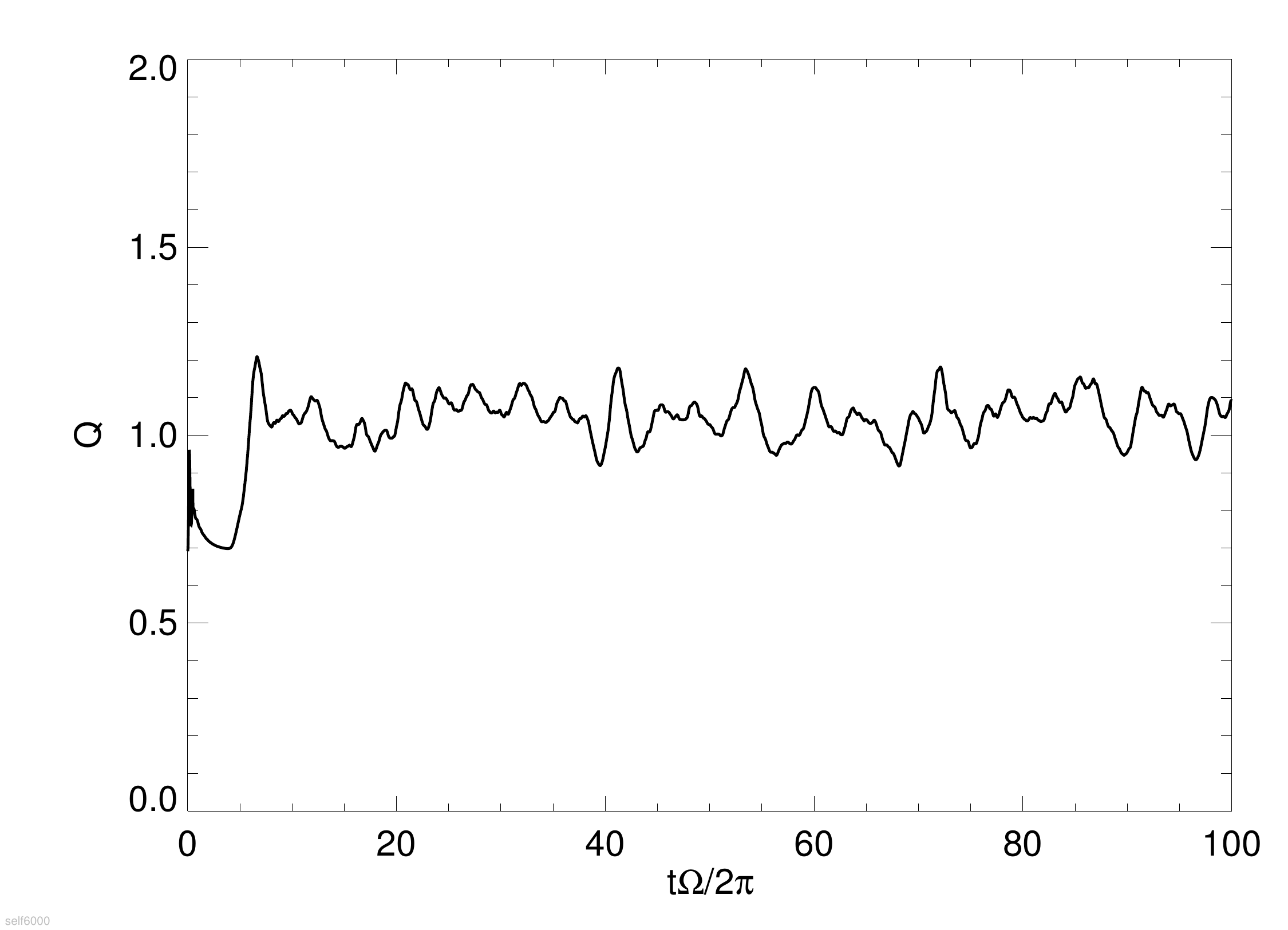}
\caption{Time evolution of vertically-integrated energies (top), vertically-integrated heating and cooling rates (middle), and the Toomre parameter near the midplane $Q_\text{mid}$ (bottom) in the fiducial run. In the top panel, the black, grey, and thin black curves are, respectively, the thermal, kinetic, and self-gravitational energies. In the middle panel, the black, grey, grey dotted, and black dotted curves are, respectively, the total cooling rate, total heating rate, irradiation heating rate, and artificial energy injection rate. For clarity, all quantities are boxcar-averaged over a single orbit.}
\label{fig:time_evolution}
\end{figure}

Figure \ref{fig:time_evolution} shows time evolution of vertically-integrated versions of energies (kinetic $E_\text{k}$, thermal $E_\text{t}$, and self-gravitational $E_\text{g}$), heating rate $q^+$ and cooling rate $q^-$ as well as the Toomre parameter near the midplane, $Q_\text{mid}$, which are defined as
\begin{align}
&\Left<E_\text{k}\Right> \equiv \Left<\frac12\rho\left(\bm{v}-\bm{v}_\text{K}\right)^2\Right>,\\
&\Left<E_\text{t}\Right> \equiv \Left<e\Right>,\\
&\Left<E_\text{g}\Right> \equiv \Left<\frac12\rho\Phi\Right>, \\
&\Left<q^+\Right> \equiv \Left<-p\nabla\cdot\bm{v}\Right> + \Left<q_\text{shock}\Right> + \Left<q_\text{irr}\Right> + \Left<q_\text{num}\Right>,\\
&\Left<q^-\Right> \equiv \Left<\frac{\p F_z}{\p z}\Right> + \Left<\frac{\p ev_z}{\p z}\Right>,\\
&Q_\text{mid} \equiv \frac{\left<\!\left<c_\text{s}\right>\!\right>_\rho\Omega}{\pi G \Sigma}.\label{eq:q}
\end{align}
We omit the radiation energy $E$ in the thermal energy $E_\text{t}$, the term $-\nabla\bm{v}:\mathsf{P}$ in the heating rate $q^+$, and the term ${\p Ev_z}/{\p z}$ in the cooling rate $q^-$ since they are negligible.
The fluctuations in the first few orbits are due to the deviation of the initial disc from a hydrostatic equilibrium. Then, around $t=5$ orbits, the axisymmetric mode of GI fully developed, which broke down into turbulent density waves. After this initial transient, the disc reached a statistically steady state around $t = 10$ orbits.

The upper panel shows that the ratio of $\Left<E_\text{k}\Right>$ to $\Left<E_\text{t}\Right>$ is about $0.67$, which is larger by a factor of a few than that in the $\beta = 10$ case in \citet{Gammie:2001hv}. {\color{black}This is because the effective cooling time in our simulation is relatively shorter ($\sim 3.17\Omega^{-1}$) as will be discussed in the next paragraph.}
It also shows that the three energies vary almost in phase. Cross-correlations of these variations reveal that the self-gravitational energy varies first, the kinetic energy second, and the internal energy third, with mutual delays of $1.8\Omega^{-1}$ and $1.1\Omega^{-1}$, respectively. The order is reasonable as a natural consequence of the causality. Similarly, the middle panel shows that the cooling rate quickly follows the heating rate, keeping a thermal balance, but with a delay of $1.5\Omega^{-1}$. The middle panel also shows that the rate of the artificial energy injection mentioned in section \ref{sec:parameters} is negligible compared to the heating or cooling rates in the steady state. The lower panel shows that the Toomre parameter took its minimum of $0.7$ when the axisymmetric mode of GI fully develops, and settled down to $1.05$ on time average in the steady state.

\begin{figure}
\includegraphics[width=7cm]{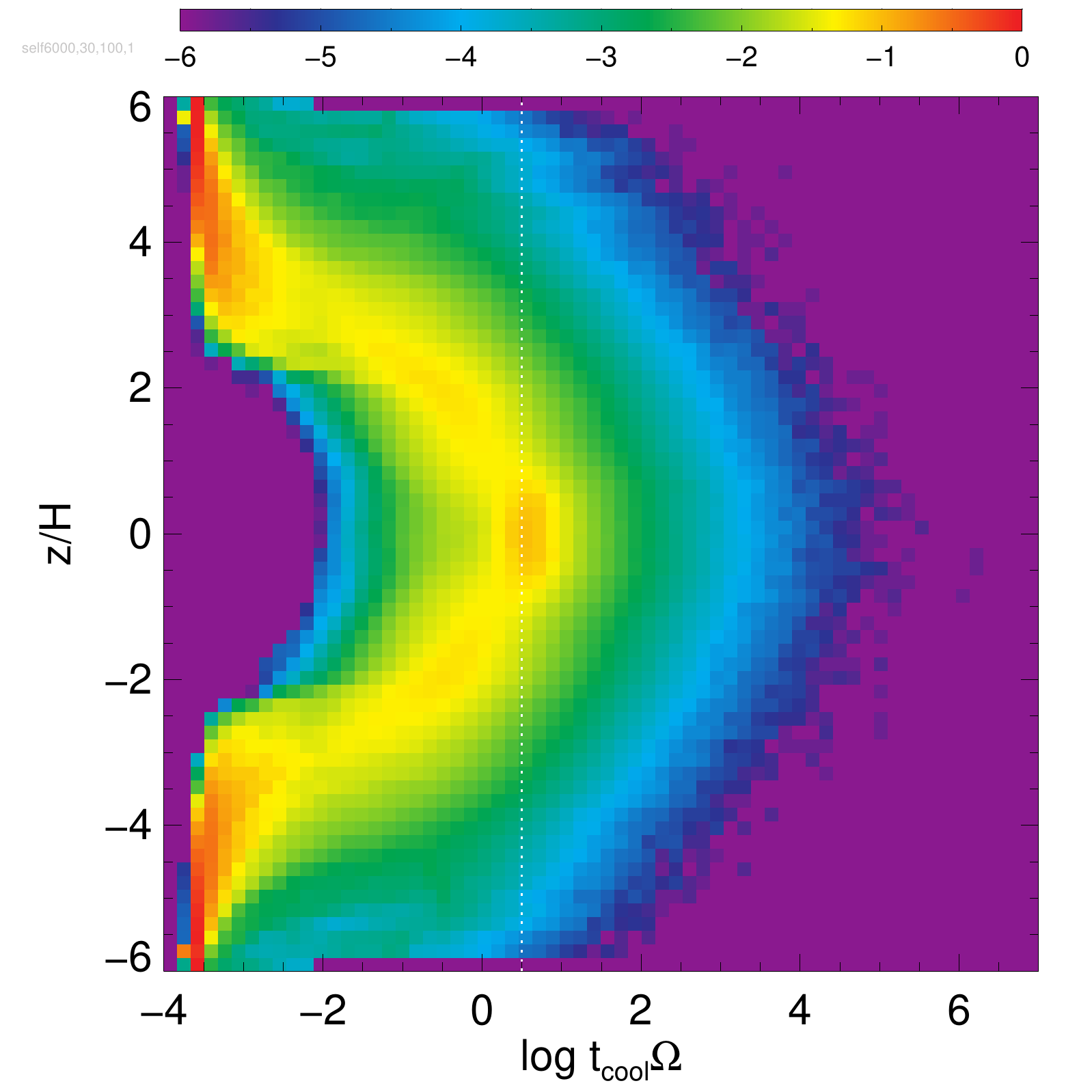}
\caption{Histogram of the cell-by-cell cooling time $t_\text{cool}\Omega$ vs. height $z/H$ in the fiducial run, compiled from all cells contained in 71 snapshots equally taken from $t = 30$ to $100$ orbits. The colour represents the total number of the cells (arbitrarily normalized) that take the corresponding values of $(t_\text{cool}\Omega,z/H)$. The vertical white dotted line represents the volume-averaged cooling time $\overline{\beta}_\text{eff} = 3.17$.}
\label{fig:cell_cooling}
\end{figure}

One may be interested in what is the $\beta$ value here, the cooling time in terms of $\Omega^{-1}$ \citep{Gammie:2001hv}. If we define an effective cooling time in the volume-averaged sense,
{\color{\ccorange}
\begin{align}
\overline{\beta}_\text{eff} \equiv \frac{\overline{\Left<e\Right>}\Omega^{-1}}{\overline{\Left<-(4\pi B(T) - cE)\kappa_\text{P}\rho\Right>}} = 3.17 \label{eq:beta_eff}
\end{align}
while if we compute a density-weighted, volume-averaged cooling time,
\begin{align}
\overline{\beta}_\text{mid} \equiv \frac{\overline{\Left<e\Right>}_\rho\Omega^{-1}}{\overline{\Left<-(4\pi B(T) - cE)\kappa_\text{P}\rho\Right>}_\rho} = 4.37. \label{eq:beta_mid}
\end{align}
} On the other hand, we can directly examine the cell-by-cell cooling time, defined as the internal energy $e$ divided by the energy exchange rate between gas and radiation $-{\kappa}_\text{P}\rho\left(4\pi B(T) - cE\right)$.
Figure \ref{fig:cell_cooling} is a two-dimensional histogram of the cell-by-cell cooling time vs. height, compiled from all cells contained in selected snapshots in the steady state. Not surprisingly, the cell-by-cell cooling time is not single-valued, but is spread over a finite range at each height. Near the midplane, the mode of the cell-by-cell cooling time is similar to the volume-averaged cooling time ($\overline{\beta}_\text{eff} = 3.17$). On the other hand, in the upper layers that absorb the stellar irradiation, the cell-by-cell cooling time is exceptionally short ($\le 10^{-3}$).
In section \ref{sec:simple_cooling}, we will see that numerical results are totally different when the simple cooling function is employed in place of the FLD radiative transfer.

\subsubsection{Hydrostatic balance}\label{sec:hydrostatic}
\begin{figure}  
\includegraphics[width=7cm]{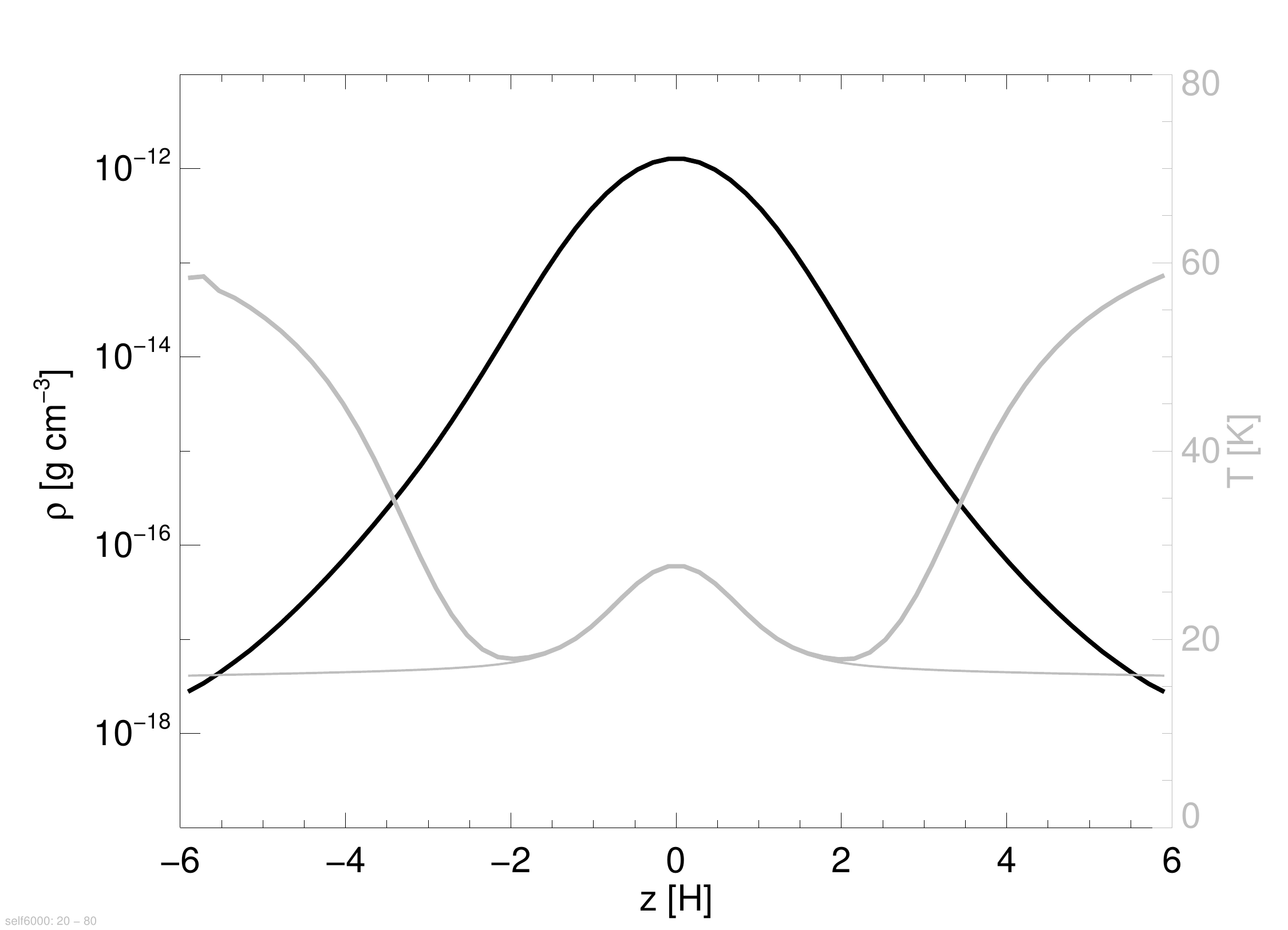}
\caption{Time-averaged vertical profiles of density (black), gas temperature (grey thick), and radiation temperature (grey thin) in the fiducial run. The axis for temperatures is shown on the right.}
\label{fig:vert_dens}
\end{figure}

Figure \ref{fig:vert_dens} shows time-averaged vertical profiles of density $\overline{\left<\rho\right>}$, gas temperature $\overline{\left<T\right>}$, and radiation temperature $\overline{\left<(E/a)^{1/4}\right>}$, where $a \equiv 4\sigma_\text{B}/c$ is the radiation constant.
The gas temperature near the midplane is low around $20$ K with a peak of $\sim 30$ K at the midplane. It is raised up to $60$ K in the upper layers by absorption of the irradiation of visible light by dust grains (see section \ref{sec:thermal_balance}).
The radiation temperature is almost equal to gas temperature near the midplane ($|z|/H < 2$), but they are apart in the upper layers where the irradiation is absorbed.
The density profile has an exponential decay at $2 \le |z|/H \le 4$ and extended tails in the irradiated hot layers ($|z|/H > 4$).

\begin{figure}
\includegraphics[width=7cm]{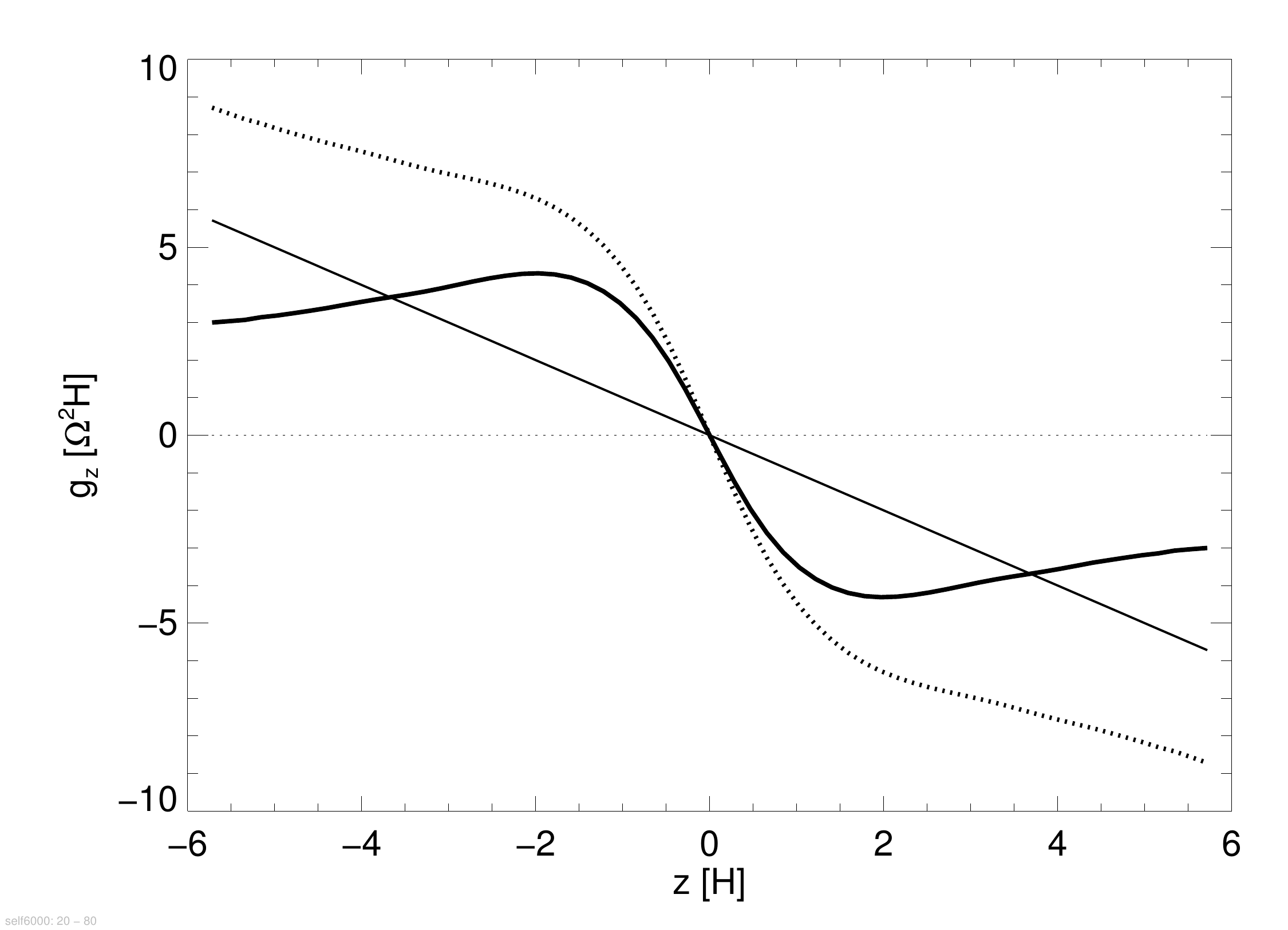}
\includegraphics[width=7cm]{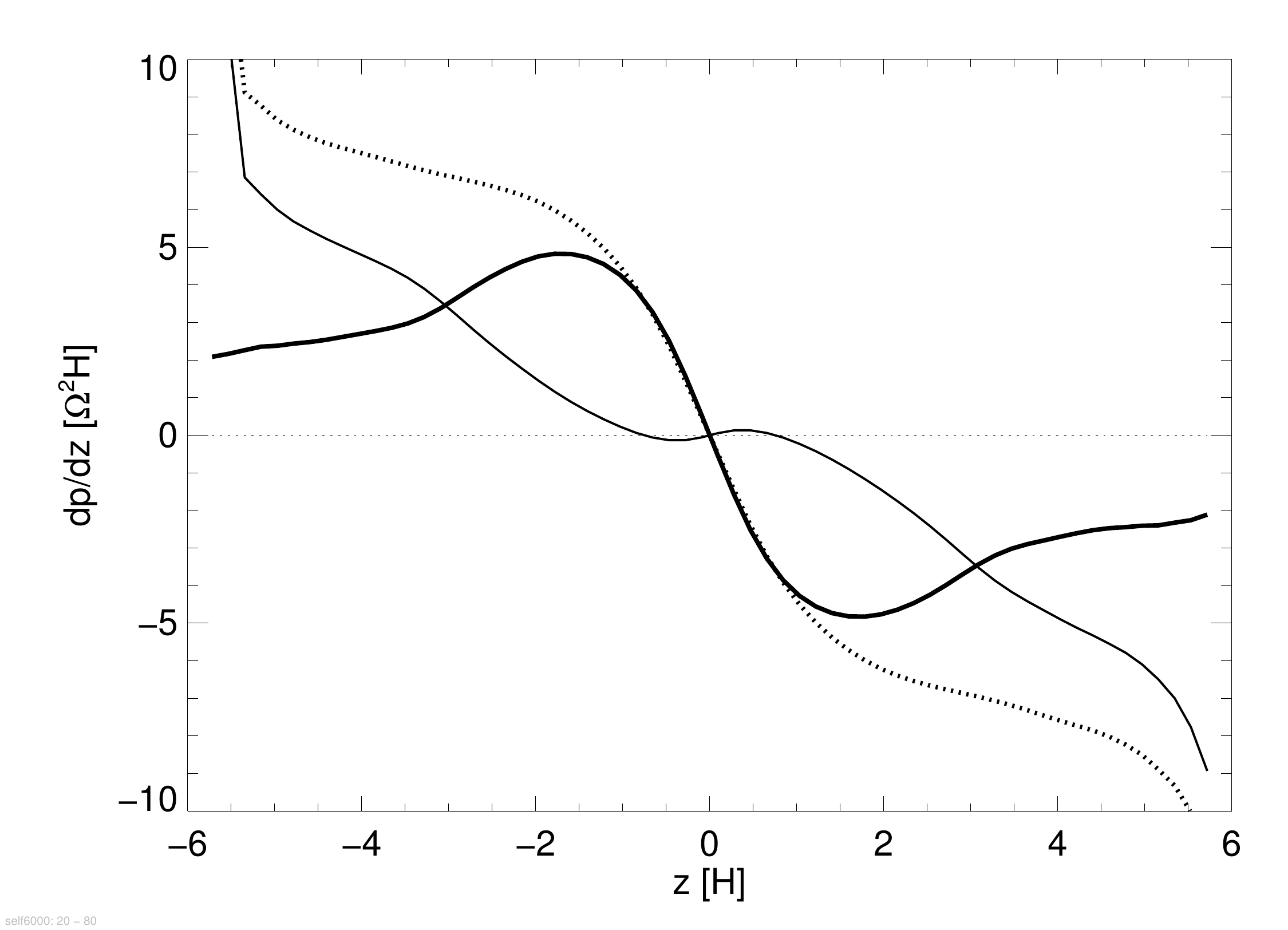}
\caption{Time-averaged vertical profiles of gravitational accelerations (upper) and pressure gradients (lower) in the fiducial run. In the upper panel, the thick, thin, and dotted curves are the self-gravitational acceleration, the external-gravitational acceleration, and their sum, respectively. In the lower panel, the thick, thin, and dotted curves are the gas pressure gradient, the dynamical pressure gradient, and their sum, respectively.}
    \label{fig:vert_accel}
\end{figure}

Figure \ref{fig:vert_accel} compares time-averaged profiles of vertical accelerations that appear in the following time-averaged version of the $z$ component of the momentum equation in a steady state,
\begin{align}
\frac{1}{\overline{\left<\rho\right>}}\frac{d}{dz}\overline{\left<p\right>} + \frac{1}{\overline{\left<\rho\right>}}\overline{\left<\frac{\partial}{\partial z}\left(\rho v_z^2\right)\right>} = -\Omega^2z - \frac{1}{\overline{\left<\rho\right>}}\overline{\left<\rho\frac{\partial\Phi}{\partial z}\right>}.\label{eq:momentum_balance}
\end{align}
We omit here the radiation force $\overline{\left<{\kappa_\text{R}\rho F_z}/{c}\right>}$ since it is negligible.
Pressure gradient terms in the left-hand side (LHS) are shown in the upper panel while vertical gravity terms in the right-hand side (RHS) are shown in the lower panel.
A hydrostatic balance holds well since the sum of the LHS terms and that of the RHS terms agree. Some anomalies near the boundaries are due to downflows arising from a hydrostatic imbalance. 

The lower panel shows that the self-gravity dominates the external gravity at $|z|/H < 4$, confirming that the disc is self-gravitating. At $|z|/H \sim 2$, the self-gravity peaks, roughly twice as large as the external gravity, which is consistent with the midplane Toomre parameter $\overline{Q}_\text{mid} = 1.05$ through the following relation:
\begin{align}
Q = \frac{c_\text{s}\Omega}{\pi G\Sigma} \sim \frac{\Omega^2H}{\pi G\Sigma} \sim 2\frac{\Omega^2 z}{d\Phi/dz},
\end{align}
where the Poisson equation is used as $d\Phi/dz \sim 4\pi G\rho H \sim 2\pi G\Sigma$.

On the other hand, the upper panel shows that the dynamical pressure gradient (the second term in LHS of equation \ref{eq:momentum_balance}) also competes the gravity as well as the thermal pressure gradient. The dynamical pressure here is created by upward flows driven by collisions of density waves near the midplane. Although the thermal pressure gradient is dominant near the midplane, the dynamical pressure gradient has non-negligible contribution, about a quarter of the total at $|z|/H \sim 2$. In the upper layers, the dynamical pressure gradient is even dominant.

\subsubsection{Shear stresses}\label{sec:stress}
\begin{figure}
\includegraphics[width=7cm]{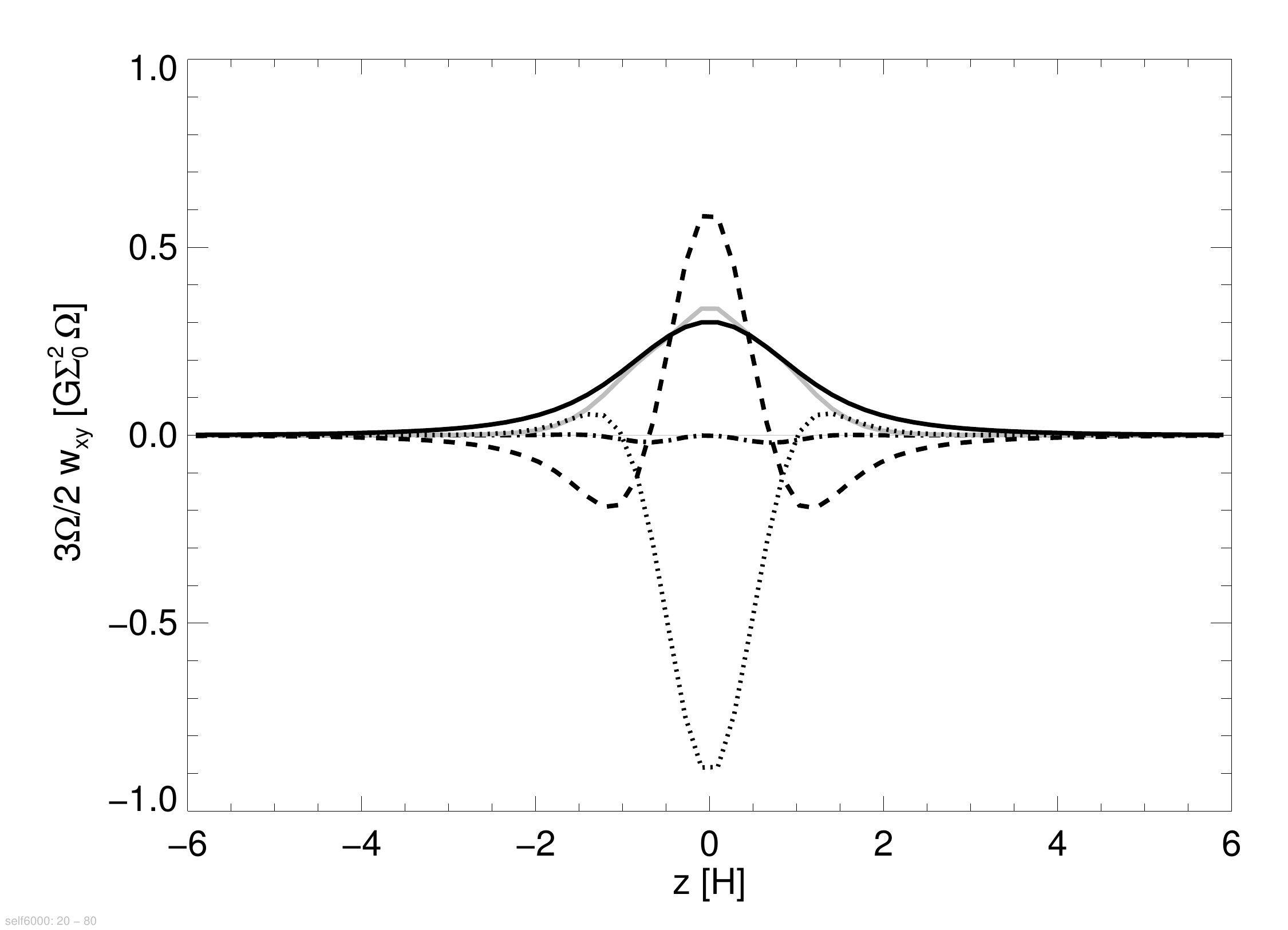}
\caption{Time-averaged vertical profiles of work done by stresses in the fiducial run. The black curve is work done by gravitational stress (the first term in RHS of equation \ref{eq:energy_selfgravity}) and grey curve is work done by the Reynolds stress. The black dashed and black dotted curves are, respectively, the second and third terms in RHS of equation (\ref{eq:energy_selfgravity}). The black dashed-dotted curve is the sum of the three terms in RHS.}
    \label{fig:vert_stress}
\end{figure}

Figure \ref{fig:vert_stress} compares time-averaged vertical profiles of works done by the shear stresses. The work done by the gravitational stress $\frac32\Omega\overline{\left<g_xg_y/4\pi G\right>}$ is the injection rate of {\color{black}$\left<E_\text{g}\right>$} into the box at each height while the work done by the Reynolds stress $\frac32\Omega\overline{\left<\rho v_x\delta v_y\right>}$ is the injection rate of {\color{black}$\left<E_\text{k}\right>$}. {\color{black}Here, $\bm{g} = (g_x,g_y,g_z) = -\nabla\Phi$ is the self-gravitational acceleration.} The Reynolds stress is slightly larger than the gravitational stress at the midplane, but the gravitational stress has a wider distribution than the Reynolds stress. When vertically-integrated, they are computed as $\frac32\Omega\overline{\Left<g_xg_y/4\pi G\Right>} = 1.83$ and $\frac32\Omega\overline{\Left<\rho v_x\delta v_y\Right>} = 1.51$, in terms of $G\Sigma_0^2\Omega H$. 

The Shakura-Sunyaev's $\alpha$, defined as the ratio of time-averaged, vertically-integrated shear stress to time-averaged, vertically-integrated thermal pressure, is commonly used to evaluate the shear stress in accretion discs \citep{Shakura73}.
{\color{\ccorange}In this paper, we define $\alpha$ as
\begin{align}
\alpha \equiv \frac{W_{xy}}{\overline{\Left<p\Right>}},\label{eq:alpha}
\end{align}
which is different from the conventional $\alpha \equiv \left|d\ln\Omega/d\ln r\right|^{-1}W_{xy}/(\Sigma c_\text{s}^2)$ in the literature by a factor of $2/3\gamma$.
}

Here, it is computed as $0.138$ and $0.105$ for the self-gravitational and Reynolds stresses, respectively.

Another way to evaluate the shear stress is to express it in terms of the mass accretion rate. Assuming a time-steady accretion, the vertically-integrated shear stress $W_{xy}$ and the mass accretion rate $\dot{M}$ are connected through
\begin{align}
\frac32\Omega W_{xy} = \frac{3}{4\pi}\Omega^2\dot{M}.\label{eq:mdot}
\end{align}
Using this relation, the mass accretion rate in the fiducial run is computed as $5.53\times10^{-6}$ $M_\odot$ yr$^{-1}$.

Figure \ref{fig:vert_stress} also tells about energetics of the self-gravitational energy. The time and horizontally-averaged version of the self-gravitational energy equation in a steady state is written as follows (see Appendix \ref{sec:selfgravitational_energy}):
\begin{align}
0 = \frac32\Omega\overline{\left<\frac{g_xg_y}{4\pi G}\right>}
  - \overline{\left<\frac{\p}{\p z}\left(\rho\Phi v_z + \dfrac{g_z}{4\pi G}\dfrac{\p\Phi}{\p t}\right)\right>} + \overline{\left<\rho\bm{v}\cdot\nabla\Phi\right>}.\label{eq:energy_selfgravity}
\end{align}
The third term is the work done by the self-gravitational force, which is the conversion rate of $\left<E_\text{g}\right>$ to $\left<E_\text{k}\right>$ at each height.
As seen in the figure, the first term (black solid) and the third term (black dotted) do not cancel, meaning that the self-gravitational energy injected via the gravitational stress is redistributed via the second term (black dashed) before converted to the kinetic energy.

\subsubsection{Thermal balance}\label{sec:thermal_balance}
\begin{figure}
\includegraphics[width=7cm]{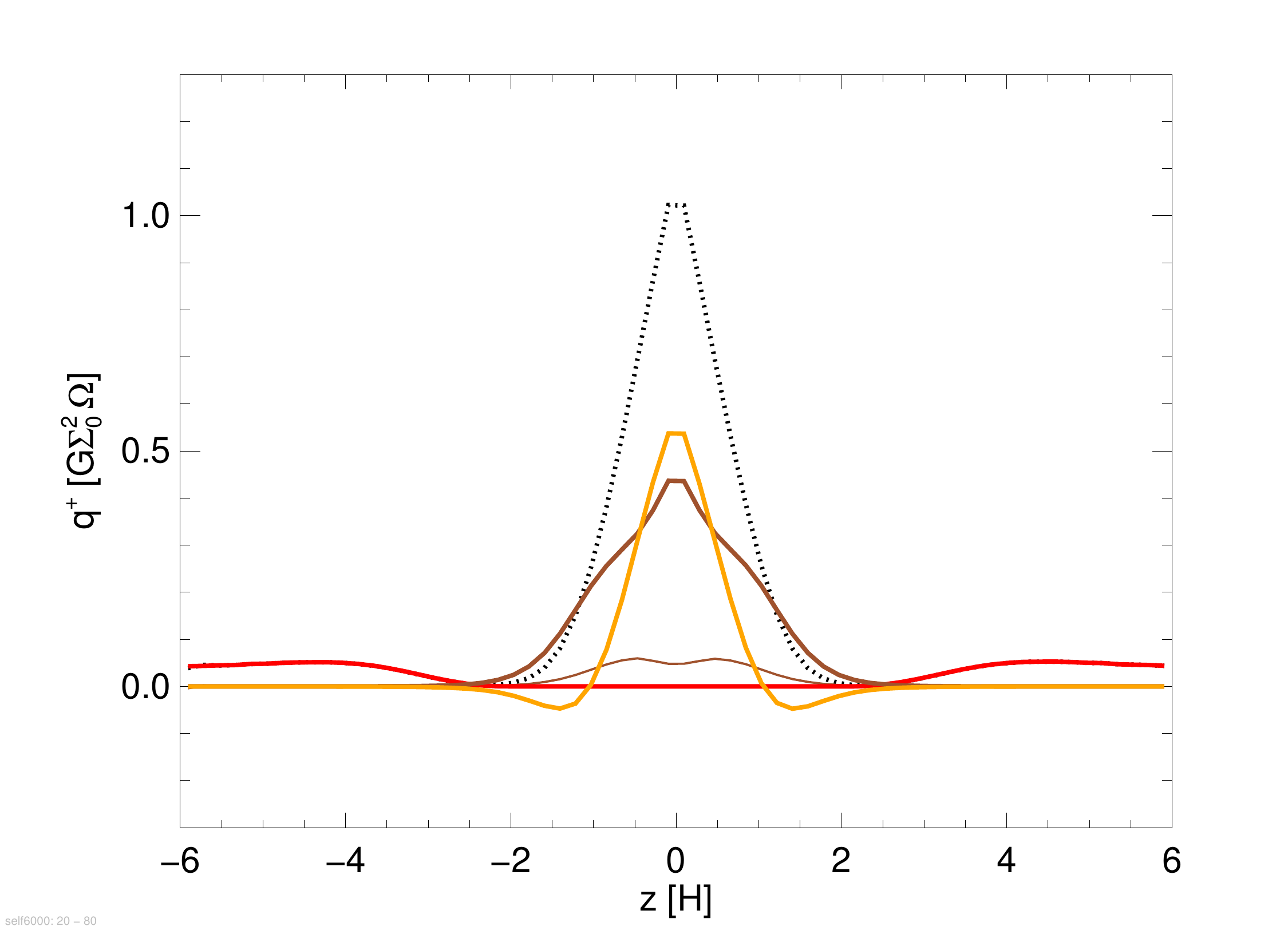}
\includegraphics[width=7cm]{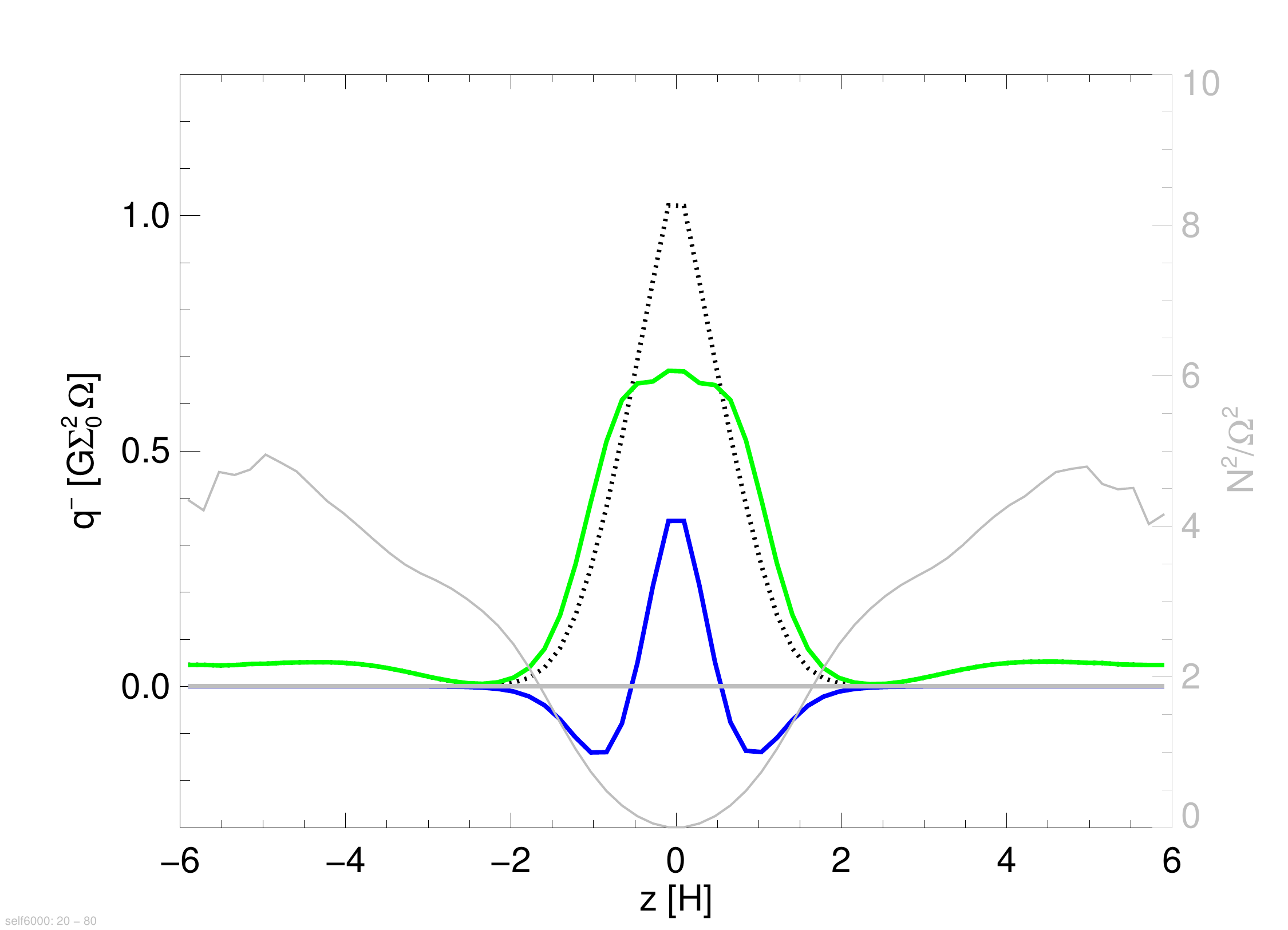}
\caption{Vertical profiles of heating rates (upper) and cooling rates (lower) in the fiducial run. In the upper panel, the red, orange, brown, and brown thin curves are, respectively, the irradiation heating, compressional heating, shock heating, and grid-scale dissipation; the black dotted curve is their sum. In the lower panel, the green and blue curves are the cooling rates associated with radiative diffusion and advection, respectively; the black dotted curve is their sum. In the lower panel, the Brunt--V\"{a}is\"{a}l\"{a} frequency squared divided by $\Omega^2$ is shown as the grey curve, whose axis is on the right.}
    \label{fig:vert_heating}
\end{figure}

The time and horizontally-averaged thermal energy equation in a steady state can be written as
\begin{align}
\overline{\left<\frac{\p ev_z}{\p z}\right>} + \overline{\left<\frac{\p F_z}{\p z}\right>} = \overline{\left<q_\text{shock}\right>} + \overline{\left<{q}_\text{num}\right>} -\overline{\left<p\nabla\cdot\bm{v}\right>} + \overline{\left<q_\text{irr}\right>}.
\end{align}
Here, heating (cooling) terms are, respectively, gathered in RHS (LHS), and are shown in the upper (lower) panel in figure \ref{fig:vert_heating}. The profiles of the total heating rate and the total cooling rate match, which means that a thermal balance well holds.

First, we examine the heating rates. The gravito-turbulence here dissipates mainly through shock heating $\overline{\left<q_\text{shock}\right>}$. However, the compressional heating (or radiation damping), $-\overline{\left<p\nabla\cdot\bm{v}\right>}$, also plays a major role; it is actually comparable to the shock heating near the midplane. We note that these two major dissipation processes are not quite dependent on the grid resolution unlike the grid-scale dissipation $\overline{\left<{q}_\text{num}\right>}$, which is minor here as shown in the figure. The irradiation heating $\overline{\left<q_\text{irr}\right>}$ only occurs above the photosphere of the visible light at $|z|/H \sim 3$. When vertically integrated, each heating rate is computed as $\overline{\Left<q_\text{shock}\Right>} = 0.888$, $-\overline{\Left<p\nabla\cdot\bm{v}\Right>} = 0.510$, $\overline{\Left<{q}_\text{num}\Right>} = 0.137$, and $\overline{\Left<q_\text{irr}\Right>} = 0.281$, in terms of $G\Sigma_0^2\Omega H$.

Here we describe how the temperature structure shown in figure \ref{fig:vert_dens} is formed. The heated upper layers emit the blackbody radiation (sometimes called reprocessed irradiation), and a part of it heats the midplane region. The base temperature in the region between the heated upper layers is determined by this reprocessed irradiation.
On the other hand, the peak temperature of $30$ K at the midplane is determined by dissipation of the turbulence (described in the previous paragraph). Therefore, if there is no dissipation of the turbulence, the temperature profile near the midplane will be flat at the base temperature. This is actually seen when the surface density is small or the grazing angle is large \citep[see sections \ref{sec:sigma} and \ref{sec:dependence_theta}; see also][]{Hirose:2011gi}. 

Next we examine the cooling rates. The radiative cooling $\overline{\left<{\p F_z}/{\p z}\right>}$ dominates the advective cooling $\overline{\left<{\p ev_z}/{\p z}\right>}$ at all heights. Actually, the advection picks up considerable amount of the dissipated energy near the midplane ($\overline{\left<{\p ev_z}/{\p z}\right>}$ is positive), but dumps all of it around $|z|/H \sim 1$ ($\overline{\left<{\p ev_z}/{\p z}\right>}$ is negative), which then the radiative diffusion takes over. When vertically-integrated, they are computed as $\overline{\Left<{\p F_z}/{\p z}\Right>} = 1.82$ and $\overline{\Left<{\p ev_z}/{\p z}\Right>} = 7\times10^{-6}$, in terms of $G\Sigma_0^2\Omega H$. The fact that the vertically-integrated advective cooling is negligible means that the dissipated energy is eventually ejected from the box via radiation only. In the lower panel, the profile of the hydrodynamical Brunt--V\"{a}is\"{a}l\"{a} frequency $N$ squared (divided by $\Omega$ squared),
\begin{align}
\frac{N^2}{\Omega^2} \equiv \frac{d\ln\overline{\left<\rho\right>}}{d\ln z} - \frac{1}{\overline{\left<\Gamma_1\right>}}\frac{d\ln\overline{\left<p\right>}}{d\ln z},
\end{align}
is also shown, where $\Gamma_1 \equiv d\ln p/d\ln\rho$ is the generalized adiabatic exponent.\footnote{$\Gamma_1$ is also precomputed as a function of $\rho$ and $e/\rho$ as one of the EOS tables.} Since it is consistently positive near the midplane (note that the axis for $N^2/\Omega^2$ is on the right), the advection is not associated with thermal convection, but is driven by the upward flows created by the collisions of the turbulent density waves. We remind readers that this upward flows also created the dynamical pressure in the hydrostatic balance (section \ref{sec:hydrostatic}).

\subsubsection{Turbulent velocity and sound velocity}

\begin{figure}  
\includegraphics[width=7cm]{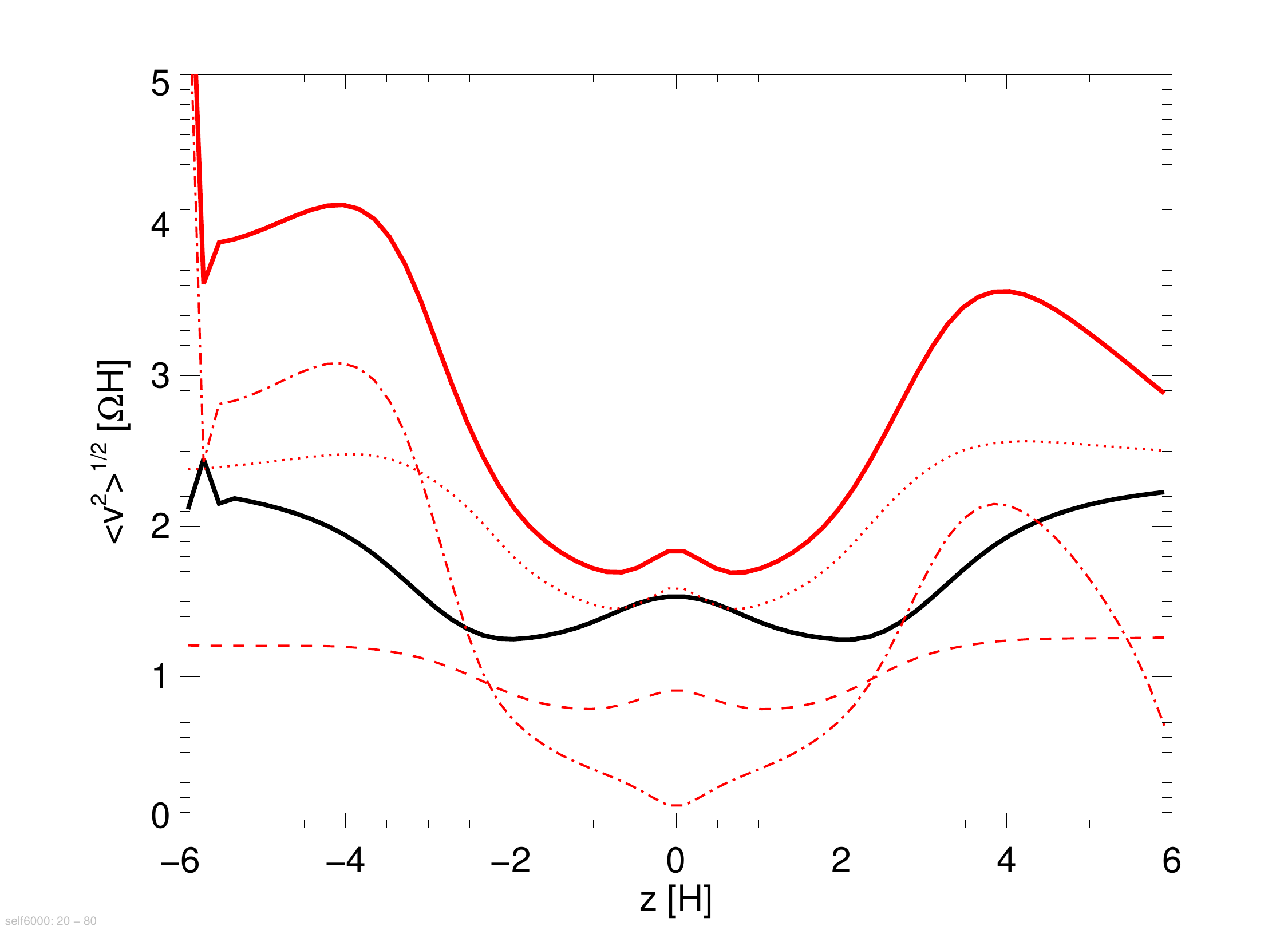}
\caption{Vertical profiles of sound velocity $\overline{\left<c_\text{s}\right>}$ (black) and turbulent velocities, $\sqrt{\overline{\left<\delta\bm{v}^2\right>}}$ (red), $\sqrt{\overline{\left<\delta v_x^2\right>}}$ (red dotted), $\sqrt{\overline{\left<\delta v_y^2\right>}}$ (red dashed), and $\sqrt{\overline{\left<\delta v_z^2\right>}}$ (red dashed-dotted), in the fiducial run.}
\label{fig:vert_vturb}
\end{figure}

\begin{figure*}
\includegraphics[width=14cm]{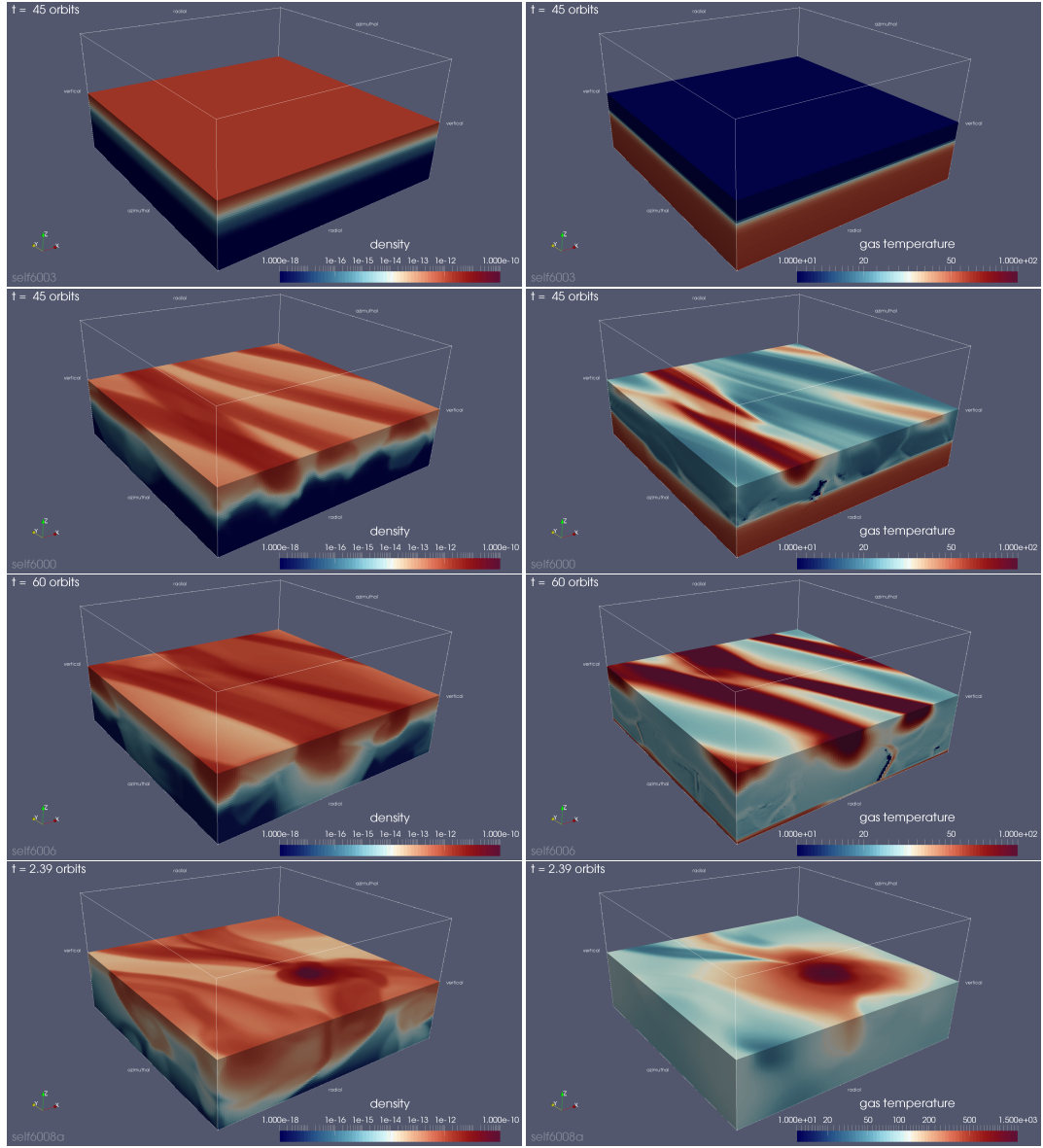}
\caption{Density (left column) and gas temperature (right column) snapshots of the lower half of the simulation box for, from the top to the bottom, $\Sigma = 60$, 100 (fiducial), 200, and 300 g cm$^{-2}$.
Note that, in the right column, the colour range in the bottom panel ($10 \le T \le 1500$ K) is different from that in other panels ($10 \le T \le 100$ K).}
\label{fig:snapshots_sigma}
\end{figure*}

{\color{black}Knowing the detailed turbulence properties are crucial to study the dust dynamics in self-gravitating protoplanetary discs \citep{Booth:2016is,2016MNRAS.459..982S}.}
Figure \ref{fig:vert_vturb} shows time-averaged vertical profiles of the turbulent velocity defined as $\delta\bm{v} \equiv (\bm{v} - \bm{v}_\text{K}) - \left<\bm{v} - \bm{v}_\text{K}\right>$ as well as {\color{black}the sound velocity $c_\text{s} \equiv \sqrt{\Gamma_1 p/\rho}$}. The figure reveals that the gravito-turbulence here is supersonic at all heights including the midplane.\footnote{The profiles of the turbulent velocities are not symmetric due to downflows near the boundaries that were consistent for some periods in this particular simulation.} {\color{black}This is consistent with \citet{Shi:2014gz}, in which the turbulent flows are supersonic when the cooling time is rather short $\beta \le 4$ (c.f. $\overline{\beta}_\text{eff} = 3.17$ in our simulation).}
The turbulent velocity varies with disc height by a factor of a few, and the $x$-component of the turbulent velocity contributes most near the midplane, which are also consistent with {\color{black}\citet{Shi:2014gz}}.

As discussed in \citet{Gammie:2001hv}, the ratio of the sound velocity to the Keplerian velocity needs to be small enough for the local model to be applicable. The Keplerian velocity in our study is computed as $v_\text{K} = {a\Omega} = ({a}/{H}){\Omega H} = 15.7{\Omega H}$.
Therefore the ratio of the sound velocity, which reads $\sim 1.5\Omega H$ in figure \ref{fig:vert_vturb}, to the Keplerian velocity is about $0.096$. This might not be small enough since \citet{Gammie:2001hv} derived that the ratio needs to be much less than 0.12 based on Fourier analysis of the surface density in his simulations. We will discuss this issue from different point of views in section \ref{sec:discussion_locality}.

\subsection{Dependence on the surface density $\Sigma$}\label{sec:sigma}

In this section, we examine dependence on the surface density $\Sigma$. Here, we change $\Sigma$ from $30$ to $300$ (specifically 30, 60, 80, 100, 150, 200, 250, and 300) g cm$^{-2}$, with all other parameters being fixed. 

When $\Sigma \le 60$ g cm$^{-2}$,
the flow is found to be laminar. In the range of $80 \le \Sigma \le 250$ g cm$^{-2}$, the gravito-turbulence is sustained.
When $\Sigma = 300$ g cm$^{-2}$, a gravitationally-bounded clump was formed during the initial transient, which increases its mass and eventually underwent a rapid collapse around $t = 2.4$ orbits. The collapse, which is actually the same physical process as the first core collapse in the star formation, couldn't be resolved with our fixed numerical grid, the simulation was stopped there (see details for Appendix \ref{sec:clump}).
In figure \ref{fig:snapshots_sigma}, we compare snapshots of density and gas temperature for $\Sigma = 60$, $100$, $200$, and $300$ g cm$^{-3}$. The snapshot for $\Sigma = 300$ g cm$^{-3}$ was taken just before the simulation was stopped.

\begin{figure}
\includegraphics[width=7cm]{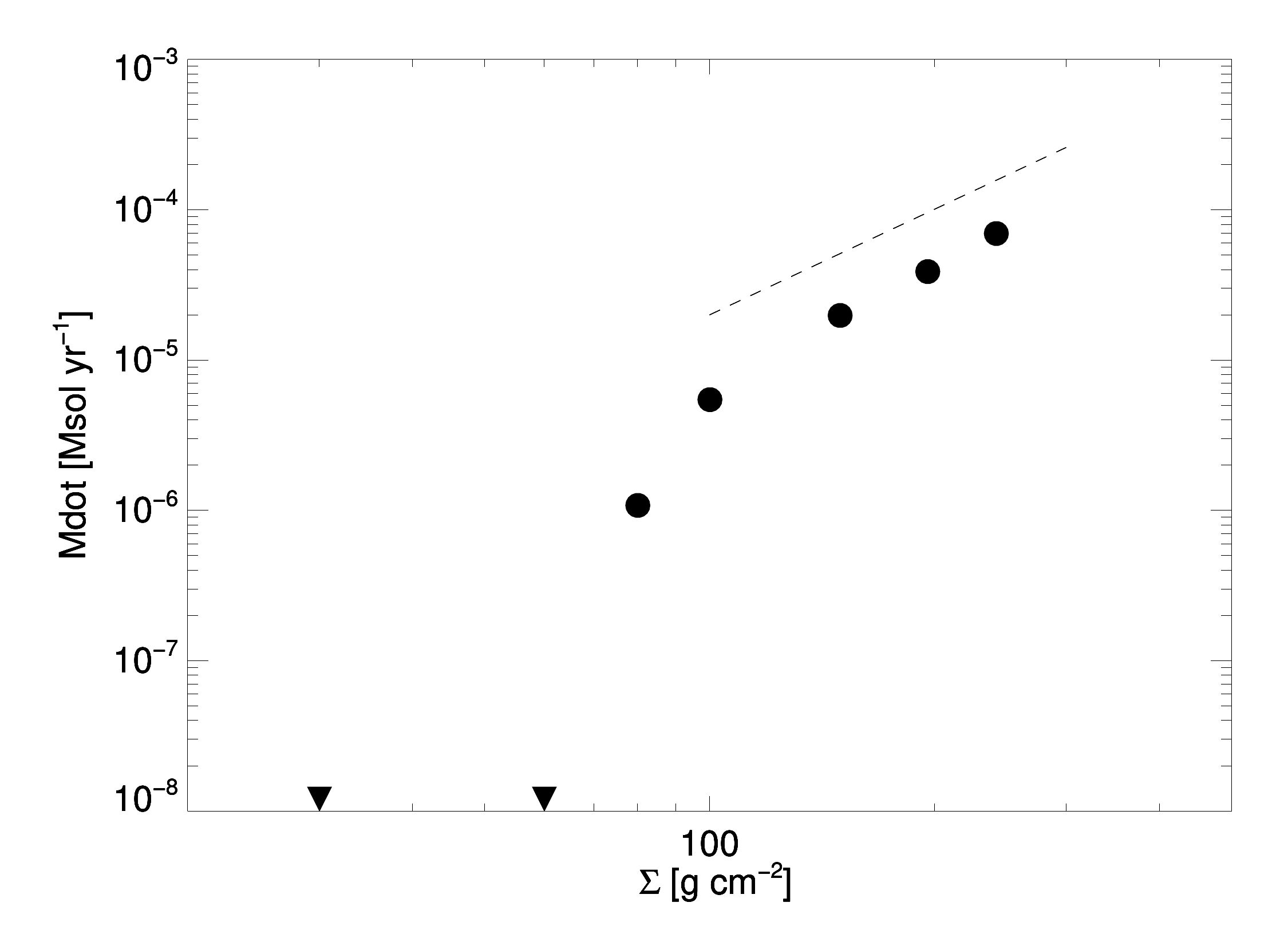}
\includegraphics[width=7cm]{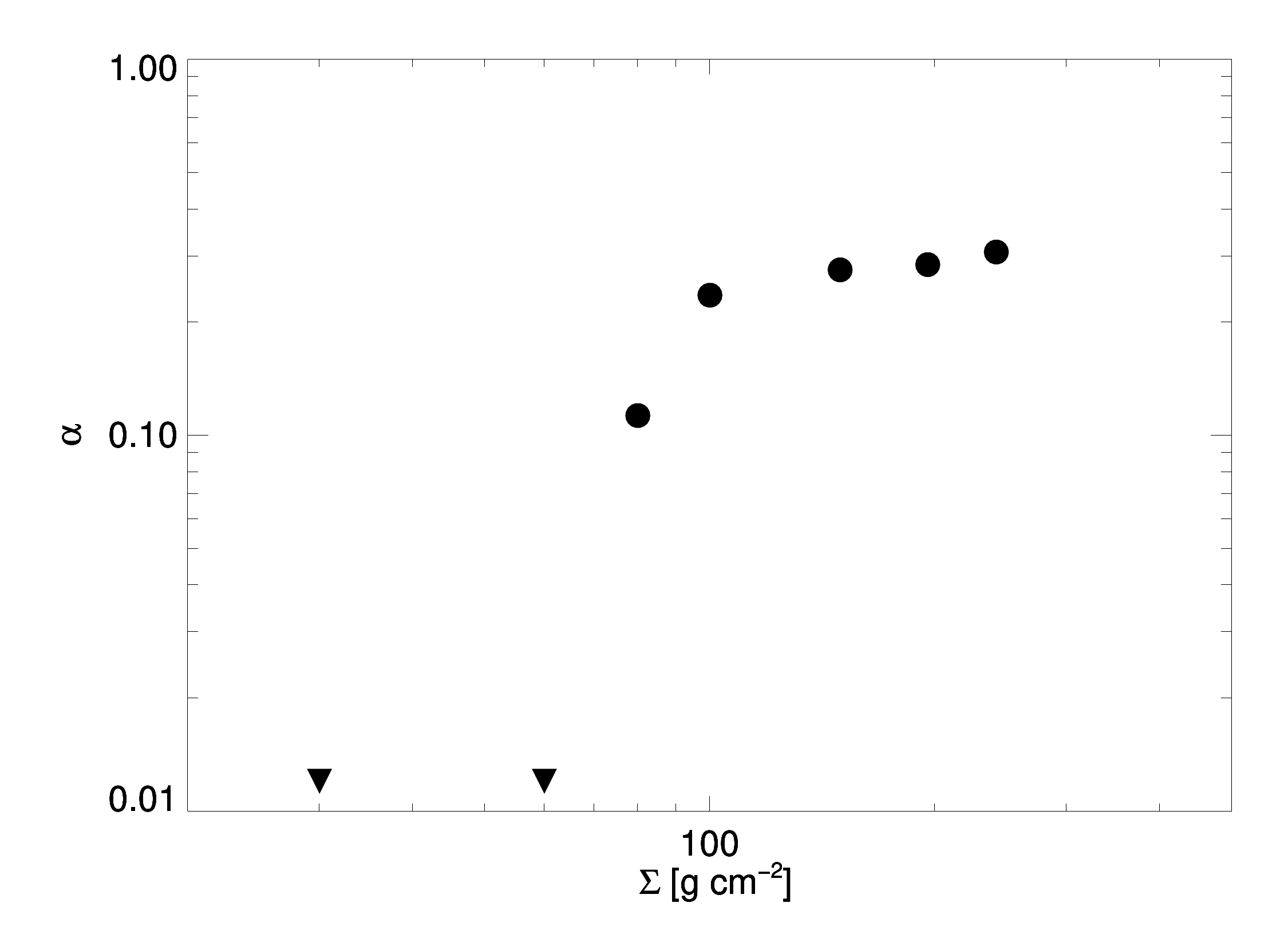}
\includegraphics[width=7cm]{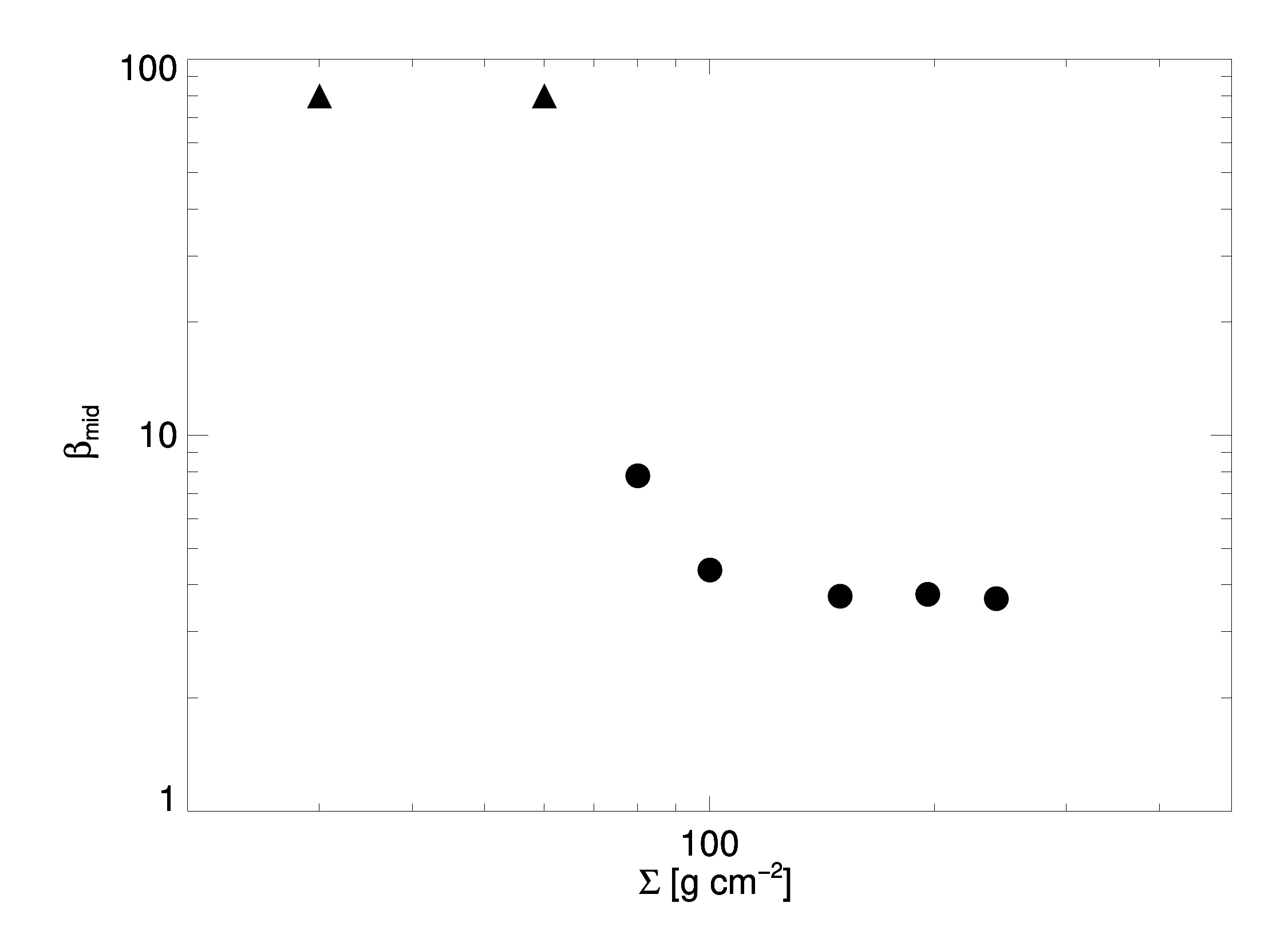}
\includegraphics[width=7cm]{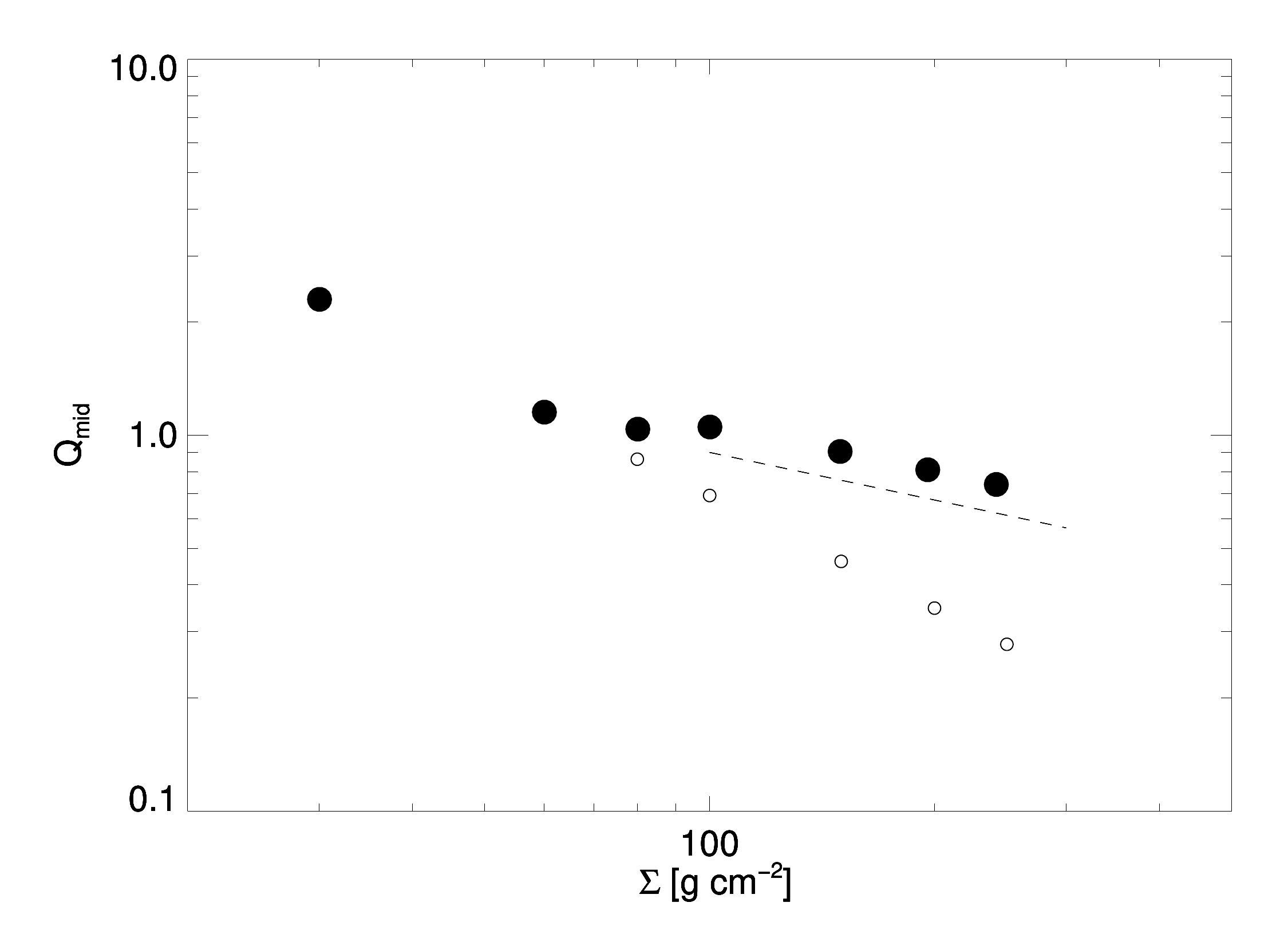}
\caption{Dependence on the surface density $\Sigma$ of, from the top to the bottom, the stress $W_{xy}$ (in terms of $\dot{M}$), Shakura-Sunyaev's $\alpha$, cooling time of the disc body $\overline{\beta}_\text{mid}$, and Toomre parameter of the disc body $\overline{Q}_\text{mid}$, for the grazing angle $\theta = 0.02$. In the top panel, the dashed line denotes $\sim \Sigma^{7/3}$. In the bottom panel, the small open symbols are the initial values, and the dotted line denotes $\sim \Sigma^{-1/3}$. The upward (downward) triangles means that the values are above (below) the displayed range.}
\label{fig:sigma_vs}
\end{figure}

Figure \ref{fig:sigma_vs} shows how various quantities depend on $\Sigma$.
The first panel shows the dependence of the time-averaged, vertically-integrated stress $W_{xy} \equiv \overline{\Left<g_xg_y/4\pi G\Right>} + \overline{\Left<\rho v_x\delta v_y\Right>}$ in terms of $\dot{M}$ via equation (\ref{eq:mdot}). When the flow is laminar in $\Sigma \le 60$ g cm$^{-2}$, the values of $\dot{M}$ is below the displayed range. When the gravito-turbulence is sustained, $\dot{M}$ is strongly correlated with $\Sigma$; specifically, $\dot{M}(\Sigma) \sim \Sigma^{7/3}$ for $100 \le \Sigma < 300$ g cm$^{-2}$, where $\dot{M}$ changes from $10^{-5}$ to $10^{-4}$ $M_\odot$ yr$^{-1}$.
In that range of $\Sigma$, the fraction of the Reynolds stress in the total stress gradually increases with $\Sigma$, from $0.45$ to $0.56$.

The Shakura \& Sunyaev's $\alpha$ is almost constant at {\color{\ccorange}$\sim 0.25$} in the gravito-turbulence regime as shown in the second panel in figure \ref{fig:sigma_vs}. 
This means that the equilibrium thermal pressure also rises as the surface density increases, in a similar way to the stress $W_{xy}$ shown in the first panel.
{\color{\ccred}We note that $\alpha$ here is different from the conventional $\alpha$ by a factor of $2/3\gamma$ (see equation \ref{eq:alpha}); the value of $\sim 0.25$ here therefore corresponds to $\sim 0.1$ in terms of the conventional $\alpha$ with $\gamma = 5/3$. Although this is rather close to the maximum $\alpha$ sustainable in the gravito-turbulence found by \citet{Rice:2005fl}, we note that the value of $\alpha$ may change with the radius (see discussion in section \ref{sec:conclusion}).
}

The third panel in figure \ref{fig:sigma_vs} shows how {\color{\ccorange}the time-averaged cooling time near the midplane $\overline{\beta}_\text{mid}$, defined in equation (\ref{eq:beta_mid}),}
changes with $\Sigma$. 
In the laminar flow range of $\Sigma$ ($\le 60$ g cm$^{-2}$), since there is little dissipation (and thus little cooling rate) {\color{\ccorange} while the thermal energy is kept finite due to the reprocessed irradiation,} $\overline{\beta}_\text{mid}$ become very large and are not displayed. When the gravito-turbulence is sustained, {\color{\ccorange}the lowest $\Sigma$ case exhibits a larger $\overline{\beta}_\text{mid}$ for a similar reason while in other cases} $\overline{\beta}_\text{mid}$ tends to be {\color{\ccorange} constant at $\sim 4$.}

The bottom panel in figure \ref{fig:sigma_vs} shows the time-averaged Toomre parameter near the midplane, $\overline{Q}_\text{mid}$, as a function of $\Sigma$. The initial ${Q}_\text{mid}$ (small open circles) is proportional to $\Sigma^{-1}$ since the initial sound velocity is the same in all cases. For $\Sigma \ge 100$ g cm$^{-2}$, where the gravito-turbulence is sustained, $\overline{Q}_\text{mid}$ is larger than the initial value because the sound velocity is raised by dissipation of the turbulence. More importantly, $\overline{Q}_\text{mid}$ decreases (as $\sim \Sigma^{-1/3}$) down to $\sim 0.7$ at $\Sigma = 250$ g cm$^{-2}$, beyond which fragmentation occurred.
This indicates that the critical value of $\Sigma$ corresponds to the minimum value of the Toomre parameter that can be realized in the gravito-turbulence. 
On the other hand, the critical value of $\Sigma$ seems not related to a specific value of the cooling time $\overline{\beta}_\text{mid}$ since it stays almost constant in the gravito-turbulence regime.

\begin{figure}
\includegraphics[width=7cm]{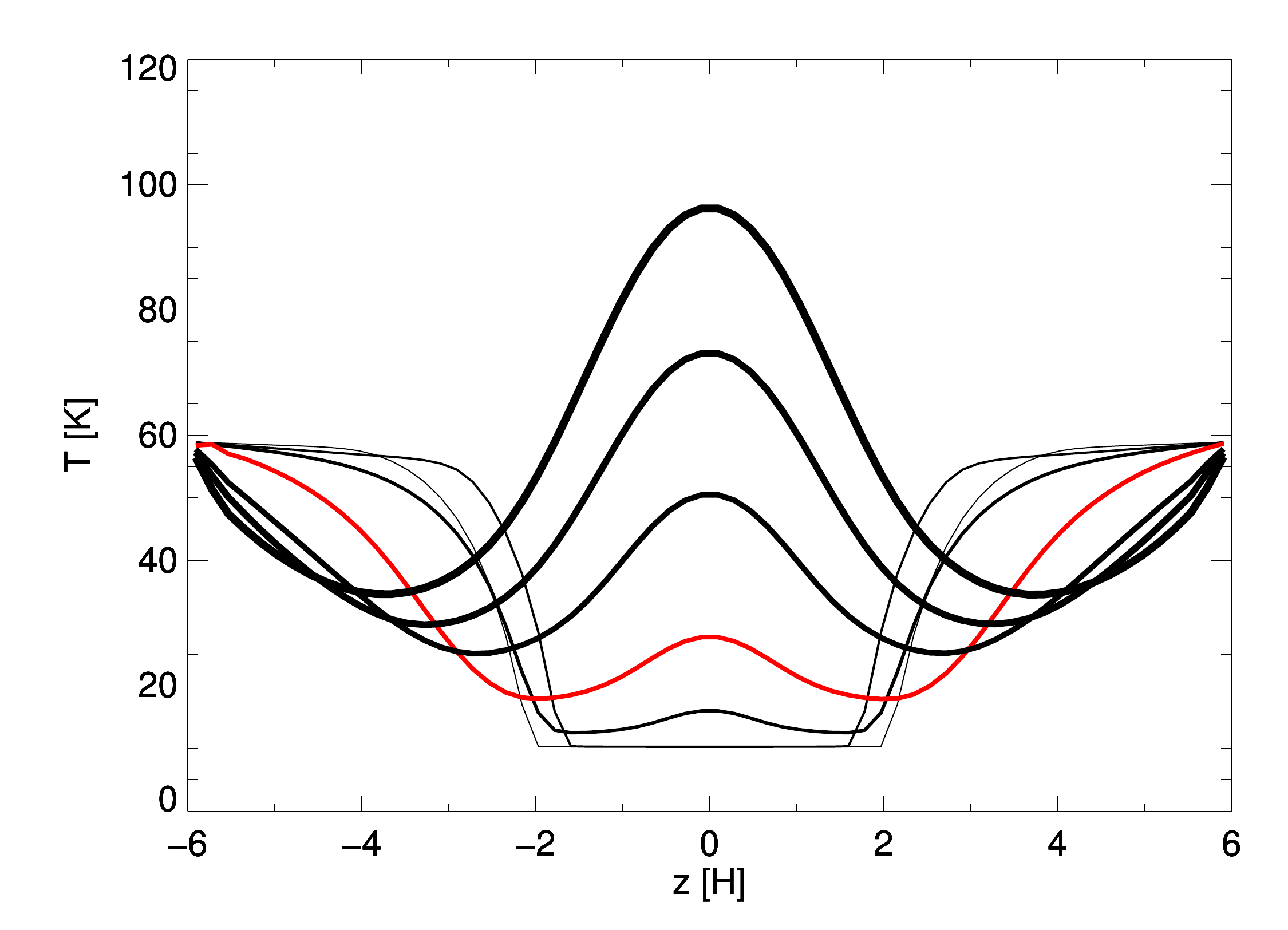}
\caption{Time-averaged vertical profiles of the gas temperature for the surface density $\Sigma = 30$, 60, 80, 100, 150, 200, and 250 g cm$^{-2}$. Thicker the curve, larger the $\Sigma$. The fiducial run ($\Sigma = 100$ g cm$^{-2}$) is coloured with red.}
    \label{fig:sigma_vs_temp}
\end{figure}

Figure \ref{fig:sigma_vs_temp} shows dependence on $\Sigma$ of the vertical profile of gas temperature.\footnote{\color{black}We did not plot the radiation temperature for clarity. As in the fiducial run (figure \ref{fig:vert_dens}), the radiation temperature always follows the gas temperature near the midplane, and continues to decrease monotonically in the upper layers.} When the flow is laminar ($\Sigma \le 60$ g cm$^{-2}$), the temperature of the disc interior is flat at the base temperature ($T \sim 10$ K) that is determined by the reprocessed irradiation from the hot upper layers ($T \sim 60$ K) \citep{Chiang97}.\footnote{The location of the boundaries between the upper layers and the interior are lower in the case of larger $\Sigma$ since downflows in the upper layers due to hydrostatic imbalance are stronger and thus the disc is more compressed.} On the other hand, when the gravito-turbulence is sustained for $\Sigma \ge 80$ g cm$^{-2}$, the interior is also heated by dissipation of the turbulence and there appears a peak at the midplane. As $\Sigma$ is increased, the midplane temperature increases, and thus the wave length of the axisymmetric mode of GI becomes longer as seen in the snapshots (the second and third panels) in figure \ref{fig:snapshots_sigma}. The midplane temperature takes a maximum value of $\sim 100$ K at $\Sigma = 250$ g cm$^{-2}$. Therefore, at least at this radius of 50 AU, heating by the gravito-turbulence will not activate MRI.

\subsection{Dependence on the grazing angle $\theta$}\label{sec:dependence_theta}
\begin{figure}
\includegraphics[width=7cm]{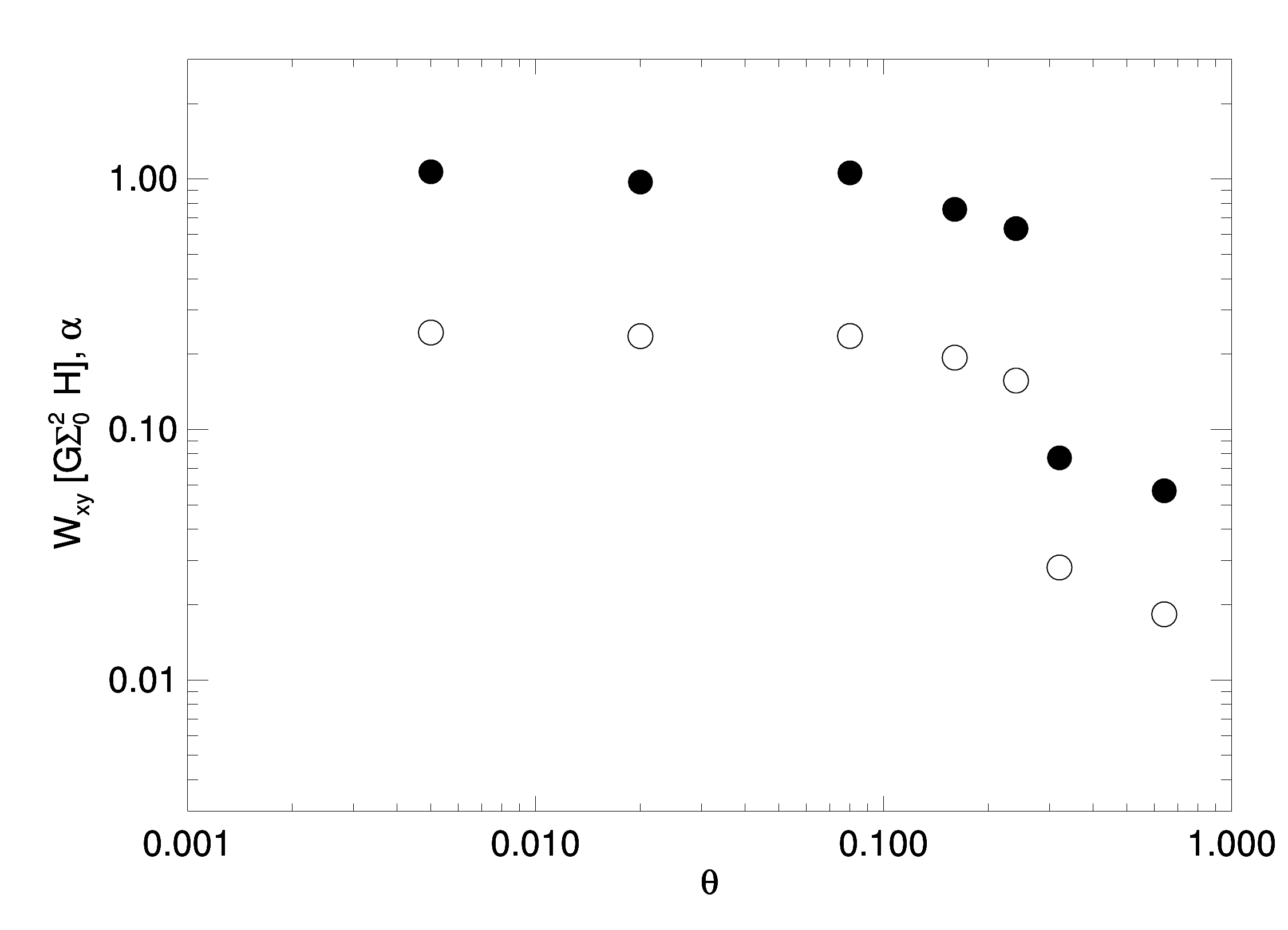}
\includegraphics[width=7cm]{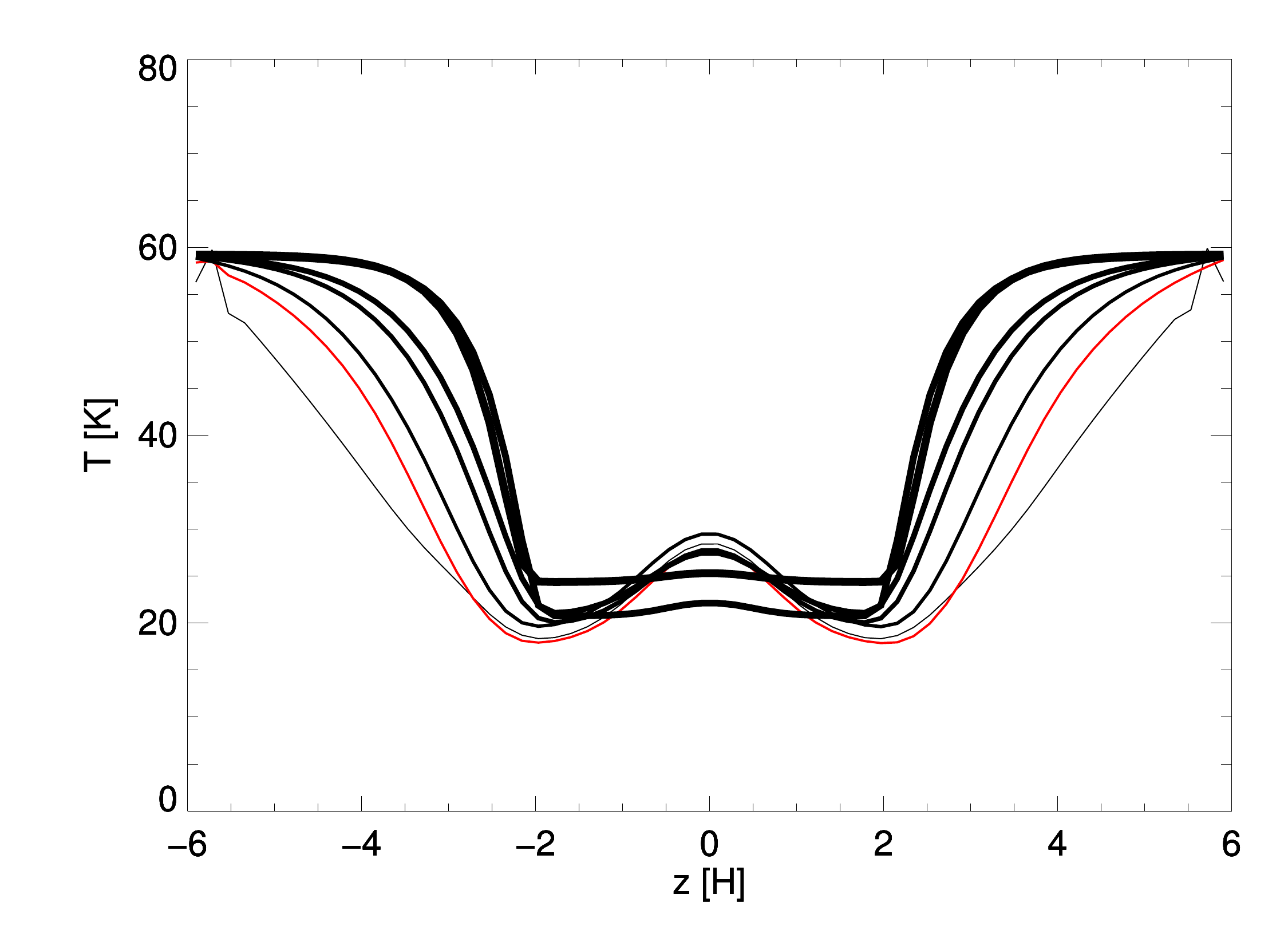}
\includegraphics[width=7cm]{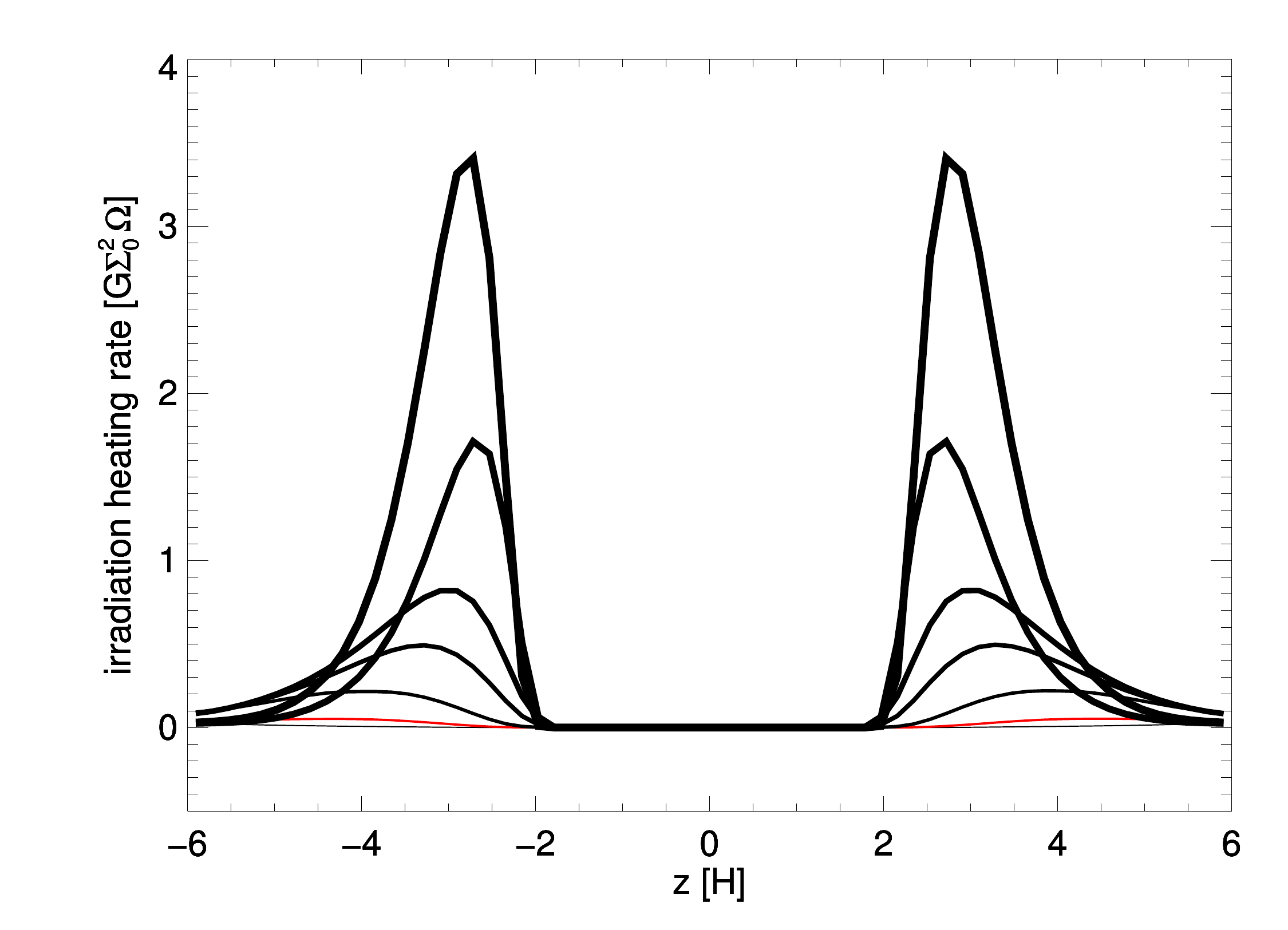}
\caption{Dependence on the grazing angle $\theta$. The top panel shows dependence of the vertically-integrated stress $W_{xy}$ (open circle) and Shakura-Sunyaev's $\alpha$ (filled circle). The middle panel shows the time-averaged gas temperature profiles for $\theta = 0.005$, 0.02 (fiducial, coloured red), 0.08, 0.16, 0.24, 0.32, and 0.64 (the thicker the curve, the larger the grazing angle $\theta$). The bottom panel is the same as the middle panel, but shows the time-averaged irradiation heating profiles.}
    \label{fig:theta_vs_stress}
\end{figure}

\begin{figure*}
\includegraphics[width=14cm]{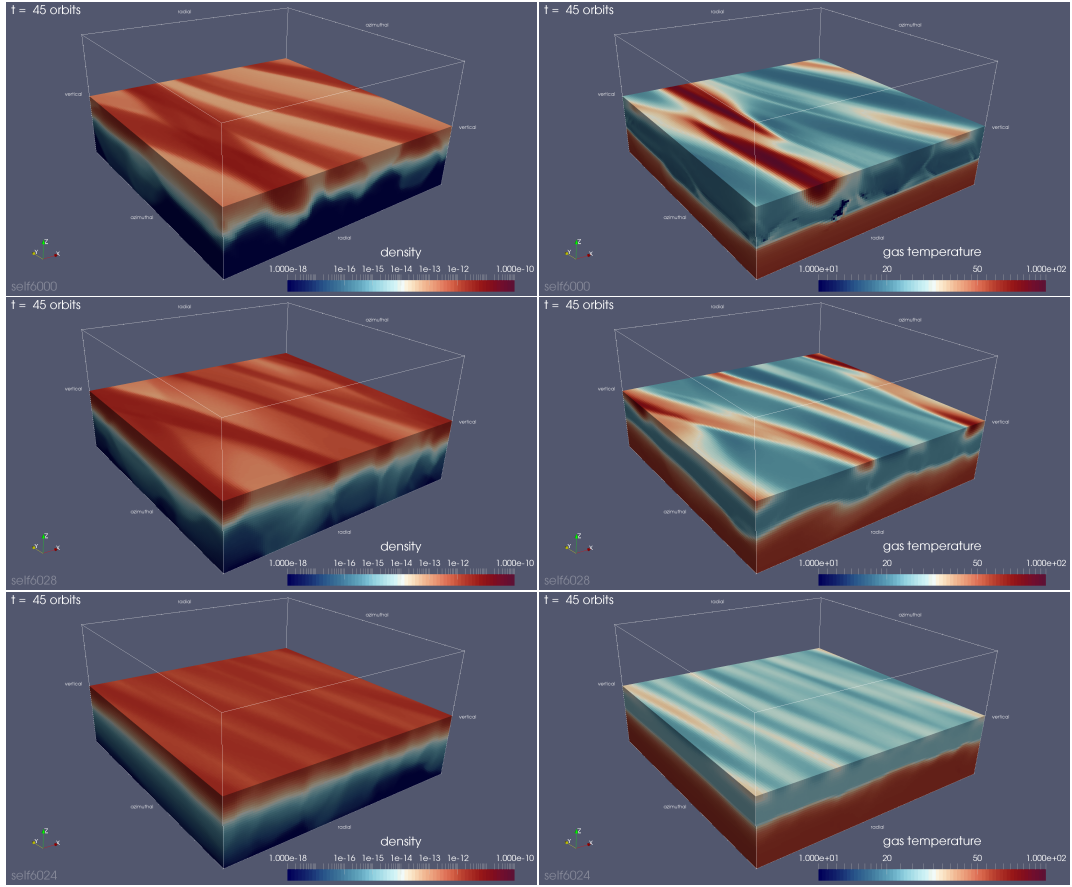}
\caption{Snapshots of the density (left) and gas temperature (right) for $\theta$ = 0.02 (fiducial; top), 0.24 (middle), and 0.64 (bottom).}
\label{fig:snapshots_theta}
\end{figure*}
In this section, we will see dependence on the grazing angle $\theta$, ranging from $0.005$ to $0.64$, for the surface density $\Sigma = 100$ g cm$^{-2}$, the same as in the fiducial run.
Note that the amount of irradiation energy injected into the simulation box per unit time is also changed as $\sim \sin\theta$ (see Appendix \ref{sec:irradiation}).
The initial condition of these simulations was taken from a snapshot in the steady state of the fiducial run. (Therefore, the unit length $H$ is unchanged.)

The top panel in figure \ref{fig:theta_vs_stress} shows dependence of the total stress $W_{xy}$ and $\alpha$.
As $\theta$ is increased, both stay almost constant until $\theta \sim 0.1$, beyond which they decrease gradually, and then drop suddenly at $\theta = 0.32$. At $\theta = 0.64$, the gravito-turbulence is very weak and the total stress $W_{xy}$ is about $1/20$ the constant value at the lower $\theta$s. Figure \ref{fig:snapshots_theta} shows snapshots of density and gas temperature for $\theta = 0.64$ and $0.24$ as well as for the fiducial run ($\theta = 0.02$).

The middle and bottom panels in figure \ref{fig:theta_vs_stress} shows, respectively, how vertical profile of gas temperature and irradiation heating changes with $\theta$. When $\theta \le 0.24$, the temperature profiles are similar, having a small peak at the midplane, which means that the main heating source there is the dissipation of the turbulence. Temperatures at $|z|/H\sim 4$ rise as $\theta$ increases since the incoming irradiation energy increases as $\sim \sin\theta$. When $\theta \ge 0.32$, the temperatures near the midplane are greatly affected by the reprocessed irradiation. At $\theta = 0.32$, the peak temperature at the midplane is reduced because the gravito-turbulence is weakened by the heat of the reprocessed irradiation. At $\theta = 0.64$, the gravito-turbulence is almost shut off and temperatures near the midplane are mainly determined by the reprocessed irradiation.

\begin{figure*}
\includegraphics[width=14cm]{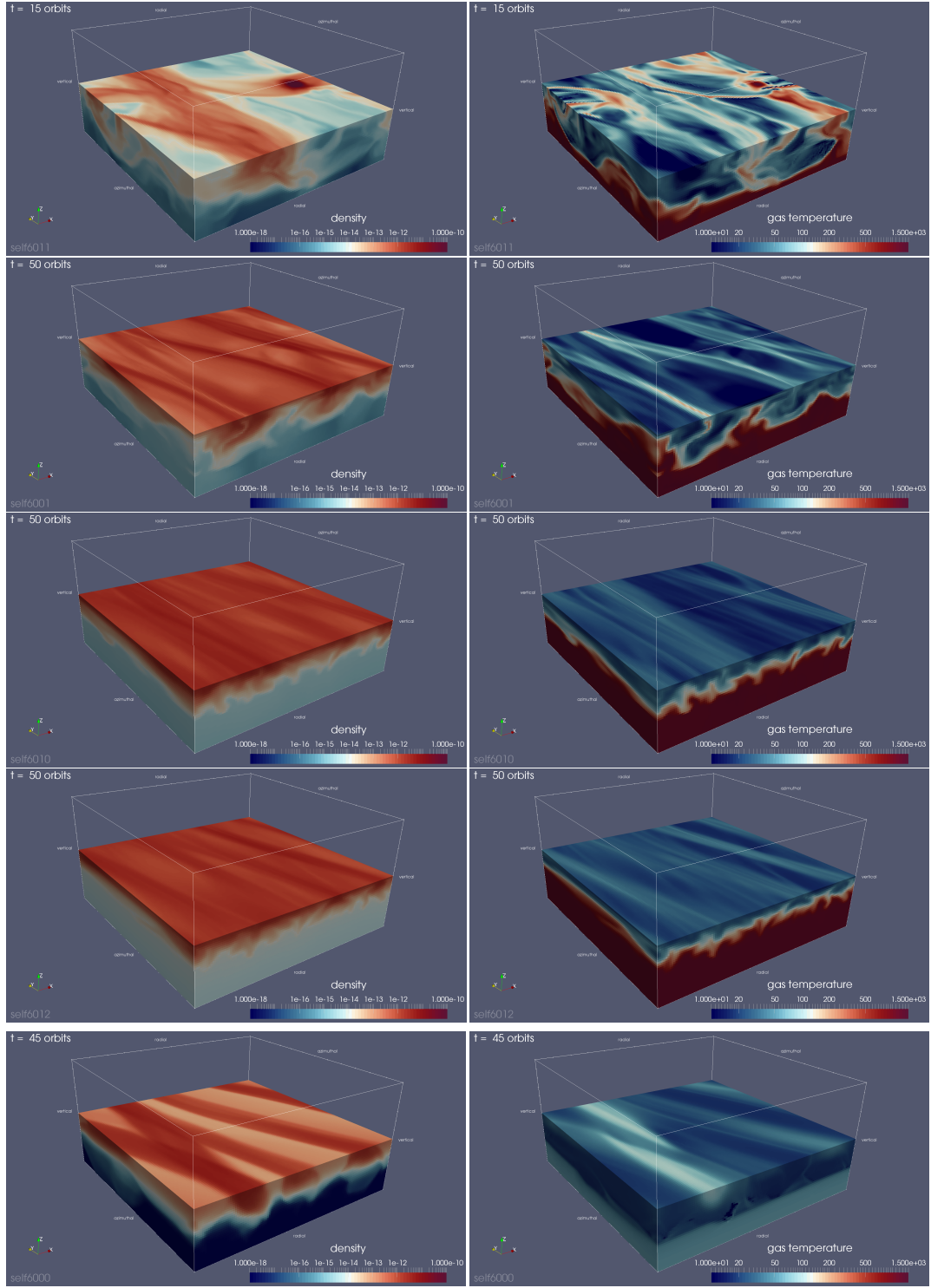}
\caption{Snapshots of the density (left) and gas temperature (right) in the simulations using the simple cooling function. From the top, the constant cooling time is $\beta = 1$, $3$, $10$, and $30$. Snapshot of the fiducial run are also shown in the bottom for reference. Note that the colour scale for temperature ranges logarithmically from 10 to $1500$ [K].}
    \label{fig:snapshots_beta}
\end{figure*}

\begin{figure}
\includegraphics[width=7cm]{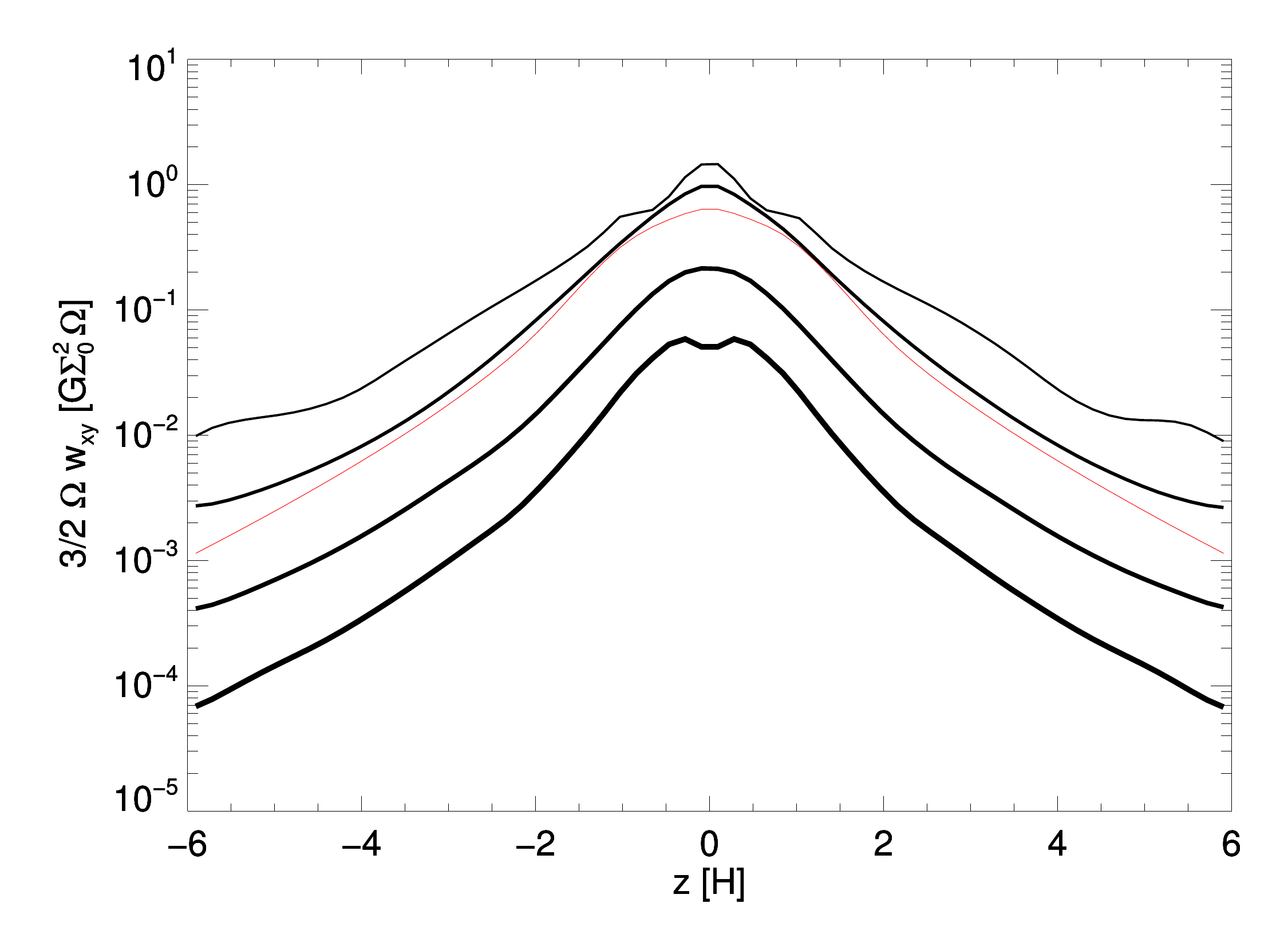}
\includegraphics[width=7cm]{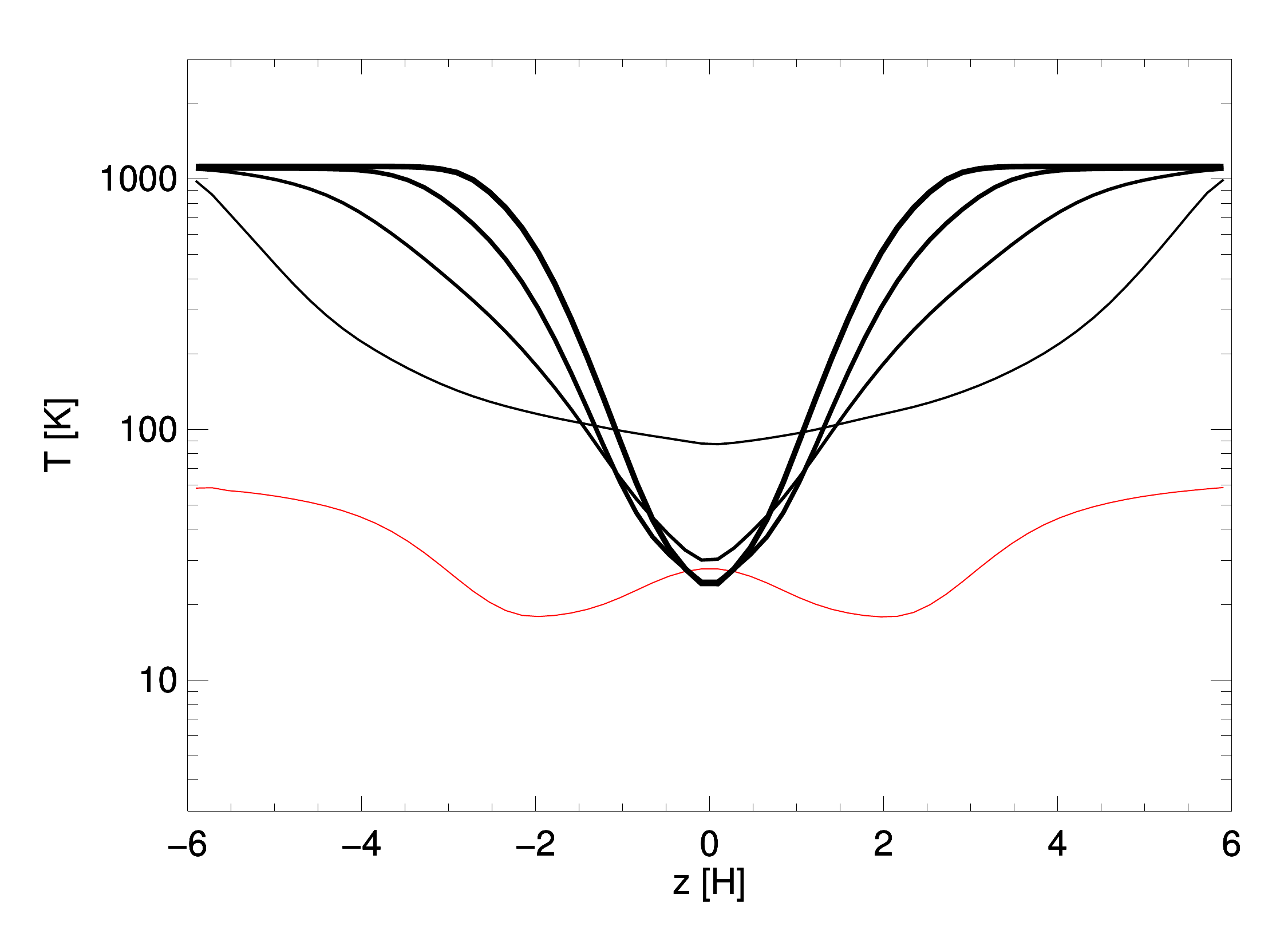}
\includegraphics[width=7cm]{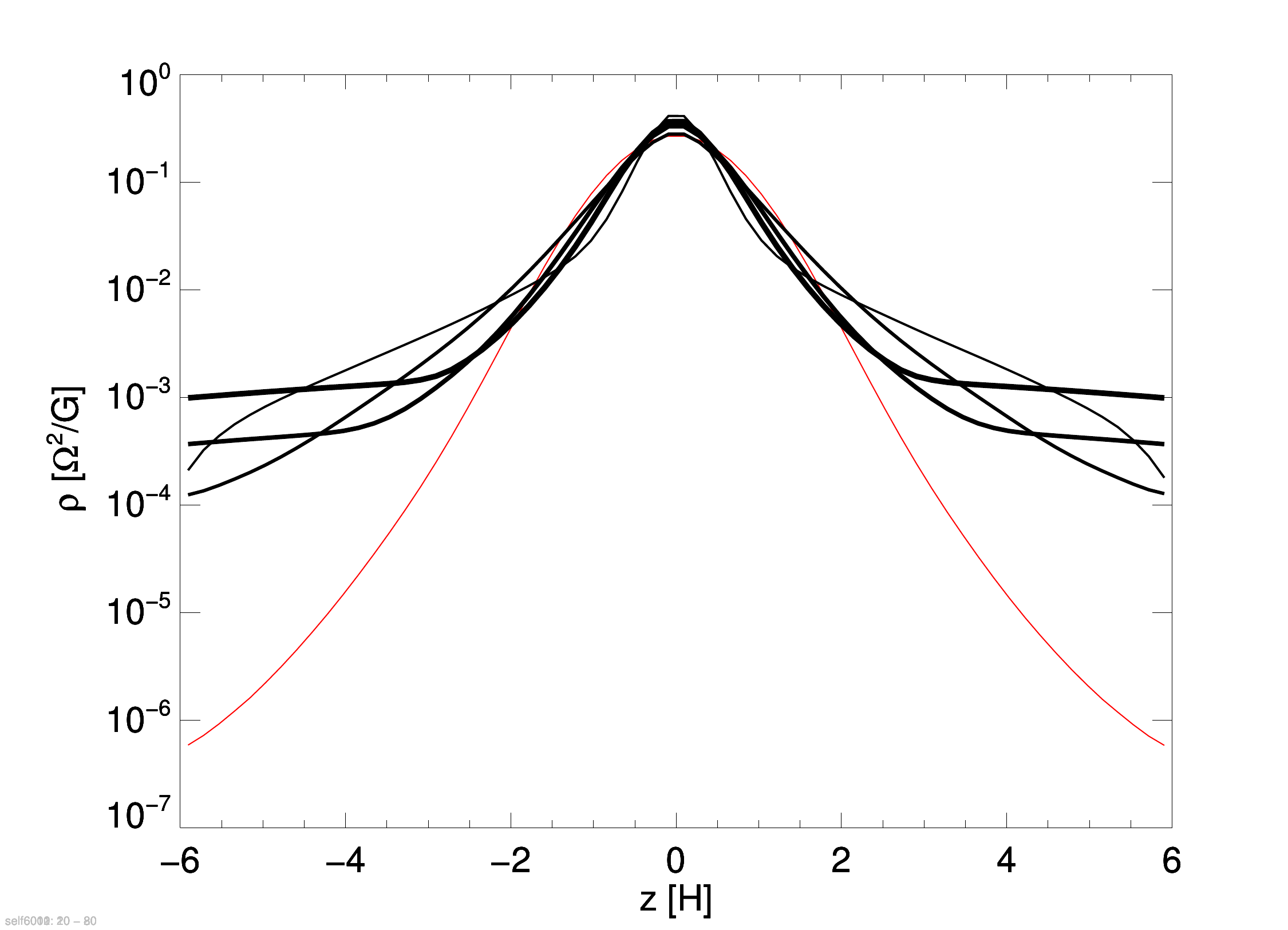}
\caption{Time-averaged vertical profiles of the stress work (top), gas temperature (middle), {\color{\ccorange}and density (bottom)} for $\beta = 1$, 3, 10, and 30 (the thicker the curve, the larger the $\beta$). The profile of the fiducial run is also shown for reference (red).}
    \label{fig:beta_vs_stress}
\end{figure}

\subsection{Cases with the simple cooling function}\label{sec:simple_cooling}
So far we have shown results using our radiative transfer solver with the FLD approximation. For comparison, we show results of simulations using the simple cooling function commonly used in gravito-turbulence simulations \citep[e.g.][]{Gammie:2001hv}. In those simulations, we simply replaced the energy equation (\ref{eq:energy_gas}) with the following equation and solve it using the predictor-corrector method:
\begin{align}
         \frac{\partial e}{\partial t} =
  -(\nabla\cdot\bm{v})p - \frac{e}{\beta\Omega^{-1}} + q_\text{irr}, \label{eq:energy_gas_simple}
\end{align}
where $\beta$ is a constant cooling time. At the same time, we dropped off the radiation energy equation (\ref{eq:energy_rad}) and the radiation force in the momentum equation (\ref{eq:motion}).
We performed four cases of $\beta = 30$, $10$, $3$, and $1$, where a snapshot of the fiducial run was used as the initial condition. After an initial transient that lasted for several orbits, they reached a steady state except for the $\beta = 1$ case. In the $\beta = 1$ case, about 10\% of the total mass was lost via the vertical boundaries during the first 10 orbits, and thus we stopped the calculation there.\footnote{The time averaging analysis was done for this first 10 orbits.}

Snapshots of density and gas temperature are compared in figure \ref{fig:snapshots_beta}.
{\color{\ccorange}Comparing with the FLD case (shown in the bottom for reference), the flow structures look quite different even in the case of $\beta = 3$, which is most close to the FLD case in terms of $\beta$ (see equation \ref{eq:beta_eff}). Especially, fine structures are seen in the cases with the simple cell-by-cell cooling function while structures look more diffusive in the FLD case, presumably due to radiative diffusion.
These indicates that 
the simple cooling function does not approximate the realistic radiative transfer.}

The top panel of figure \ref{fig:beta_vs_stress} shows profiles of time-averaged {\color{black}total} stress for the four cases as well as the FLD case.
The shape of the profiles is similar to that of the FLD case except for the $\beta = 1$ case. The profile of the $\beta = 3$ case is quantitatively similar to the FLD case{\color{\ccorange}, in which $\overline{\beta}_\text{eff} = 3.17$}. In the case of $\beta = 1$, the profile is irregular due to fragmentation as seen in the snapshot (figure \ref{fig:snapshots_beta}), indicating that the fragmentation criterion is similar to the \citet{Gammie:2001hv}'s condition (equation \ref{eq:gammie}) when the irradiation is included.

The middle panel of figure \ref{fig:beta_vs_stress} shows profiles of time-averaged gas temperature. Unlike the stress in the above, the profiles of temperature of the four cases are completely different from the profile of the FLD case.
As $\beta$ decreases, temperatures in the upper layers decreases because the cooling time is shorter, but the midplane temperature increases according to the increase of stresses. More importantly, temperatures except the midplane are consistently much higher than the FLD case, indicating that the simple cooling function with a constant $\beta$ value does not represent the realistic cooling regardless of the $\beta$ value. We remind readers that the cell-by-cell cooling time in the FLD run is actually quite short ($\beta \ll 1$) except the midplane (figure \ref{fig:cell_cooling}). {\color{black}In the cases of $\beta \ge 3$, it is seen that the gas temperatures in the upper layers are saturated just above $10^3$ K. This is because increase in the internal energy is used to dissociate H$_2$ molecules, rather than to increase the translational energy of the molecules, at that temperature (see $\Gamma_1(\rho,T)$ for the upper panel in figure \ref{fig:opacity_table}).}

{\color{\ccorange}
The bottom panel of figure \ref{fig:beta_vs_stress} shows profiles of time-averaged density.
In the simple cooling function cases, corresponding to the high temperatures at higher altitudes shown in the middle panel, the density scale height there is generally larger than that in the FLD case. On the other hand,
the density profiles near the midplane are similar (except the case of $\beta = 1$). This is the reason why the stress profiles are similar as shown in the top panel
because the profiles of gravitational potential are mostly determined by the density profiles near the midplane.
}

\section{Discussion}\label{sec:discussion}

\subsection{Numerical convergence of vertical profiles of stress work and gas temperature}\label{sec:discussion_vertical_profile}
\begin{figure}
\includegraphics[width=7cm]{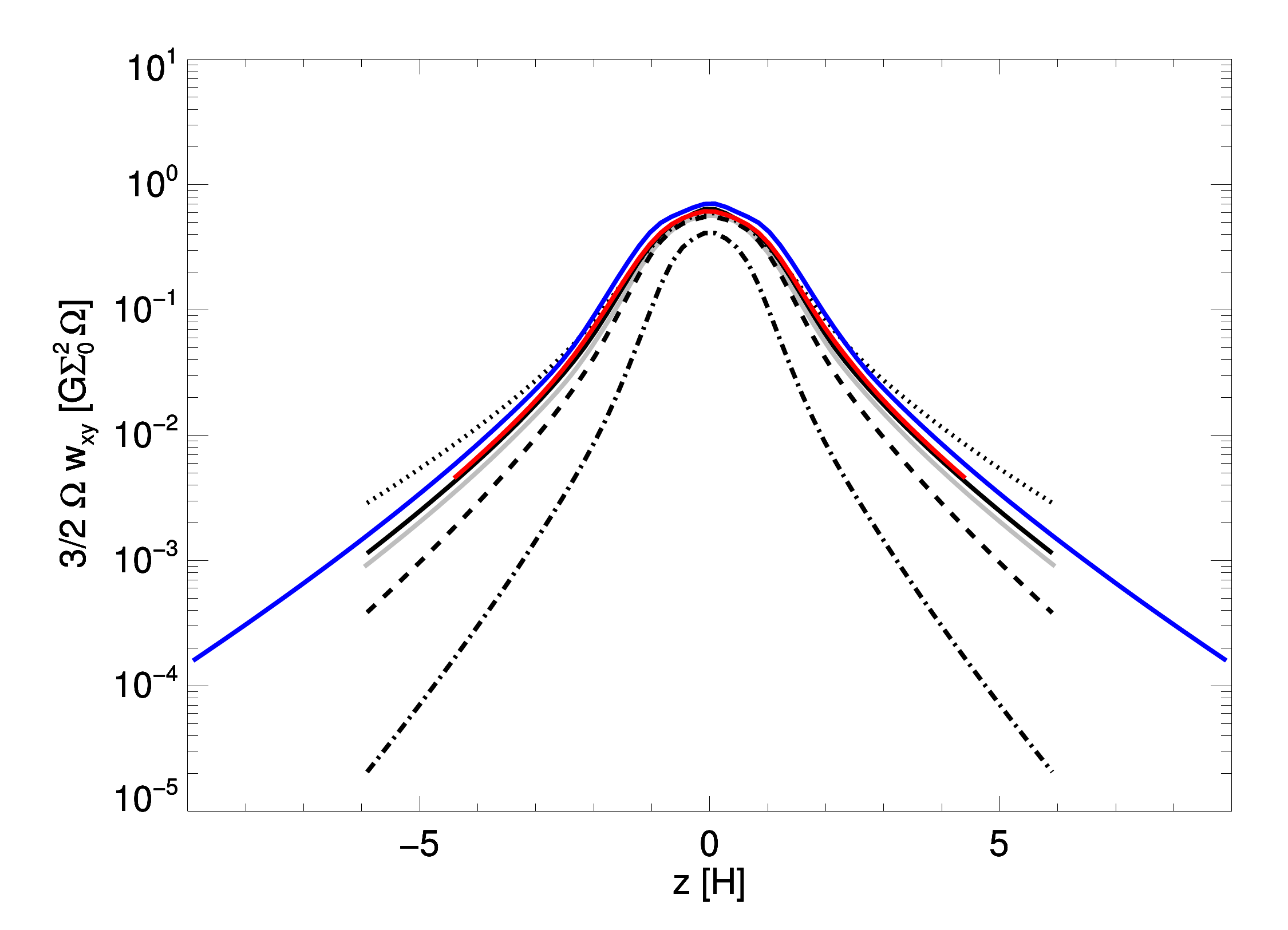}
\includegraphics[width=7cm]{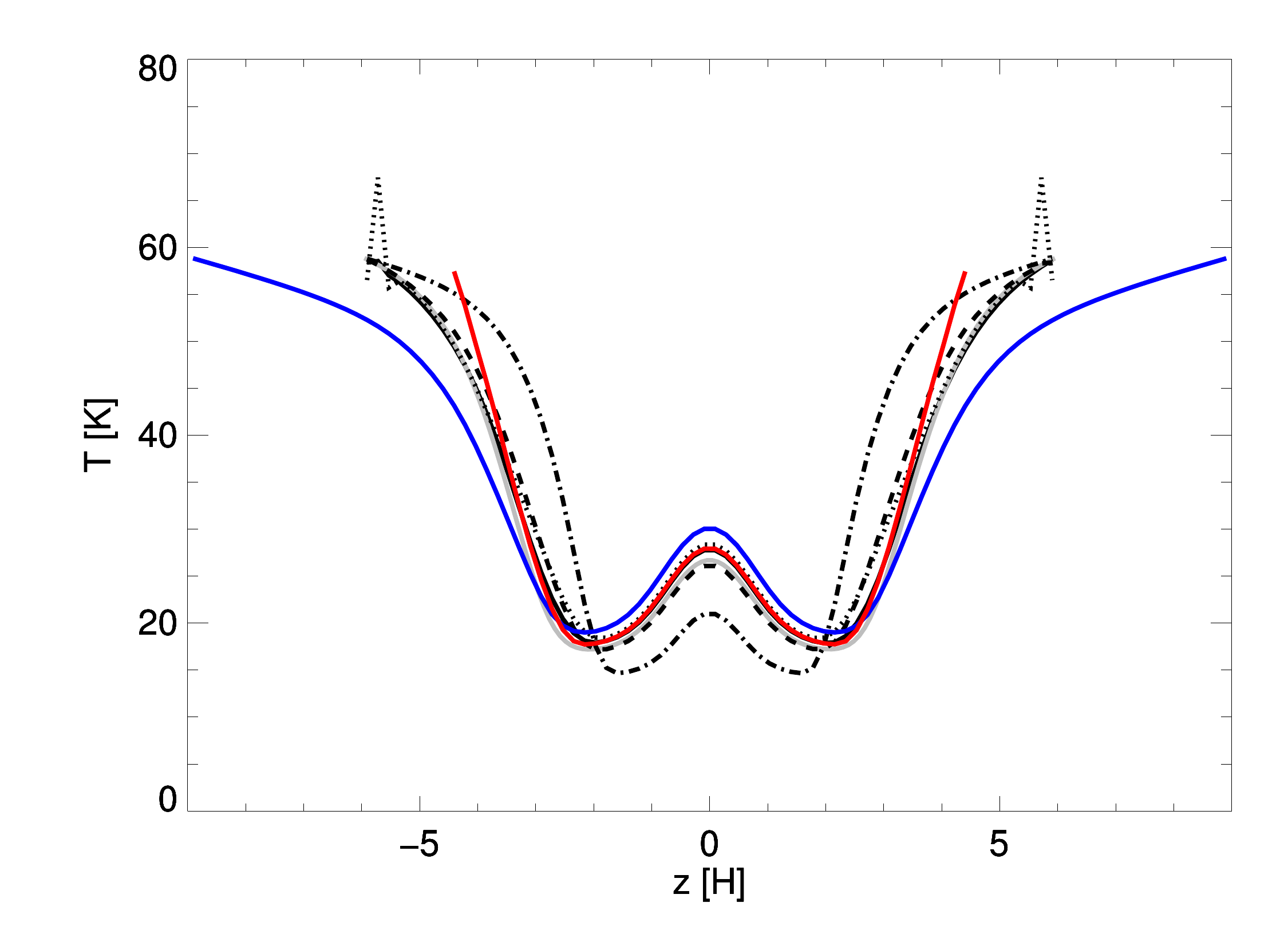}
\caption{Dependence of the profiles of the stress work $3/2\Omega\left<w_{xy}\right>$ (upper panel) and the gas temperature (lower panel), on the box size and the resolution. The black curves compare the horizontal box size; $L_x/H$ = $12$ (dash-dotted), $18$ (dashed), $24$ (fiducial; solid), and $36$ (dotted). The solid curves compare the vertical box size and the resolution for $L_x/H=24$; $L_z/H$ = $9$ (red), $12$ (fiducial = black, doubled resolution = grey), and $18$ (blue).}
\label{fig:vs_lx}
\end{figure}

In figure \ref{fig:vs_lx}, the upper panel shows dependence of the vertical profile of the stress work on the horizontal box size $L_x (= L_y)$, the vertical box size $L_z$, and the resolution. The solid curves correspond to the cases that have the same box size as the fiducial ($L_x/H=24$), but are different in the vertical box size (blue: $L_z/H=18$, red: $L_z/H=9$) or in the resolution (grey: two times finer than the fiducial). Since they show similar profiles, we see that the vertical box size or the resolution does not affect much the result. On the other hand, the black curves compare the cases that have different horizontal box sizes. The smallest box run (dash-dotted: $L_x/H=12$) shows consistently smaller stress work than others. The other cases ($L_x/H \ge 18$) show similar profiles near the midplane, but the stress work in the upper layers ($z/H \ge 2$) increases as $L_x/H$ is increased.

The different amounts of the stress work in the upper layers do not affect temperatures there since they are mainly determined by the irradiation heating. This can be seen by comparing the black curves in the lower panel, where the vertical profiles of gas temperature are compared using the same notation. On the other hand, comparing the solid curves in the lower panel, it is seen that the temperature profile in the upper layers is affected by the vertical box size. This is because we assume that the irradiation always begins to be absorbed at the vertical boundaries, and thus the temperatures there are virtually fixed at the same value ($\sim 60$ K). In summary, at least for the gravito-turbulence in the main body of the disc, which we are interested in, we may conclude that our results are firm.

\subsection{Dependence on the surface density}\label{discussion_dependence_sigma}
\begin{figure}
\includegraphics[width=7cm]{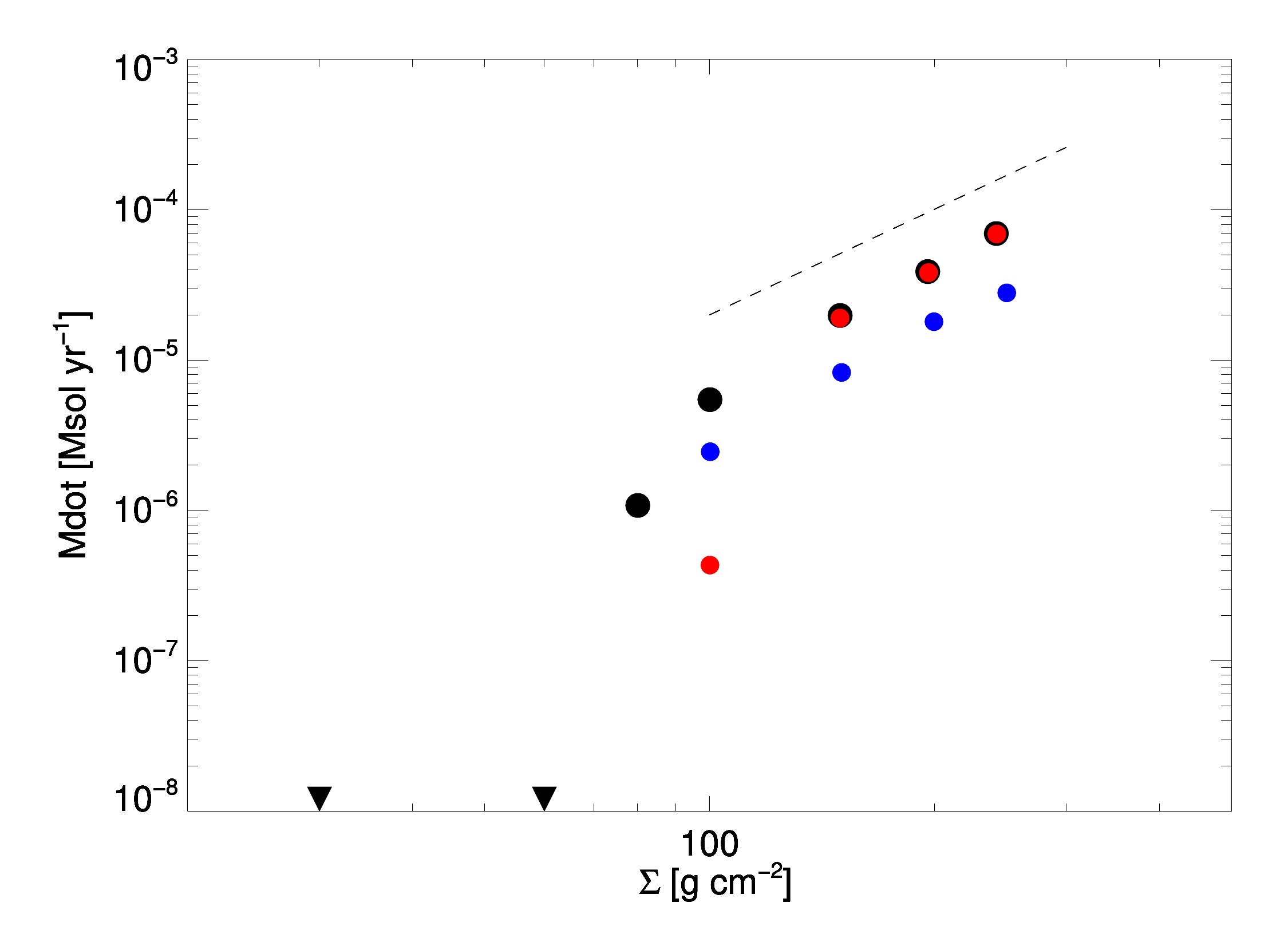}
\includegraphics[width=7cm]{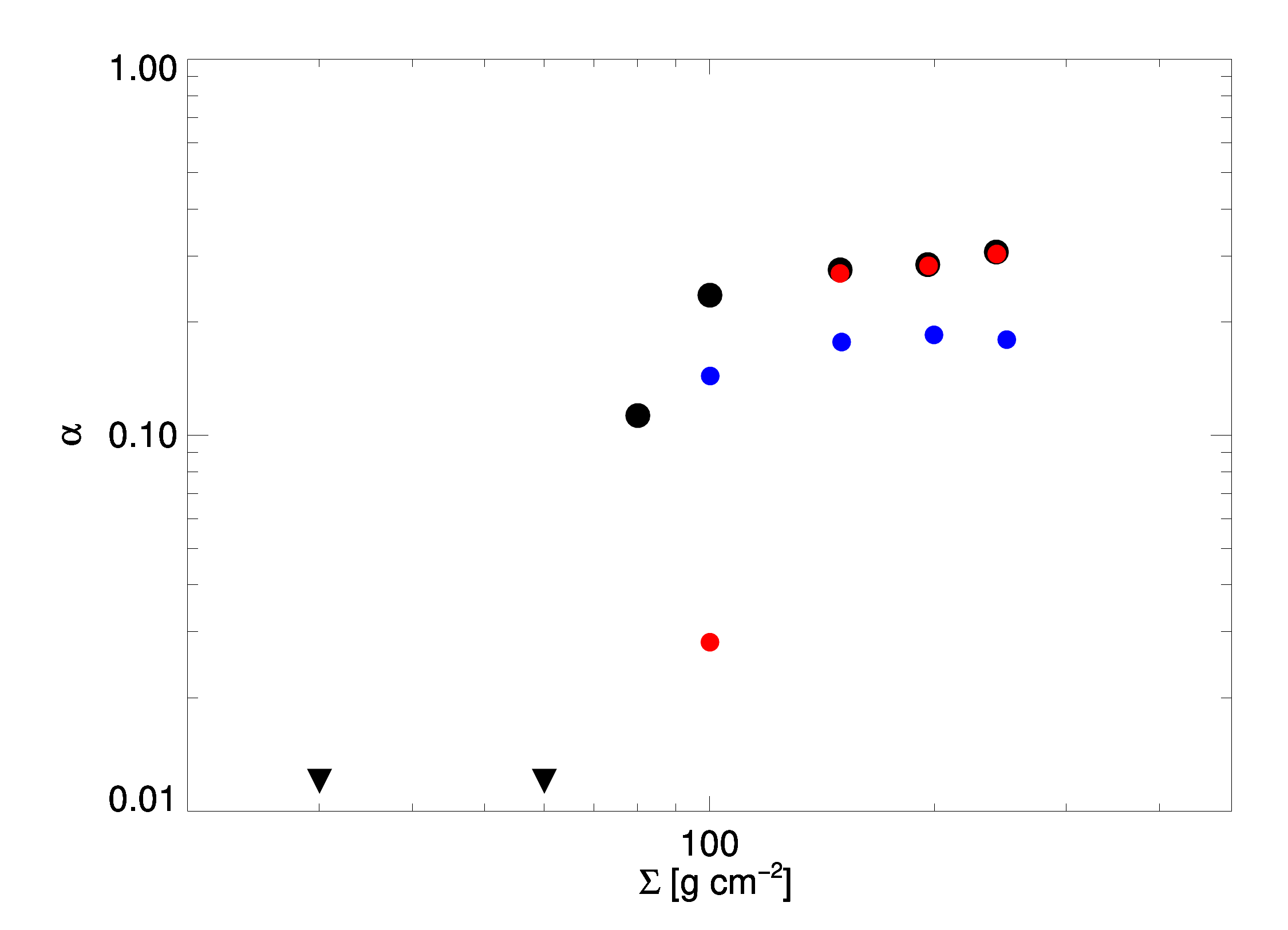}
\includegraphics[width=7cm]{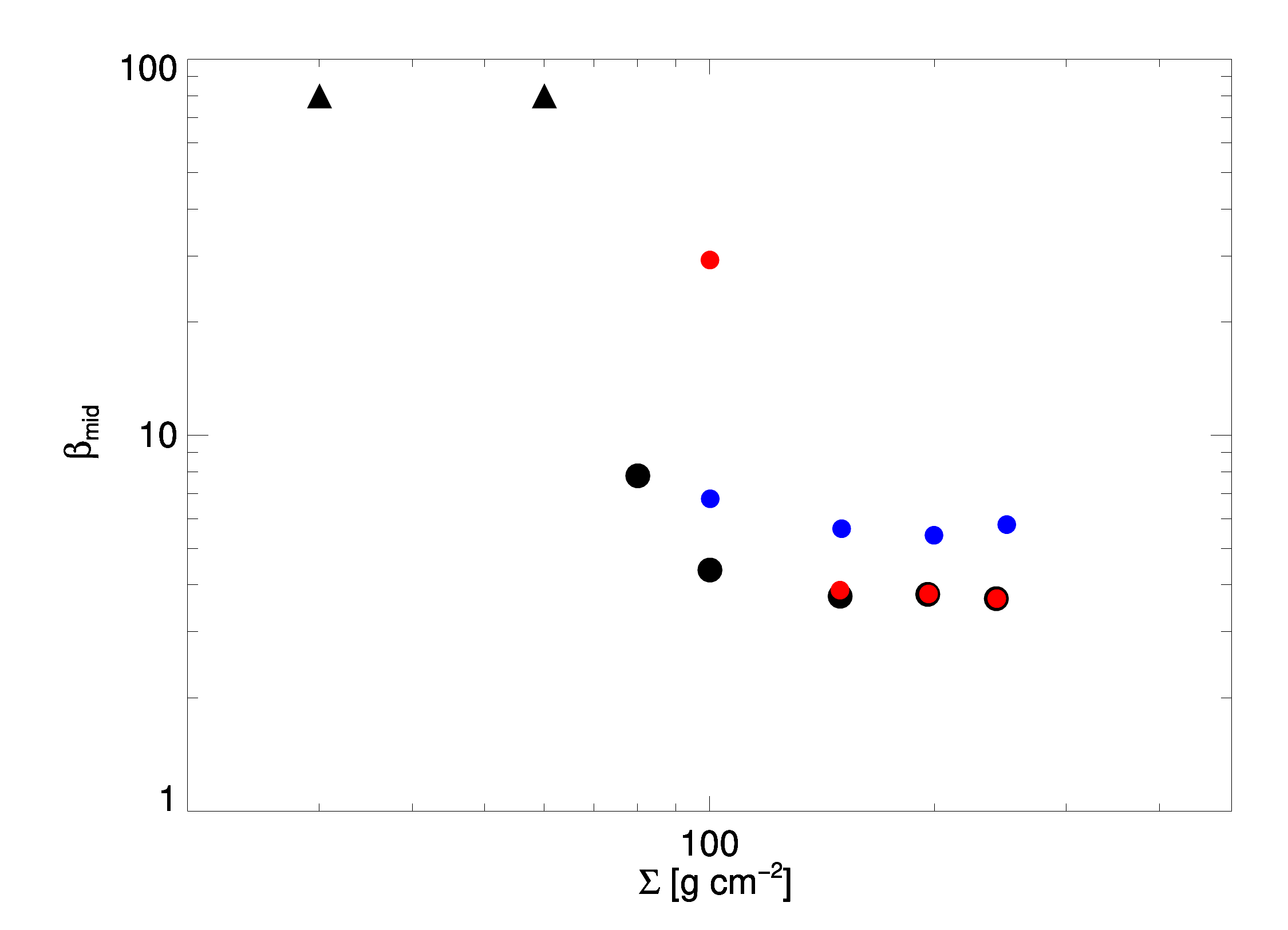}
\includegraphics[width=7cm]{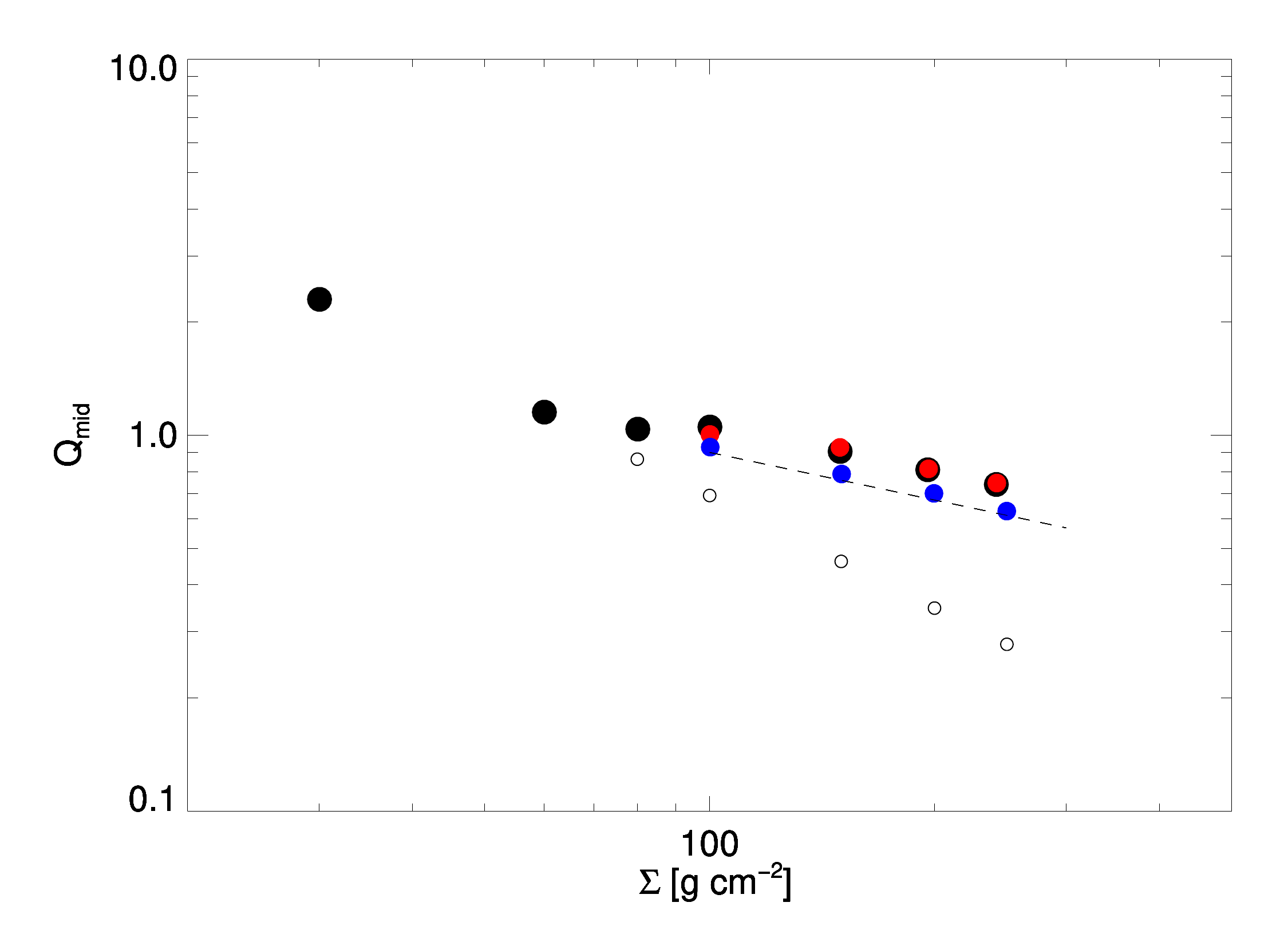}
\caption{Same as figure \ref{fig:sigma_vs}, but also plotted are runs using a half-sized box (blue) and runs with a larger grazing angle $\theta = 0.32$ (red).}
\label{fig:sigma_vs_cmp}
\end{figure}

In section \ref{sec:sigma}, we explored dependence of some key quantities on the surface density and found that i) $W_{xy}(\Sigma) \sim \Sigma^{7/3}$, ii) $\alpha$ and $\overline{\beta}_\text{mid}$ stays almost constant, and iii) $\overline{Q}_\text{mid}(\Sigma) \sim \Sigma^{-1/3}$, in the regime of gravito-turbulence. These properties seem robust since we confirm that they hold also for a different grazing angle ($\theta = 0.32$) and for a different box size ($L_x/H = 12$), as shown in figure \ref{fig:sigma_vs_cmp}.
(The turbulence is weaker when $\Sigma \le 100$ g cm$^{-2}$ for $\theta = 0.32$ since the irradiation affects the midplane temperature more when $\Sigma$ is smaller.) Although the smaller box size ($L_x/H = 12$) case shows smaller values in $W_{xy}$ (and thus smaller $\alpha$s and larger $\overline{\beta}_\text{mid}$s), which is a box size effect as discussed in the previous section, the slope is the same as others, and $\overline{Q}_\text{mid}(\Sigma)$ is even quantitatively similar to others.

\begin{figure}
\includegraphics[width=7cm]{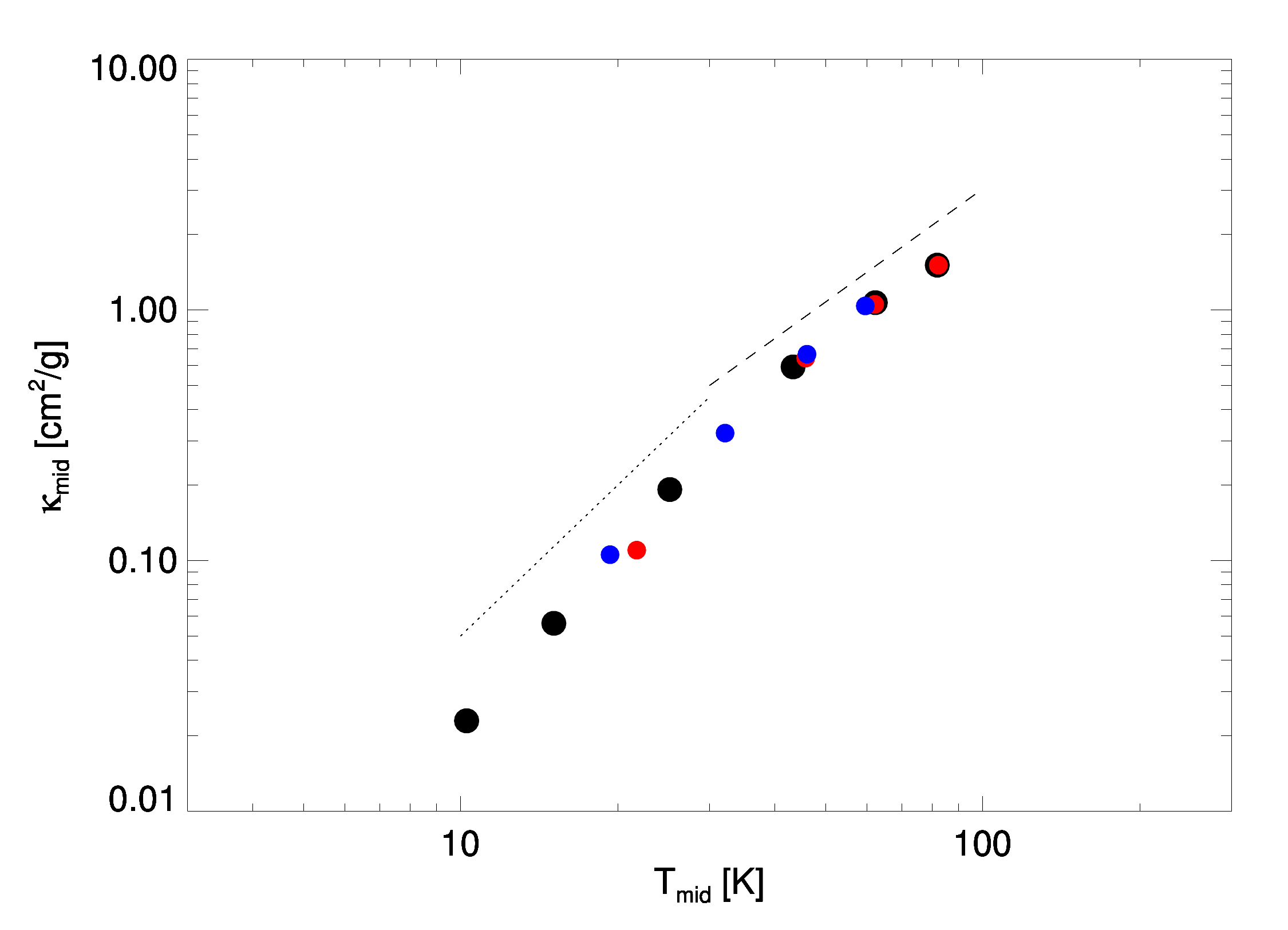}
\caption{The dependence of the midplane opacity on the midplane temperature. The notations are the same as in figure \ref{fig:sigma_vs_cmp}. The dashed line denotes $\sim T_\text{mid}^{3/2}$ {\color{\ccorange} while the dotted denotes $\sim T_\text{mid}^{2}$.}}
\label{fig:kappa_vs_tmp}
\end{figure}

{\color{\ccred}
Here, we examine what determines those scalings on $\Sigma$ found in our simulations. First, we note the weak scaling of the Toomre parameter on $\Sigma$. This is contrasted with the assumption of $Q \sim 1$ in the literature that discuss scalings in the gravito-turbulence \citep[c.f.][]{Levin:2007eza,Clarke:2009id,Rafikov:2009cda,Paardekooper:2012dta}. Given the weak scaling of the Toomre parameter on $\Sigma$, we have $\overline{\Left<T\Right>}_\rho \sim \Sigma^{4/3}$ from equation (\ref{eq:toomre}). Then, the cooling time $\overline{\beta}_\text{mid}$, approximated as the thermal energy content divided by the cooling rate by radiative diffusion,\footnote{In our simulations, the primary cooling mechanism is still radiative diffusion although advection also contributes to cooling as discussed in section \ref{sec:thermal_balance}.} can be written as a function of $\Sigma$ as
\begin{align}
\overline{\beta}_\text{mid} \sim \frac{\Sigma\overline{\Left<T\Right>}_\rho}{\overline{\Left<T\Right>}_\rho^4/\left(\overline{\Left<\kappa\Right>}_\rho\Sigma\right)} \sim \Sigma^{4b/3 - 2},
\end{align}
where we assumed $\overline{\Left<\kappa\Right>}_\rho \sim \overline{\Left<T\Right>}_\rho^b$.
We see that the rather constant $\overline{\beta}_\text{mid}$ ($\sim \Sigma^0$) found in our simulations corresponds to $b = 3/2$. This is roughly confirmed in figure \ref{fig:kappa_vs_tmp}, where the solutions of $\overline{\Left<T\Right>}_\rho \gtrsim 30$ K correspond to those following the scalings shown in figure \ref{fig:sigma_vs_cmp}.
Thus, the constant $\overline{\beta}_\text{mid}$ can be explained as a consequence of $\overline{Q}_\text{mid}(\Sigma) \sim \Sigma^{-1/3}$ and $\overline{\Left<\kappa\Right>}_\rho \sim \overline{\Left<T\Right>}_\rho^{3/2}$. 
We note that the same constant $\beta \sim \Sigma^0$  is also derived with the assumption of $Q\sim 1$ and $\kappa \sim T^2$ at given radius \citep{Paardekooper:2012dta}.

The scaling $\overline{\Left<\kappa\Right>}_\rho \sim \overline{\Left<T\Right>}_\rho^{3/2}$
for $\overline{\Left<T\Right>}_\rho \le 100$ K
found in our simulations is a result of mixing $\kappa_\text{R}(\rho,T) \sim T^2$ for $T \le 100$ K and $\kappa_\text{R}(\rho,T) \sim T^0$ for $T \ge 100$ K \citep[e.g.][see also the inset in the lower panel in figure \ref{fig:opacity_table}]{Johnson:2003kl}. This is because the temperature $T$ can be locally larger than $100$ K even when its horizontal average $\left<T\right>$ is less than $100$ K, as seen in the second and third panels in the right column in figure \ref{fig:snapshots_sigma}.

As discussed in \citet{Gammie:2001hv}, a thermal balance condition that the cooling rate equals to the stress work $3/2\Omega W_{xy} = 3/2\Omega\alpha\overline{\Left<p\Right>}$ relates the cooling time $\beta$ and $\alpha$ as
\begin{align}
\beta = \frac{\overline{\Left<p\Right>}/(\Gamma_1 -1)}{3/2\alpha\overline{\Left<p\Right>}} = \frac{1}{\alpha},
\end{align}
where $\Gamma_1 = 5/3$ is substituted. Therefore, $\overline{\beta}_\text{mid} \sim \Sigma^0$ readily leads to $\alpha \sim \Sigma^0$. Finally, the scaling of stress work (or mass accretion rate) can be derived as
\begin{align}
W_{xy} \sim \alpha\Sigma\overline{\Left<T\Right>}_\rho \sim \Sigma^{7/3}.
\end{align}

Note that the scalings shown in the above are obtained from our simulations at a single radius of $a = 50$ AU. To see whether our scalings are universal, we need to perform simulations at different $\Omega$s (i.e. different radii) (Hirose \& Shi, in preparation).
}

\subsection{Locality of angular momentum transport and shearing box}\label{sec:discussion_locality}
\begin{figure}
\includegraphics[width=7cm]{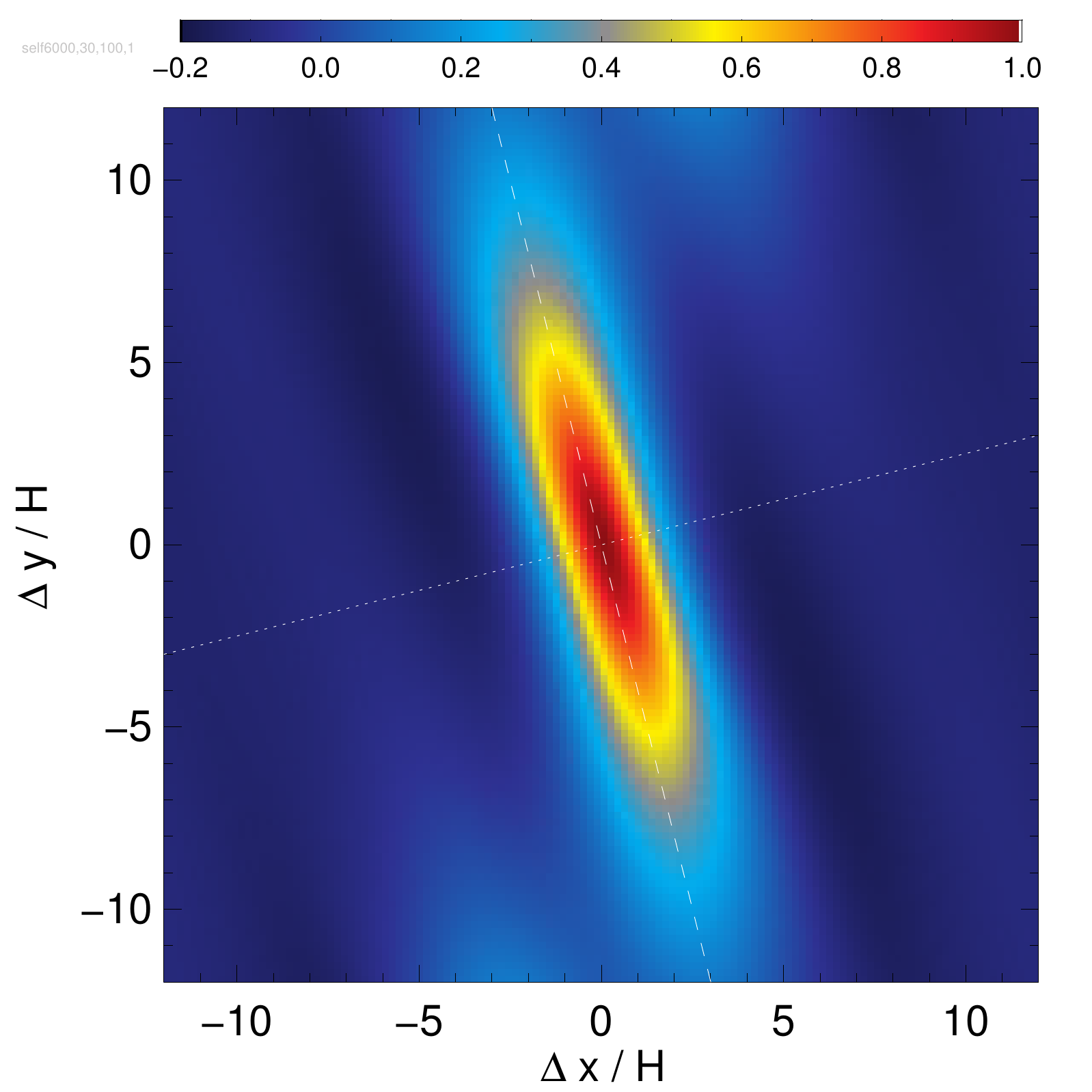}
\includegraphics[width=7cm]{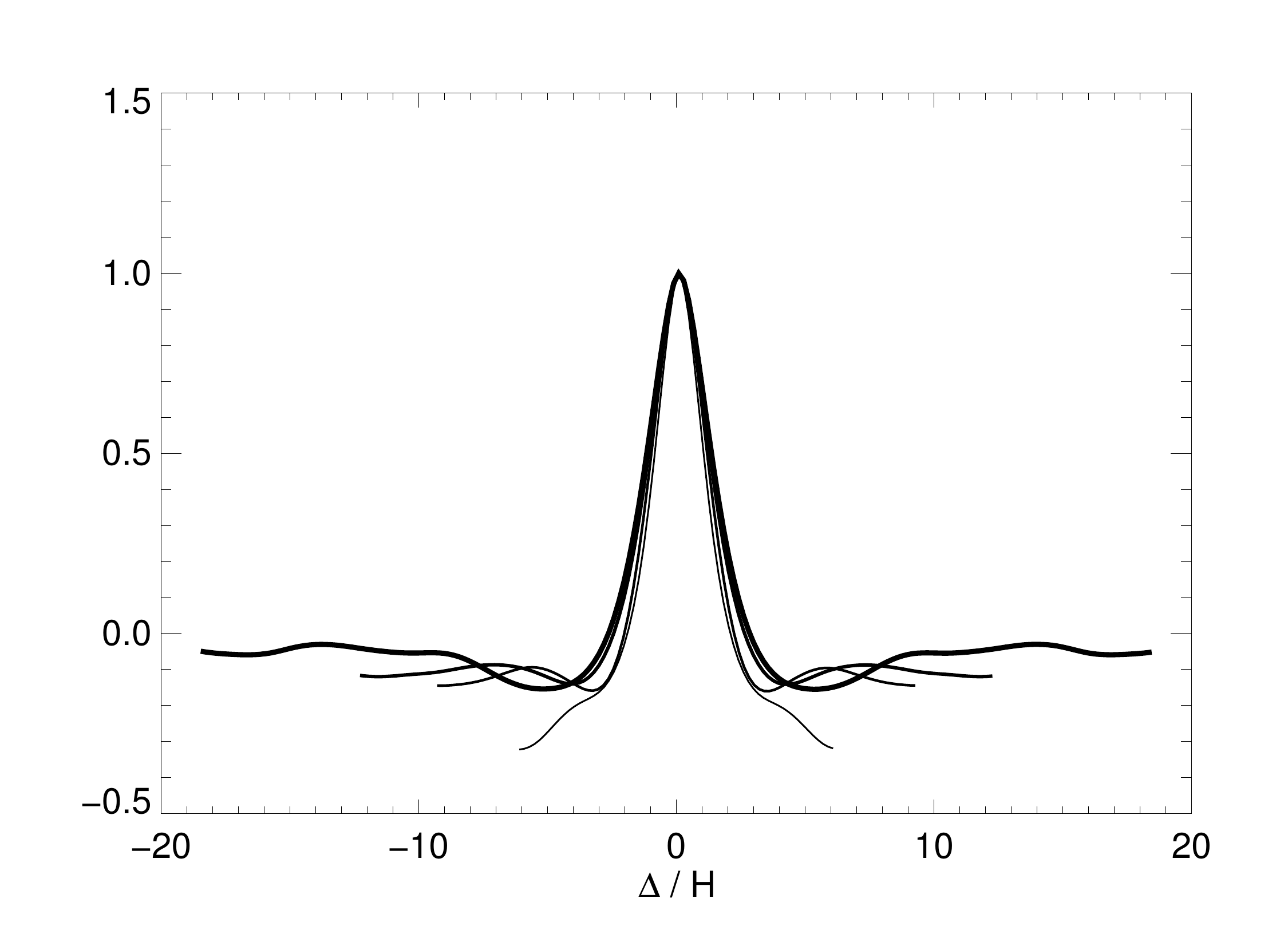}
\includegraphics[width=7cm]{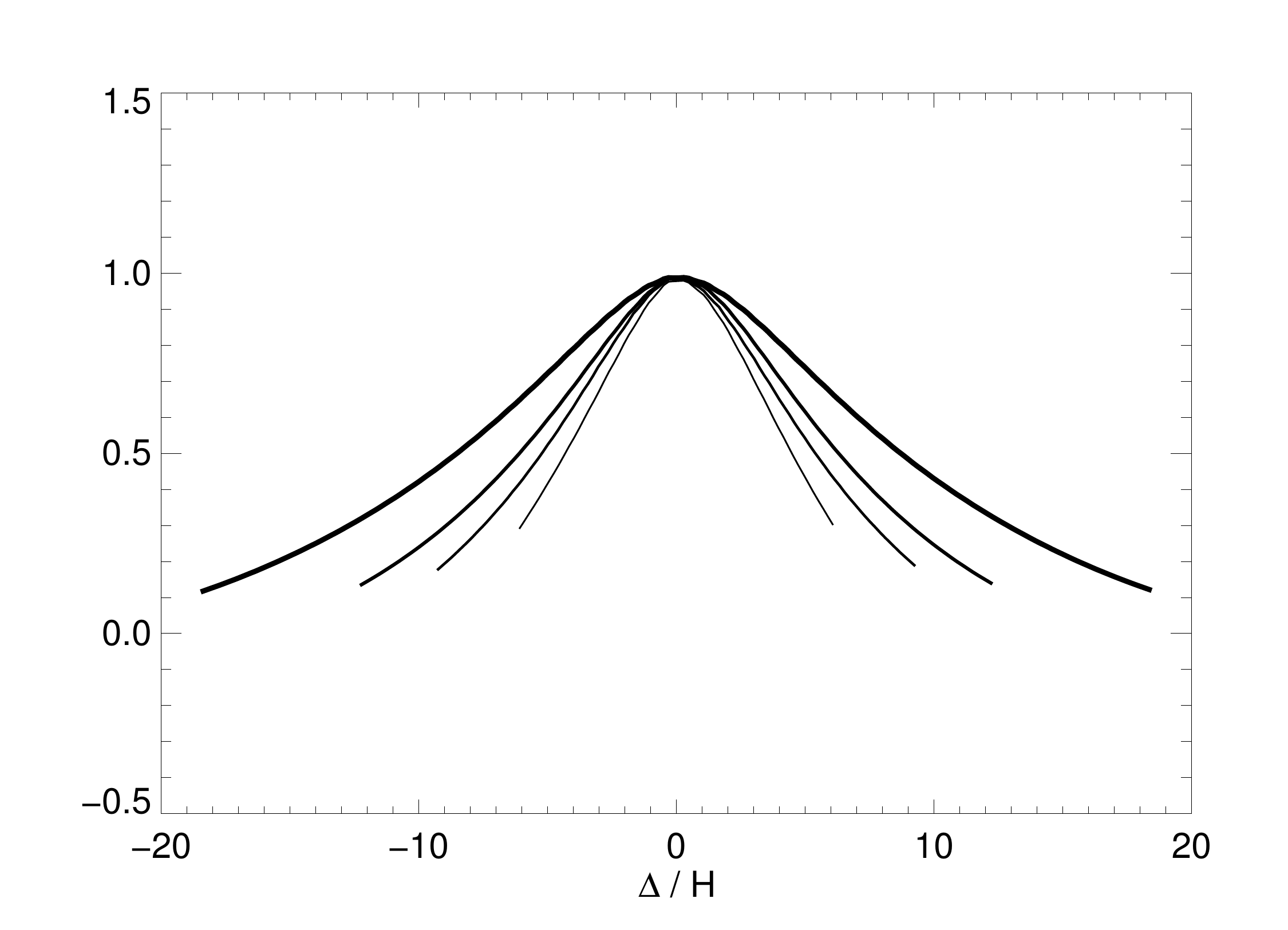}
\caption{Autocorrelation function of the density fluctuation $\delta\rho \equiv \rho - \left<\rho\right>$, obtained from 71 snapshots equally taken from $t=30$ to $t=100$ orbits. The top panel shows the midplane slice in the fiducial run. The middle and bottom panels shows one-dimensional cuts of the midplane slice along the long and short axes (such shown as the dashed and the dotted lines, respectively, in the top panel) for $L_x/H=12$, 18, 24, and 32. The thick the curve the larger the $L_x/H$.}
\label{fig:autocorrelation}
\end{figure}

Here, we examine the locality of angular momentum transport in the gravito-turbulence in our simulations.
As pointed out by \citet{Balbus:1999gw}, the shearing boundary conditions automatically exclude the global angular momentum transport by self-gravity. Therefore, to be self-consistent, the locality of angular momentum transport must be satisfied.

{\color{\ccred}
Physically, the locality of angular momentum transport is determined by the disc mass $M_\text{disc}$ in terms of the stellar mass $M_*$, which was studied by global simulations using the $\beta$ cooling prescription or more realistic cooling. Generally, the local description of angular momentum transport is valid when $M_\text{disc}/M_* \lesssim 0.25$ \citep{Lodato:2004js,2006ApJ...651..517B,Cossins:2009gt} while the non-local transport cannot be negligible when $M_\text{disc}/M_* \gtrsim 0.1$ \citep{Cai:2008bf,Harsono:2011ef} and becomes significant when $M_\text{disc}/M_* \gtrsim 0.5$ \citep{Lodato:2005cw,Forgan:2010kr}.
The disc mass does not appear in local shearing box simulations since they are independent from the global disc model. Here, we evaluate corresponding disc mass from the surface density $\Sigma$ simply as $M_\text{disk} \approx \pi a^2\Sigma$, as often done in the literature \citep[e.g. equation (2) in][]{Kratter:2016dwa}. Then, the ratio of the disc to stellar mass is $0.088$ in the fiducial case ($\Sigma = 100$ gcm$^{-2}$) and 0.27 in the maximum surface density case ($\Sigma = 300$ gcm$^{-2}$). Given the ambiguity in evaluating $M_\text{disc}$ from $\Sigma$, the non-local angular momentum transport might not be negligible in most of our cases and could be significant in the largest $\Sigma$ cases.
}

Next, we numerically examine the locality of angular momentum transport by computing a 3-dimensional auto-correlation function of density fluctuation \citep[e.g.][]{Gammie:2001hv},
\begin{align}
\xi(\bm{r}) \equiv \frac{\int \delta\rho(\bm{r}+\bm{r}')\delta\rho(\bm{r}')d\bm{r}'}{\int \delta\rho^2(\bm{r}')d\bm{r}'},
\end{align}
where $\delta\rho(x,y,z) = \rho(x,y,z) - \left<\rho\right>(z)$.
Figure \ref{fig:autocorrelation} shows the midplane slice $\xi(x,y,z=0)$ for the fiducial run ($L_x/H = 24$) as well as one-dimensional cuts along the long and short axes (such as shown in the plot of the midplane slice) for $L_x/H=12$, 18, 24, and 32. If we look at the cuts along the short axis, the locality looks fairly good except $L_x/H=12$. On the other hand, along the long axis, convergence cannot be seen although the locality improves as $L_x$ increases (in the sense that the minimum value decreases). Apparently, we need even a larger box than $L_x/H=32$ to assure the locality of angular momentum transport. 

However, we also have another limitation on the box size, which is the validity of using the local shearing box. In the fiducial run, the box, whose half size is $L_x/2 = 12H = 38$ AU, is located at the distance, $a = 50$ AU, from the central star. Therefore, increasing the box size even larger than the fiducial run ($L_x/H=24$) makes no sense in terms of the validity of using the local shearing box.

We can see this problem more quantitatively by taking the ratio of the wavelength of the fastest growing mode of axisymmetric GI $\lambda = 2\pi Qc_\text{s}/\Omega$ and the distance from the central star $a$, which scales as
\begin{align}
\frac{\lambda}{a} &= \frac{1}{a}2\pi Q\sqrt{\gamma\frac{R}{\mu}T}\sqrt{\frac{a^3}{GM_*}}\nonumber\\
&= 0.53 Q\left(\frac{\gamma}{\mu}\right)\left(\frac{T}{15\text{ K}}\right)^{\frac12}\left(\frac{a}{50\text{ AU}}\right)^{\frac12}\left(\frac{M_*}{M_\odot}\right)^{-\frac12}.\label{eq:lambda}
\end{align}
We would like to have this ratio as small as possible so that $\lambda \ll L_x \ll a$ is satisfied. Then, we can take a box whose size $L_x$ is much larger than $\lambda$ (so that the locality of angular momentum transport is satisfactory), but still smaller than the distance from the star $a$ (so that the locality of the shearing box is guaranteed). To do that, we need to make both $T$ and $a$ small. However, it would be difficult to lower the ratio $\lambda/a$ less than order of $0.1$ because the ratio only weakly depends on the two parameters (the exponents are both $\frac12$) and they are usually anti-correlated.
The bottom line is that the parameters that we chose for the fiducial run were the best we could do for both the locality of angular momentum transport and the validity of the shearing box. This problem was not explicitly mentioned in the previous studies of gravito-turbulence since they usually used dimensionless quantities or ignoring the irradiation.

\section{Summary}\label{sec:conclusion}

We explored the gravito-turbulence in irradiated protoplanetary discs using radiation hydrodynamics simulations. We used a stratified shearing box located at the radius of $a=50$ AU, which is irradiated by the central star of $T_*=4000$ K, $M_*=1M_\odot$, and $R_*=1R_\odot$. Under these conditions, we found that:
\begin{enumerate}
\item Gravito-turbulence is sustained for a finite range of the surface density, {\color{\ccorange} $80 \le \Sigma \le 250$ gcm$^{-2}$}.
The flow is laminar below the range and fragmentation occurs above the range.
\item In the regime of gravito-turbulence, the Toomre parameter $\overline{Q}_\text{mid}$ decreases monotonically from $\sim 1$ to $\sim 0.7$ as the surface density increases while an effective cooling time is rather constant at $\overline{\beta}_\text{mid}\sim ${\color{\ccorange}4}. 
\item In the gravito-turbulence, successive collisions of turbulent density waves contribute to hydrostatic balance via the dynamic pressure as well as thermal balance via the advection cooling. The turbulence dissipates through both shock heating and compressional heating.
\item The irradiation heating does not affect much the gravito-turbulence of the main body of the disc unless the grazing angle is as large as $\theta \ge 0.32${\color{\ccorange}, where the turbulence is orders of magnitude smaller in terms of $\alpha$}.
\item The simple cooling function used in the previous studies of gravito-turbulence does not approximate the realistic radiative transfer in the irradiated discs, regardless of the cooling time $\beta$.
\end{enumerate}

{\color{\ccred}The point (ii) indicates, for a fixed radius in protoplanetary discs, that there is a minimum Toomre parameter that can sustain the gravito-turbulence and that fragmentation is determined by the Toomre parameter \citep[c.f.][]{Takahashi:2016bx}.
As discussed in \citet{Kratter:2016dwa}, fragmentation may be driven by either cooling (by decreasing the temperature) or accretion (by increasing the surface density). The criterion found here, which is based on the surface density dependence, may apply to the latter type of fragmentation. On the other hand, the cooling time $\beta$ is expected to strongly depend on the radius \citep{Clarke:2009id,Clarke:2009gd,Paardekooper:2012dta}. Therefore, studies on the radial dependence will be required to address the criterion for the cooling-driven fragmentation criterion.
}


{\color{\ccred}We also discussed locality of angular momentum transport as well as locality of the simulation box in our simulations. The fact that both are not quite satisfactory indicates limitations on using shearing box simulations for gravito-turbulence in realistic  protoplanetary discs.}

\section*{Acknowledgments}
{\color{\ccred}We thank the anonymous referee for his/her useful comments to improve the manuscript.}
We thank Kengo Tomida, who kindly provided us with the EOS tables that were used in our simulations.
Numerical calculations were carried out partly on Cray XC30 at CfCA, National Astronomical Observatory of Japan, and partly on Cray XC40 at YITP in Kyoto University.
SH was supported by Japan JSPS KAKENH 15K05040 and the joint research project of ILE, Osaka University.
JS was supported in part by the National Science Foundation under grant PHY-1144374, "A Max-Planck/Princeton Research Center for Plasma Physics" and grant PHY-0821899, "Center for Magnetic Self-Organization".








\appendix

\section{Irradiation heating rate}\label{sec:irradiation}
In this section, we describe how we compute the irradiation heating rate $q_\text{irr}$ by a simple ray tracing method. We inject a single ray into every single cell located at the top and bottom surfaces of the simulation box along the direction $\bm{\Omega} = (\cos\theta,0,\mp\sin\theta)$. A single ray represents the irradiation flux entering the box through the surface of a single cell. Then the initial energy flux [erg s$^{-1}$] of each ray is $F_\text{irr}\sin\theta\Delta S$, where $F_\text{irr} = \sigma_\text{B}T_*^4({R_*}/{a})^2$ and $\Delta S$ is the surface area of the cell.

As a ray travels through the box, it is attenuated by absorption, which in turn heats the gas.\footnote{We assume that a ray that escapes from the box through a radial boundary re-enters the box through the opposite side of the box.} When it enters $n$-th cell, its energy flux [erg s$^{-1}$] is written as
\begin{align}
  f^{n} = (F_\text{irr}\Delta S\sin\theta) \prod_{n'=1}^{n-1}e^{-\rho\kappa_\text{P$_*$}^{n'}\Delta l^{n'}},
\end{align}
where $\Delta l^{n'}$ is the path length within the $n'$-th cell, and $\kappa_\text{P$_*$}^{n'} = \kappa_\text{P$_*$}^{n'}(\rho,T)$ is the star-temperature Planck-mean opacity in the cell. Here, $\rho\kappa_\text{P$_*$}$ is fixed for simplicity. The rate at which the gas in the cell is heated by the ray is then computed as
\begin{align}
  q_\text{irr}^{n} = \dfrac{f^{n-1} - f^{n}}{\Delta S\Delta z}
  = \frac{F_\text{irr}\sin\theta}{\Delta z}(1 - e^{-\rho\kappa_\text{P$_*$}^{n}\Delta l^{n}})\prod_{n'=1}^{n-1}e^{-\rho\kappa_\text{P$_*$}^{n'}\Delta l^{n'}},
\end{align}
where $\Delta z$ is the thickness of the cell. The (total) irradiation heating rate $q_\text{irra}$ in the cell is computed as a sum of contributions by the all rays considered. 

\section{Equation of self-gravitational energy}\label{sec:selfgravitational_energy}
The self-gravitational energy equation is written as \citep[e.g.][]{Balbus:1999gw}
\begin{align}
  &\frac{\p}{\p t}\left(\rho\Phi + \frac{1}{8\pi G}|\nabla\Phi|^2\right) +
  \nabla\cdot\left(\rho\Phi\bm{v}
  - \dfrac{\nabla\Phi}{4\pi G}\dfrac{\p\Phi}{\p t}\right) = \rho\bm{v}\cdot\nabla\Phi,
\end{align}
or equivalently
\begin{align}
  &\frac{\p}{\p t}\left(\rho\Phi + \frac{1}{8\pi G}|\nabla\Phi|^2\right) +
  \nabla\cdot\left(\rho\Phi\bm{v} + v_y\dfrac{\nabla\Phi}{4\pi G}\dfrac{\p\Phi}{\p y}
  - \dfrac{\nabla\Phi}{4\pi G}\dfrac{D\Phi}{Dt}\right) = \rho\bm{v}\cdot\nabla\Phi,\\
&\quad \dfrac{D}{Dt} \equiv \dfrac{\p}{\p t} + v_y\dfrac{\p}{\p y}.
\end{align}
In a steady state, the time and horizontally-averaged version of the equation is written as
\begin{align}
0 = -\overline{\left<\dfrac{\p}{\p x}\left(\rho\Phi v_x + \dfrac{g_x}{4\pi G}\dfrac{\p\Phi}{\p t}\right)\right>} - \overline{\left<\frac{\p}{\p z}\left(\rho\Phi v_z + \dfrac{g_z}{4\pi G}\dfrac{\p\Phi}{\p t}\right)\right>}
  + \overline{\left<\rho\bm{v}\cdot\nabla\Phi\right>}.\label{eq:appendix_energy}
\end{align}

Applying the shearing periodic boundary conditions, the first term in RHS is rewritten as \citep[see][]{Balbus:1999gw}
\begin{align}
-\overline{\left<\dfrac{\p}{\p x}\left(\rho\Phi v_x + \dfrac{g_x}{4\pi G}\dfrac{\p\Phi}{\p t}\right)\right>} &= -\overline{\left<\dfrac{\p}{\p x}\left(\rho\Phi v_x + v_y\dfrac{g_xg_y}{4\pi G} + \dfrac{g_x}{4\pi G}\dfrac{D\Phi}{Dt}\right)\right>} \nonumber\\
  &=-\overline{\left<\dfrac{\p}{\p x}\left(v_y\dfrac{g_xg_y}{4\pi G}\right)\right>} \nonumber\\
  &=-\frac{\left.v_y\frac{g_xg_y}{4\pi G}\right|_{x_+} - \left.v_y\frac{g_xg_y}{4\pi G}\right|_{x_-}}{L_x}\nonumber\\
  &=\frac32\Omega\left.\frac{g_xg_y}{4\pi G}\right|_{x^+}\nonumber\\
  &=\frac32\Omega\overline{\left<\dfrac{g_xg_y}{4\pi G}\right>},
\end{align}
where, $|_{x^\pm}$ denotes averaging in the $y$ direction at each height on the plane of $x=x_\pm$.
The last equality would be validated when the box size is large enough compared with the typical length of the turbulence.

Thus, the equation \ref{eq:appendix_energy} is written as
\begin{align}
0 = \frac32\Omega\overline{\left<\frac{g_xg_y}{4\pi G}\right>}
  - \overline{\left<\frac{\p}{\p z}\left(\rho\Phi v_z + \dfrac{g_z}{4\pi G}\dfrac{\p\Phi}{\p t}\right)\right>} + \overline{\left<\rho\bm{v}\cdot\nabla\Phi\right>}.
\end{align}

\section{Evolution of the gravitationally bounded clump}\label{sec:clump}
\begin{figure}
\includegraphics[width=7cm]{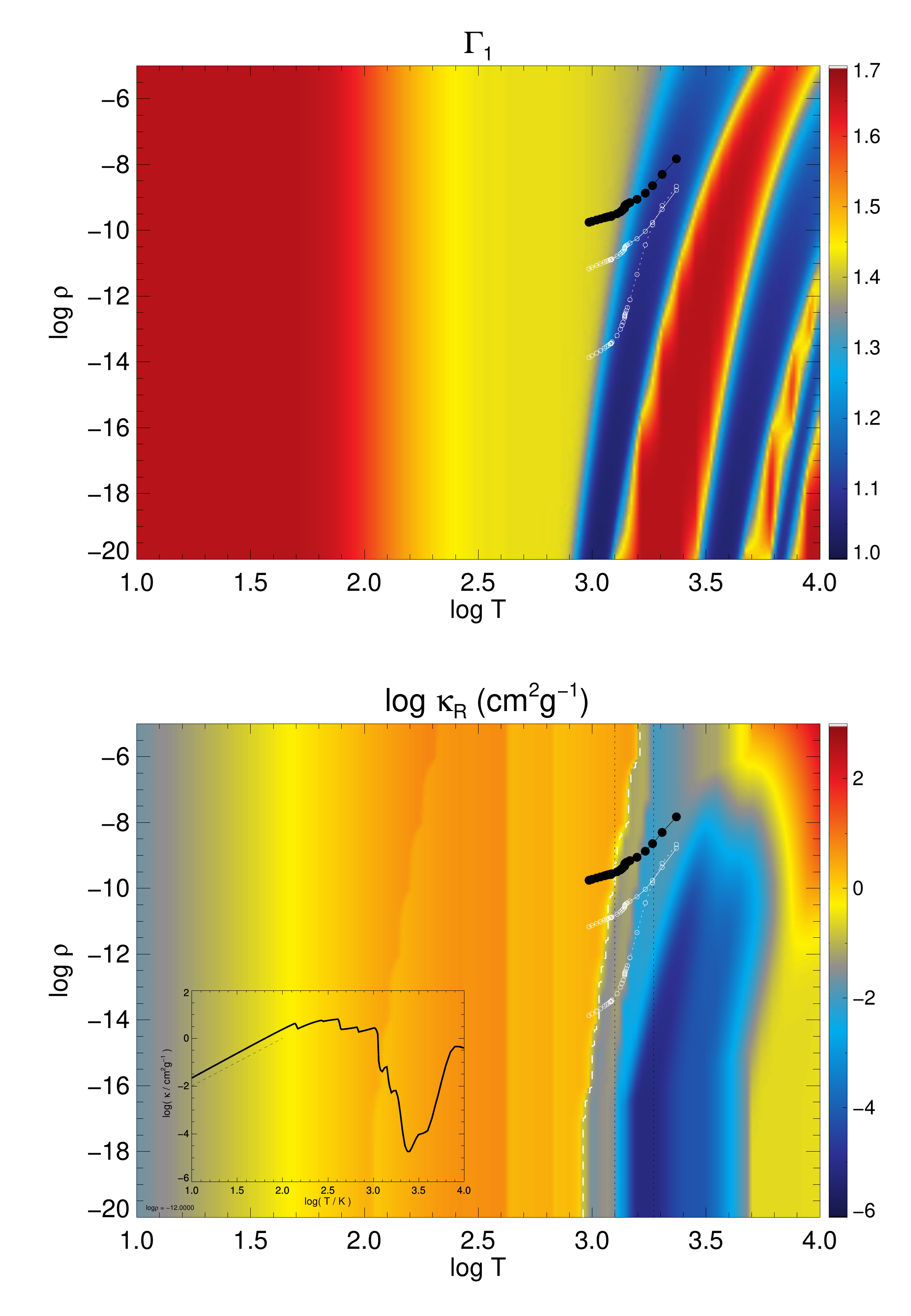}
\caption{Time evolution tracks of density (black), pressure (multiplied by $10^{-12}$) (white solid), and $\rho^{\Gamma_1}$ (white dotted) vs. gas temperature of the cell at the centre of the gravitationally-bounded clump in the case of $\Sigma = 300$ g cm$^{-2}$. The background colour shows the adiabatic exponent $\Gamma_1(T,\rho)$ (upper) and the Rosseland mean opacity $\kappa_\text{R}(T,\rho)$ (lower). In the lower panel, the vertical dashed lines denote the temperature range in which pressure increase is much less than that expected in the adiabatic evolution ($\sim \rho^{\Gamma_1}$) while the white dashed line denotes the dust sublimation temperatures. The inset in the lower panel shows a cross section, $\log\kappa_\text{R}(\log T,\log\rho=-12)$, where the dashed line denotes $\sim T^2$.
    }
\label{fig:opacity_table}
\end{figure}

{\color{black}
In this section, we describe the rapid collapse of a gravitationally-bounded clump observed in the case of $\Sigma = 300$ g cm$^{-2}$.
In figure \ref{fig:opacity_table}, we plotted the evolutionary track of $(T(t),\rho(t))$ of the cell at the centre of the clump, on the colour contour of the Rosseland-mean opacity $\kappa_\text{R}(T,\rho)$ as well as on that of the adiabatic exponent $\Gamma_1(T,\rho)$, from $t = 2.07$ to $t = 2.39$ orbits. The track evolves from the lower left (low $T$ and low $\rho$) to the upper right (high $T$ and high $\rho$).
As seen from intervals between the marks, which are placed for every 0.01 orbits on the track, the density increase is accelerated when the adiabatic exponent $\Gamma_1$ decreases from $4/3$ to $\sim 1.1$ due to dissociation of H$_2$ molecules. This is exactly the same physical process as the first core collapse in the star formation \citep[e.g.][]{Tomida:2012fb}.

We can learn more about the collapse by comparing the two extra tracks plotted in white in the same figure; one is $(T(t),p(t)\times10^{-12})$ and the other is $(T(t),\rho(t)^{\Gamma_1(t)})$. We see that the pressure increase is actually much smaller than that expected in the adiabatic evolution ($\sim \rho^{\Gamma_1}$) for $3.1 \le \log T \le 3.25$. This means that the cooling is very effective for that temperature range. Since the beginning of the range $\log T \sim 3.1$ exactly corresponds to the dust sublimation temperature, the effective cooling should come from large reduction of Rosseland-mean opacities due to the dust sublimation. Therefore, the rapid collapse of the clump was first triggered by the dust sublimation, and then followed up by the dissociation of H$_2$ molecules.
}

\bsp	
\label{lastpage}
\end{document}